\pdfoutput=1
\newcommand*{\ATLASLATEXPATH}{latex/}
\documentclass[cernpreprint, texlive=2016, UKenglish]{\ATLASLATEXPATH atlasdoc}

\usepackage{\ATLASLATEXPATH atlaspackage}
\usepackage{\ATLASLATEXPATH atlasbiblatex}

\usepackage{\ATLASLATEXPATH atlasphysics}

\graphicspath{{logos/}{figures/}}

\usepackage{verbatim}
\usepackage{tikz}
\usepackage{rotating}
\usetikzlibrary{arrows,shapes}
\usetikzlibrary{automata,positioning}

\newcommand{\CalibLineSize}{0.75mm}
\newcommand{\PhysLineSize}{1.3mm}

\tikzset{
    >=stealth,
    vertex/.style = {
        circle,
        fill            = black,
        outer sep = 2pt,
        inner sep = 1pt,
    }
}

\AtlasTitle{Operation and performance of the ATLAS Tile Calorimeter in Run~1}

\AtlasAbstract{%
The Tile Calorimeter is the hadron calorimeter covering the central region of the ATLAS experiment at the Large Hadron Collider. Approximately 10\,000 photomultipliers collect light from scintillating tiles acting as the active material sandwiched between slabs of steel absorber. This paper gives an overview of the calorimeter's performance during the years 2008--2012 using cosmic-ray muon events and proton--proton collision data at centre-of-mass energies of 7 and 8\,TeV with a total integrated luminosity of nearly 30\,fb$^{-1}$. The signal reconstruction methods, calibration systems as well as the detector operation status are presented. The energy and time calibration methods performed excellently, resulting in good stability of the calorimeter response under varying conditions during the LHC Run~1. Finally, the Tile Calorimeter response to isolated muons and hadrons as well as to jets from 
proton--proton collisions is presented. The results demonstrate excellent performance in accord with specifications mentioned in the Technical Design Report.
}

\AtlasRefCode{TCAL-2017-01}

\PreprintIdNumber{CERN-EP-2018-062}

\AtlasDate{\today}

\AtlasJournal{EPJC}
\AtlasJournalRef{Eur.\ Phys.\ J.\ C78 (2018) 987}
\AtlasDOI{10.1140/epjc/s10052-018-6374-z}

\AtlasCoverSupportingNote{Supporting note for the Tile Calorimeter Run~1 paper}{https://cds.cern.ch/record/2302787}
\AtlasCoverEgroupEditors{atlas-tcal-2017-01-contact-editors@cern.ch}

\AtlasCoverEgroupEdBoard{atlas-tcal-2017-01-editorial-board@cern.ch}

\hypersetup{pdftitle={ATLAS document},pdfauthor={The ATLAS Collaboration}}

\begin{document}

\maketitle

\tableofcontents

\clearpage

\section{Introduction}\label{sec:intro}

ATLAS~\cite{bib:detectorPaper} is a general-purpose detector designed to reconstruct events from colliding hadrons at the Large Hadron Collider (LHC)~\cite{bib:LHCDetPaper}. 
The hadronic barrel calorimeter system of the ATLAS detector is formed
by the Tile Calorimeter (TileCal),
which provides essential input to the measurement of the jet energies
and to the reconstruction of the missing transverse momentum.
The TileCal, which surrounds the barrel electromagnetic calorimeter, consists of tiles of plastic scintillator regularly spaced between low-carbon steel absorber plates. Typical thicknesses in one period are 3\,mm of the scintillator and 14\,mm of the absorber parallel to the colliding beams' axis, with the steel:scintillator volume ratio being 4.7:1. The calorimeter is divided into three longitudinal segments; one central long barrel (LB) section with 5.8\,m in length ($|\eta| < 1.0$), and two extended barrel (EB) sections ($0.8 < |\eta| < 1.7$) on either side of the barrel each 2.6\,m long.\footnote{ATLAS uses a right-handed coordinate system with its origin at the nominal interaction point (IP) in the centre of the detector and the $z$-axis along the beam pipe. The $x$-axis points from the IP to the centre of the LHC ring, and the $y$-axis points upward. Cylindrical coordinates $(r,\phi)$ are used in the transverse plane, $\phi$ being the azimuth angle around the $z$-axis. The pseudorapidity is defined in terms of the polar angle $\theta$ as $\eta=-\ln\tan(\theta/2)$.}
Full azimuthal coverage around the beam axis is achieved with 64 wedge-shaped modules, each covering $\Delta\phi = 0.1$ radians. 
The Tile Calorimeter is located at an inner radial distance of 2.28\,m
from the LHC beam-line, and has three radial layers with depths of
1.5, 4.1, and $1.8\lambda$ ($\lambda$ stands for the nuclear interaction length\footnote{Nuclear interaction length is defined as the mean path length to reduce the flux of relativistic primary hadrons to a fraction 1/e.}) for the LB, and 1.5,
2.6, and $3.3\lambda$ for the EB\@. The amount of material in front of the TileCal corresponds to $2.3\lambda$ at $\eta=0$~\cite{bib:detectorPaper}. 
A detailed description of the ATLAS
TileCal is provided in a dedicated Technical Design
Report~\cite{bib:TileTDR}; the construction, optical instrumentation and installation into the ATLAS detector are described in Refs.~\cite{bib:MechanicalInstallation,bib:tileOpticalInstrumentation}.

The TileCal design is driven by its ability to reconstruct hadrons, jets, and missing transverse momentum within the physics programme intended for the ATLAS experiment.
For precision measurements involving the reconstruction of jets, the
TileCal is designed to have a stand-alone energy resolution for jets of $\sigma/E = 50\%/\sqrt{E \mathrm{(GeV)}} \oplus 3\%$ \cite{bib:detectorPaper, bib:TileTDR}. 
To be sensitive to the full range of energies expected in the LHC lifetime, the response is expected to be linear within 2\% for jets up to 4\,TeV.
Good energy resolution and calorimeter coverage are essential for precise
missing transverse momentum reconstruction. A special Intermediate Tile Calorimeter (ITC) system is installed between the LB and EB to correct for energy losses in the region between the two calorimeters.

This paper presents the performance of the Tile Calorimeter during the first phase of LHC operation. Section~\ref{sec:ExperimentalSetup} describes the experimental data and simulation used throughout the paper.
Details of the online and offline signal reconstruction are provided in Section~\ref{sec:SignalReco}.
The calibration and monitoring of the approximately 10\,000 channels and data acquisition system are described in Section~\ref{sec:calib}. 
Section~\ref{sec:DQ} explains the system of online and offline data quality checks applied to the hardware and data acquisition systems.
Section~\ref{sec:InSitu} validates the full chain of the TileCal
calibration and reconstruction using events with single muons and
hadrons. The performance of the calorimeter is summarised in Section~\ref{sec:Conclusion}.

\subsection{The ATLAS Tile Calorimeter structure and read-out electronics}
\label{sec:Tilecal}
\begin{figure}[htp]                                                                                      
\centering
\includegraphics[width=1.00\textwidth]{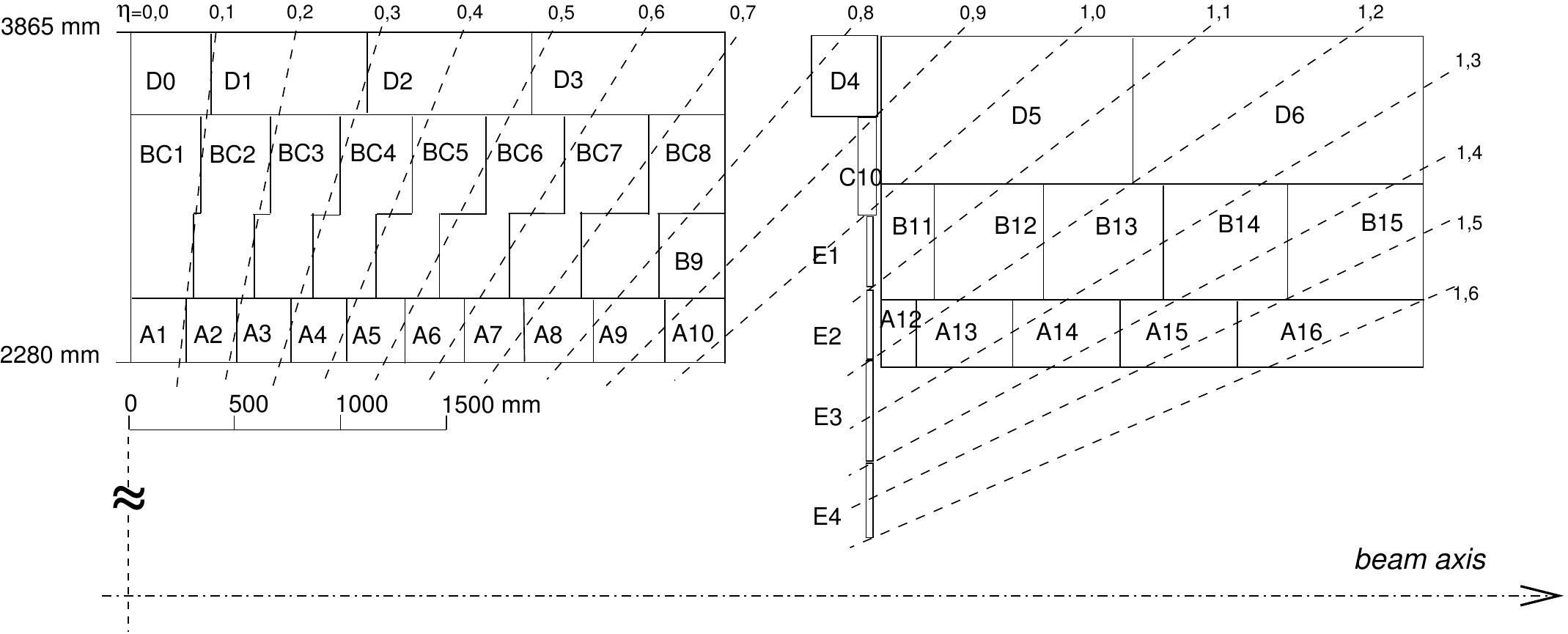}
\caption{The layout of the TileCal cells, denoted by a letter (A to E) plus an integer number. The A-layer is closest to the beam-line. The naming convention is repeated on each side of $\eta = 0$. \label{fig:cellLayout}}
\end{figure}

The light generated in each plastic scintillator is collected at two
edges, and then transported to photomultiplier tubes (PMTs) by
wavelength shifting (WLS) fibres~\cite{bib:tileOpticalInstrumentation}. The read-out cell geometry is
  defined by grouping the fibres from individual tiles on the
  corresponding PMT\@. A typical cell is read out on each side (edge)
  by one PMT, each corresponding to one channel. The dimensions of the
cells are $\Delta\eta \times \Delta\phi = 0.1 \times 0.1$ in the first
two radial layers, called layers A and BC (just layer B in the EB),
and $\Delta\eta \times \Delta\phi = 0.2 \times 0.1$ in the third layer,
referred to as layer D. The projective layout of cells and naming
convention are shown in Figure~\ref{fig:cellLayout}. The so-called ITC
cells (D4, C10 and E-cells) are located between the LB and EB, and
provide coverage in the range $0.8 < |\eta| < 1.6$. 
Some of the C10 and D4 cells have reduced thickness or special geometry 
in order to accommodate services and read-out electronics for other ATLAS 
detector systems~\cite{bib:TileTDR,bib:tileReadiness}.
%
The gap (E1--E2) and crack (E3--E4) cells
are only composed of scintillator and
are exceptionally read out by only one PMT\@. For Run~1, eight crack
scintillators were removed per side, to allow for routing of fibres
for 16 Minimum Bias Trigger Scintillators (MBTS), used to trigger on
events from colliding particles, as well as to free up the  necessary electronics channels for read-out of the MBTS\@. The MBTS scintillators are also read out by the TileCal EB electronics.

The PMTs and front-end electronics are housed in a steel girder at the outer radius of each module in 1.4\,m long aluminium units that can be fully extracted while leaving the remaining module in place, and hence are given the name of electronics drawers.  Each drawer holds a maximum of 24 channels, two of which form a super-drawer.
There are nominally 45 and 32 active channels per super-drawer in the LB and EB, respectively. Each channel consists of a unit called a PMT block, which contains the light-mixer, PMT tube and voltage divider, and a so-called 3-in-1 card~\cite{bib:CISRef, bib:pmtBlock}.
This card is responsible for fast signal shaping in two gains (with a bi-gain ratio of 1:64), the slow integration of the PMT signal, and provides an input for a charge injection calibration system. 

The maximum height of the analogue pulse in a channel is proportional to the amount
of energy deposited by the incident particle in the corresponding cell. 
The shaped signals are sampled and digitised every 25\,ns by 10-bit ADCs~\cite{bib:digi}. The sampled data are temporarily stored in a pipeline memory until a trigger Level-1 signal is received. Seven samples, centred around the pulse peak, are obtained. A gain switch is used to determine which gain information is sent to the back-end electronics for event processing. By default the high-gain signal is used, unless any of the seven samples saturates the ADC, at which point the low-gain signal is transmitted. 

Adder boards receive the analogue low-gain signal from the 3-in-1 cards and sum the signal from six 3-in-1 cards within $\Delta\eta \times \Delta\phi = 0.1 \times 0.1$ before transmitting it to the ATLAS hardware-based trigger system as a trigger tower.

The integrator circuit measures PMT currents (0.01\,nA to 1.4\,$\mu$A) over a long time window of 10--20\,ms with one of the six available gains, and is used for calibration with a radioactive caesium source and to measure the rate of soft interactions during collisions at the LHC~\cite{bib:ATLASLumiMeasurement}. It is a low-pass DC amplifier that receives less than 1\% of the PMT current, which is then digitised by a 12-bit ADC card (which saturates at 5\,V)~\cite{bib:intScienceDirect}. 
  
Power is supplied to the front-end electronics of a single
super-drawer by means of a low-voltage power supply (LVPS) source, which is positioned in an external steel box mounted just outside the electronics super-drawer. 
The high voltage is set and distributed to each individual PMT using dedicated boards positioned inside the super-drawers located with the front-end electronics.

The back-end electronics is located in a counting room approximately 100\,m away from the ATLAS detector. 
The data acquisition system of the Tile Calorimeter is split into four partitions, the ATLAS A-side ($\eta > 0$) and C-side ($\eta <  0$) for both the LB and EB, yielding four logical partitions: LBA, LBC, EBA, and EBC\@.  
Optical fibres transmit signals between each super-drawer and the back-end trigger, timing and control (TTC) and read-out driver (ROD~\cite{bib:ROD}) crates. There are a total of four TTC and ROD crates, one for each physical partition. 
The ATLAS TTC system distributes the LHC clock, trigger decisions, and configuration commands to the front-end electronics. 
If the TTC system sends the trigger acceptance command to the front-end electronics, the corresponding digital signals for all channels of the calorimeter are sent to the ROD via optical links, where the signal is reconstructed for each channel.

\section{Experimental set-up}\label{sec:ExperimentalSetup}

The data used in this paper were taken by the Tile
Calorimeter system using the full ATLAS data acquisition chain. In addition to the TileCal, there are also other ATLAS subsystems used to assist in particle identification, track, momentum, and energy reconstruction.
The inner detector is composed of a silicon pixel detector (Pixel), a
semiconductor tracker (SCT), and a transition radiation tracker (TRT). Together they provide tracking of charged particles for $|\eta| < 2.5$, with a design resolution of $\sigma_{p_\mathrm{T}}/p_{\mathrm{T}} = 0.05\% \cdot p_{\mathrm{T}} \mathrm{(GeV)} \oplus 1\%$~\cite{bib:detectorPaper}.
The electromagnetic lead/liquid-argon barrel (EMB~\cite{bib:EMB}) and endcap (EMEC~\cite{bib:EMEC}) calorimeters provide coverage for $|\eta| < 3.2$. The energy resolution of the liquid-argon (LAr) electromagnetic calorimeter is designed to be $\sigma_E/E = 10\%/\sqrt{E \mathrm{(GeV)}} \oplus 0.7\%$. 
The hadronic calorimetry in the central part of the detector ($|\eta|
< 1.7$) is provided by the TileCal, which is described in detail in
Section~\ref{sec:intro}.
In the endcap region ($1.5 < |\eta| < 3.2$) hadronic calorimetry is
provided by a LAr/copper sampling calorimeter (HEC~\cite{bib:HEC})
behind a LAr/lead electromagnetic calorimeter with accordion geometry, while in the forward region ($3.2 < |\eta| < 4.9$)
the FCal~\cite{bib:FCal} provides electromagnetic (the first module with
LAr/copper) and hadronic (the second and third module with LAr/tungsten) calorimetry.
The muon spectrometer system, the outermost layer of the ATLAS detector, is composed of monitored drift tubes, and cathode strip chambers for the endcap muon track reconstruction for $|\eta| < 2.7$. Resistive plate chambers (RPCs) and thin gap chambers (TGCs) are used to trigger muons in the range $|\eta| < 2.4$. 
ATLAS has four superconducting magnet systems. In the central region, a 2\,T solenoid placed between the inner detector and calorimeters is complemented with 0.5\,T barrel toroid magnets located outside of TileCal. Both endcap regions encompass their own toroid magnet placed between TileCal and muon system, producing the field of 1.0\,T\@.

A three-level trigger system~\cite{bib:AtlasTrigger} was used by ATLAS in Run~1 to
reduce the event rate from a maximum raw rate of 40\,MHz to 200\,Hz,
which is written to disk. The Level 1 Trigger (L1) is a hardware-based decision using the energy collected in coarse regions of the
calorimeter and hits in the muon spectrometer trigger system. The High
Level Trigger (HLT) is composed of the Level 2 Trigger (L2) and the
Event Filter (EF). The HLT uses the full detector information in the regions of interest defined by L1. The reconstruction is further refined in going from L2 to the EF, with the EF using the full offline reconstruction algorithms.
A trigger chain is defined by the sequence of algorithms used in going from L1 to the EF.  
Events passing trigger selection criteria are separated into different streams according to the trigger category for which the event is triggered. 
Physics streams are composed of triggers that are used to identify physics objects (electrons, photons, muons, jets, hadronically-decaying $\tau$-leptons, missing transverse momentum) in collision data.
There are also calibration streams used by the various subsystems for calibration and monitoring purposes, which take data during empty bunch crossings in collision runs or in dedicated calibration runs.
Empty bunch crossings are those with no proton bunch and are separated from any filled bunch by at least five bunch crossings to ensure signals from collision events are cleared from the detector. 
The calibration and monitoring data are explained in more detail in the next sections.

\subsection{ATLAS experimental data}
The full ATLAS detector started recording events from cosmic-ray muons in 2008 as a part of the detector commissioning~\cite{bib:tileReadiness, bib:cosmicRayMuonsATLAS}.
Cosmic-ray muon data from 2008--2010 are used to validate test beam and in situ calibrations, and to study the full calorimeter in the ATLAS environment; these results are presented in Section~\ref{subsec:cosmics}.


The first $\sqrt{s} =  7$\,TeV proton--proton ($pp$) collisions were recorded in March 2010, and started a rich physics programme at the LHC\@. 
In 2011 the LHC $pp$ collisions continued to be at $\sqrt{s} = 7$\,TeV, but the instantaneous luminosity increased and the bunch spacing decreased to 50\,ns. Moving to 2012 the centre-of-mass energy increased to 8\,TeV. 
In total, nearly 30\,fb$^{-1}$ of proton collision data were delivered to ATLAS during Run~1. 
A summary of the LHC beam conditions is shown in Table~\ref{tab:dataPeriods} for 2010--2012, representing the collision data under study in this paper. 
In ATLAS, data collected over long periods of time spanning an LHC
fill or generally stable conditions are grouped into a ``run'', while the
entire running period under similar conditions for several years is
referred to as a ``Run''. 
Data taken within a run are broken down into elementary units called luminosity blocks, corresponding to up to one minute of collision data for which detector conditions or software calibrations remain approximately constant. 

\begin{table}[tbp]
\begin{center}
\begin{tabular}{| l | c | c | c | }
\hline
                                       & 2010    &   2011    & 2012 \\
\hline\hline
Maximum beam energy [TeV]                & 3.5                        &  3.5                      &   4 \\
Delivered integrated luminosity          &   48.1\,pb$^{-1}$    &   5.5\,fb$^{-1}$    &  22.8\,fb$^{-1}$ \\
ATLAS analysis integrated luminosity     &   45.0\,pb$^{-1}$    &   4.7\,fb$^{-1}$    &  20.3\,fb$^{-1}$ \\ 
Minimum bunch spacing  [ns]              &  150                     &  50$^{a}$      &  50$^{a}$ \\
Maximum number of bunches                & 348                      & 1331$^{b}$        & 1380 \\
Mean number of interactions per bunch crossing    & 4 (1)         &  17 (9)                  &  36 (20) \\
Maximum instantaneous luminosity [$10^{33}$ cm$^{-2}$s$^{-1}$]  & 0.2  & 3.8         &     7.5    \\
\hline
\end{tabular}
\end{center}
\caption{Summary of proton collision data presented in this
  paper. The ATLAS analysis integrated luminosity corresponds to the
  total integrated luminosity approved for analysis, passing all data
  quality requirements ensuring the detector and reconstruction
  software is properly functioning. 
  The maximum and the average (listed in 
  parentheses) of the distribution of the mean number of interactions per 
  bunch crossing are given. 
  $^{a}$Additional special
  runs with low integrated luminosity used for commissioning purposes
  were taken with a minimal bunch spacing of
  25\,ns. $^{b}$Additional special runs were taken with low
  integrated luminosity where the number of colliding bunches was increased to 1842 in 2011. \label{tab:dataPeriods}} 
\end{table}

ATLAS also recorded data during these years with lower-energy proton collisions (at $\sqrt{s}$ = 900\,GeV, 2.76\,TeV), and data containing lead ion collisions. 
Nevertheless, this paper focuses on the results obtained in $pp$ collisions at $\sqrt{s} = 7$ and 8\,TeV.

\subsection{Monte Carlo simulations}
Monte Carlo (MC) simulated data are frequently used by performance and
physics groups to predict the behaviour of the detector.
It is crucial that the MC simulation closely matches the actual data, so those relying on simulation for algorithm optimisations and/or searches for new physics are not misled in their studies.

The MC process is divided into four steps: event generation, simulation, digitisation, and reconstruction. Various event generators were used in the analyses as described in each subsection.
The ATLAS MC simulation~\cite{bib:SimSoftware} relies on the {\GEANT}4 toolkit~\cite{bib:geant4v1} to model the detector and interactions of particles with the detector material. 
During Run~1, ATLAS used the so-called QGSP\_BERT physics model to describe the hadronic interactions with matter, where at high energies the hadron showers are modelled using the Gluon String Plasma model, and the Bertini intra-nuclear cascade model is used for lower-energy hadrons~\cite{bib:Geant4PhysicsLists}.
The input to the digitisation is a collection of hits in the active scintillator material, characterised by the energy, time, and position. The amount of energy deposited in scintillator is divided by the calorimeter sampling fraction to obtain the channel energy~\cite{bib:tileSampleFrac}. In the digitisation step, the channel energy in GeV is converted into its equivalent charge using the electromagnetic scale constant (Section~\ref{sec:calib}) measured in the beam tests. The charge is subsequently translated into the signal amplitude in ADC counts using the corresponding calibration constant (Section~\ref{subsec:CIS}). The amplitude is convolved with the pulse shape and digitised each 25\,ns as in real data. The
electronic noise is emulated and added to the digitised samples as
described in Section~\ref{sec:electronicNoise}. Pile-up
(i.e.\ contributions from additional minimum-bias interactions 
occurring in the same bunch crossing as the hard-scattering collision or in nearby ones),
are simulated with \PYTHIAV{6}~\cite{bib:Pythia6} in 2010--2011 and \PYTHIAV{8}~\cite{bib:Pythia8} in 2012, and mixed at realistic rates with the hard-scattering process of interest during the digitisation step.
Finally, the same reconstruction methods, detailed in Section~\ref{sec:SignalReco}, as used for the data are applied to the digitised samples of the simulations.

\section{Signal reconstruction}\label{sec:SignalReco}

\begin{figure}[tp]
\centering
\includegraphics[width=0.49\textwidth]{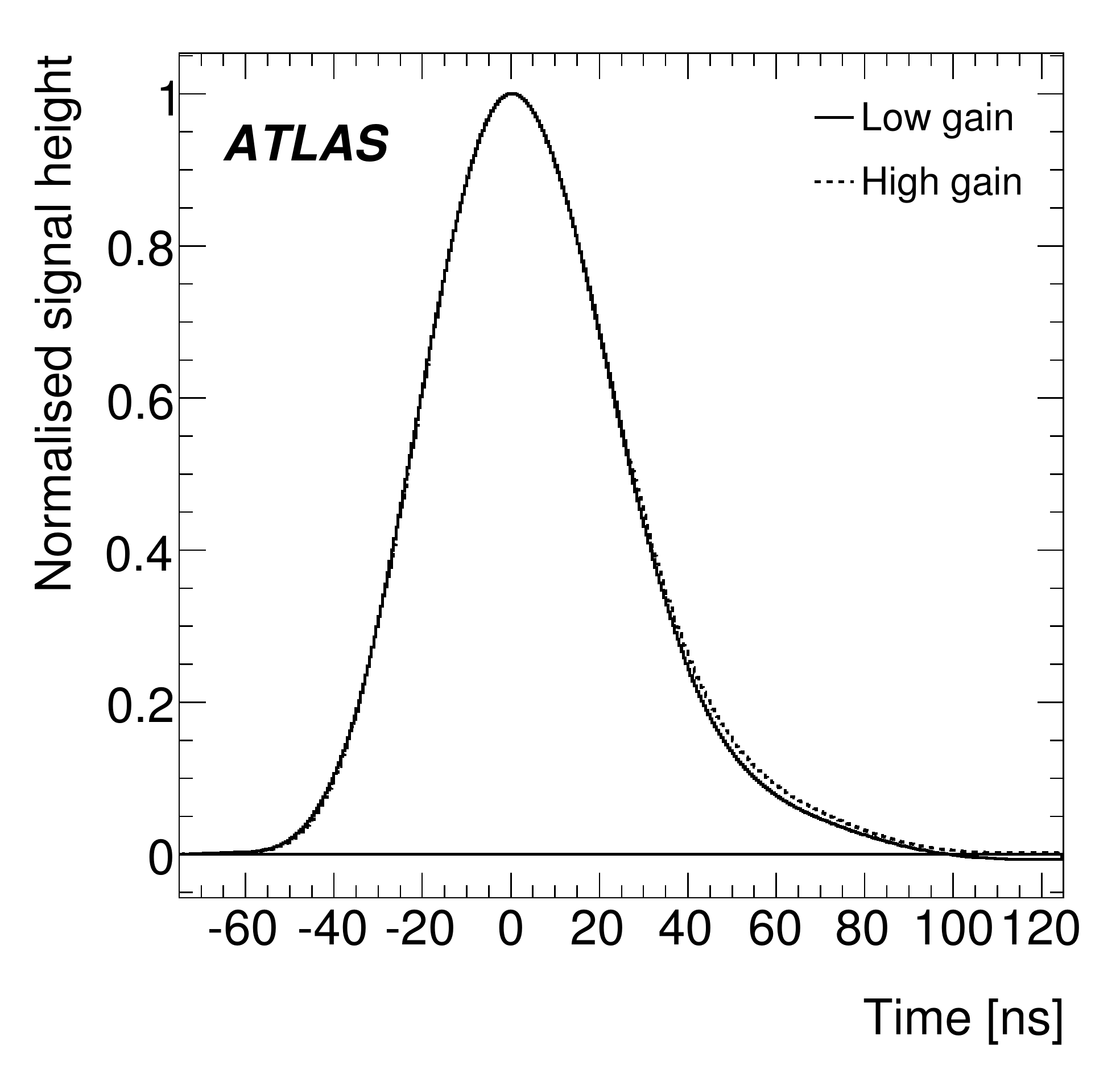}
\caption{The reference pulse shapes for high gain and low gain, shown in arbitrary units~\cite{bib:tileReadiness}. \label{fig:pulseShapes}}
\end{figure}

The electrical signal for each TileCal channel is reconstructed from seven consecutive digital samples, taken every 25\,ns.  
Nominally, the reconstruction of the signal pulse amplitude, time, and pedestal is made using the Optimal Filtering (OF) technique~\cite{bib:OF}. This technique weights the samples in accordance with a reference pulse shape. 
The reference pulse shape used for all channels is taken as the average pulse shape from test beam data, 
with reference pulses for both high- and low-gain modes, each of which is shown in Figure~\ref{fig:pulseShapes}. 
The signal amplitude ($A$), time phase ($\tau$), and pedestal ($p$) for a channel are calculated using the ADC count of each sample $S_i$ taken at time $t_i$:

\begin{equation}
A  = \sum_{i=1}^{n=7} a_i S_i , \qquad  A\tau =  \sum_{i=1}^{n=7} b_i S_i , \qquad p = \sum_{i=1}^{n=7} c_i S_i 
\label{eq:OF}
\end{equation}

where the weights ($a_i$, $b_i$, and $c_i$) are derived to minimise the resolution of the amplitude and time, with a set of weights extracted for both high and low gain. Only electronic noise was considered in the minimisation procedure in Run~1.

The expected time of the pulse peak is calibrated such that for
particles originating from collisions at the interaction point the
pulse should peak at the central (fourth) sample, synchronous with the LHC 40\,MHz clock. 
The reconstructed value of $\tau$ represents the small time phase in~ns between the expected pulse peak 
and the time of the actual reconstructed signal peak, arising from fluctuations in particle travel time and uncertainties in the electronics read-out. 

Two modes of OF reconstruction were used during Run~1, an
iterative and a non-iterative implementation. In the iterative method, the pulse shape is recursively fit when the difference between maximum and minimum sample is above a noise threshold. The initial time phase is taken as the time of the maximum sample, and subsequent steps use the previous time phase as the starting input for the fit. Only one iteration is performed assuming a pulse with the peak in the central sample for signals below a certain threshold.
For events with no out-of-time pile-up (see Section~\ref{sec:pileupNoise}) 
this iterative method proves
successful in reconstructing the pulse peak time to within
0.5\,ns. This method is used when reconstructing events occurring
asynchronously with the LHC clock, such as cosmic-ray muon data and
also to reconstruct data from the 2010 proton collisions. With an
increasing number of minimum-bias events per bunch crossing, the
non-iterative method, which is more robust against pile-up, is used.
The time phase was fixed for each individual channel and only
a single fit to the samples was applied in 2011--2012 data.

In real time, or online, the digital signal processor (DSP) in the ROD
performs the signal reconstruction using the OF technique, and
provides channel energy and time to the HLT\@.
The conversion between signal amplitude in ADC counts and energy units of MeV is done by applying channel-dependent calibration constants which are described in the next section. 
The DSP reconstruction is limited by the use of fixed point arithmetic, which has a precision of 0.0625 ADC counts (approximately 0.75\,MeV in high gain), 
and imposes precision limitations for the channel-dependent calibration constants. 

The offline signal is reconstructed using the same iterative or
non-iterative OF technique as online. In 2010 the raw data were
transmitted from the ROD for offline signal reconstruction, and the
amplitude and time computations from the ROD were used only for
the HLT decision.  
From 2011 onward, with increasing instantaneous luminosity the output
bandwidth of the ROD becomes saturated, and only channels for which
the difference between the maximum and minimum $S_i$ is larger than
five ADC counts (approximately 60\,MeV) have the raw data transmitted from the ROD for the offline signal reconstruction; otherwise the ROD signal reconstruction results are used for the offline data processing.

\begin{figure}[tp]
\centering
\begin{minipage}{.45\textwidth}
  \centering
 \vspace{0.05\textwidth}%
 \includegraphics[width=0.99\textwidth]{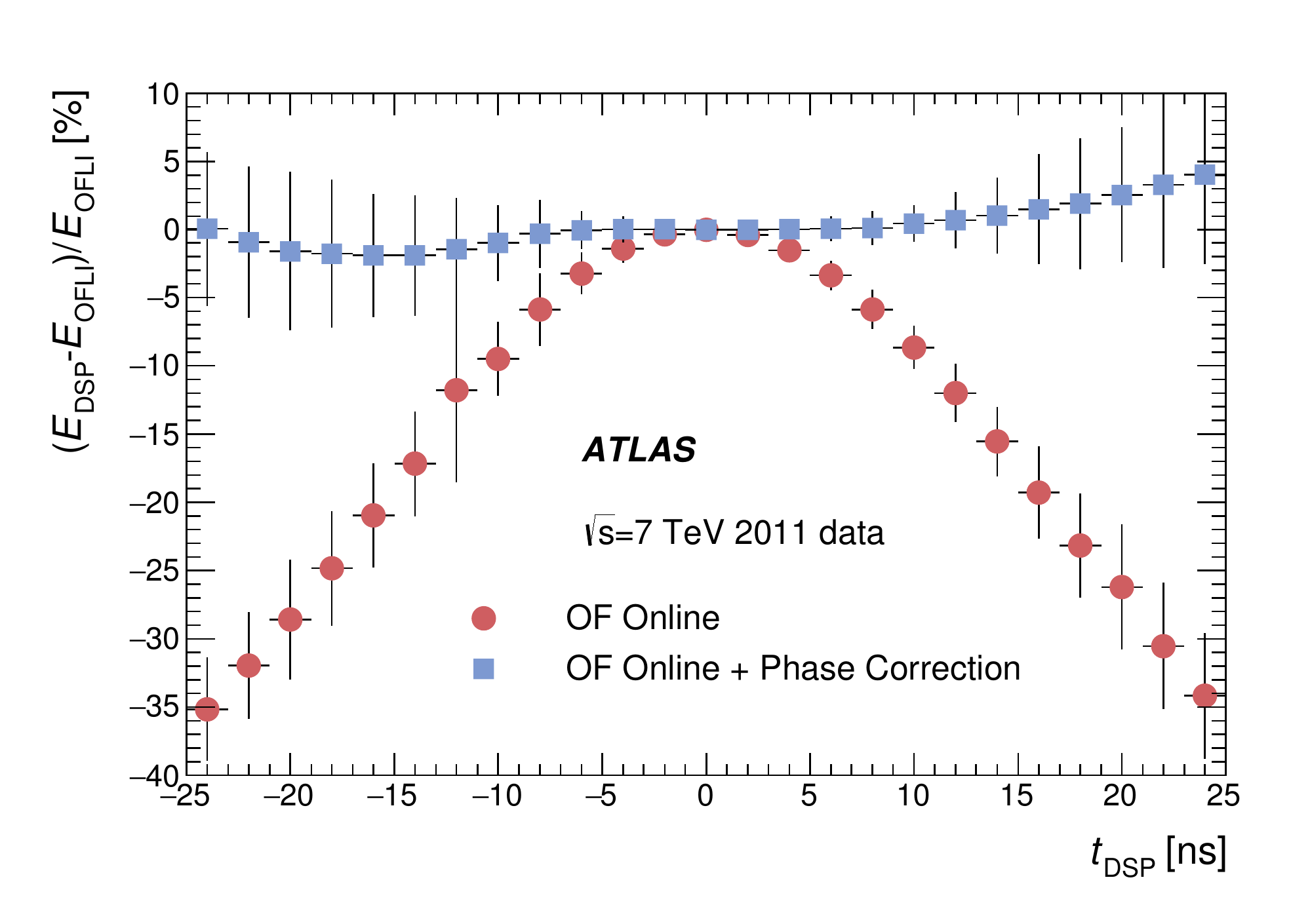}
  \caption{The relative difference between the online channel energy ($E_{\mathrm{DSP}}$) calculated using the non-iterative OF method and the offline ($E_{\mathrm{OFLI}}$) channel energy reconstruction using the iterative OF method, as a function of the phase computed by the DSP ($t_{\mathrm{DSP}}$) with no correction (circles) and with application of the parabolic correction (squares) as a function of phase ($\tau$). The error bars are the standard deviations (RMS) of the relative difference distribution. Data are shown for collisions in 2011.\label{fig:signalRecoParabolicCorr}}
\end{minipage} \hspace{0.05\textwidth}%
\begin{minipage}{.45\textwidth}
  \centering
  \includegraphics[width=0.99\textwidth]{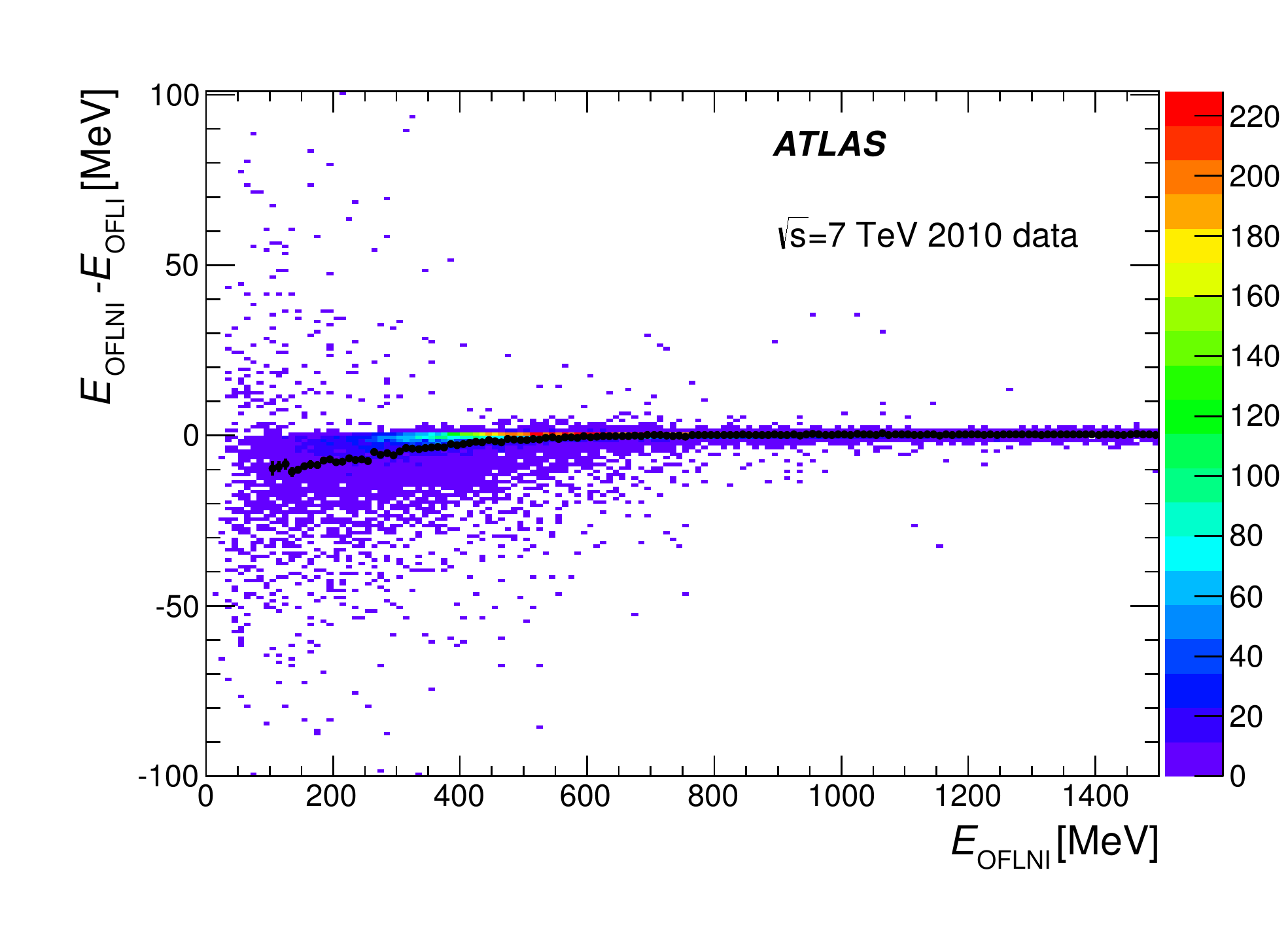}
  \caption{The absolute difference between the energies reconstructed using the optimal filtering reconstruction method with the non-iterative ($E_{\mathrm{OFLNI}}$) and iterative ($E_{\mathrm{OFLI}}$) signal reconstruction methods as a function of energy. The black markers represent mean values of $E_{\mathrm{OFLNI}}-E_{\mathrm{OFLI}}$ per a bin of $E_{\mathrm{OFLNI}}$. The parabolic correction is applied to $E_{\mathrm{OFLNI}}$. The data shown uses high $\pT$ ($> 20$\,GeV) isolated muons from $\sqrt{s}=7$\,TeV collisions recorded in 2010.\label{fig:signalRecoParabolicCorrVsE}}
\end{minipage}%
\end{figure}

The reconstructed phase $\tau$ is expected to be small, but for any
non-zero values of the phase, there is a known bias when the non-iterative pulse reconstruction is used that causes the reconstructed amplitude to be underestimated. 
A correction based on the phase is applied when the phase is reconstructed within half the LHC bunch spacing and the channel amplitude is larger than 15\,ADC counts, to reduce contributions from noise.
Figure~\ref{fig:signalRecoParabolicCorr} shows the difference between the non-iterative energy reconstructed in the DSP without (circles) and with (squares) this parabolic correction, relative to the iterative reconstruction calculated offline for data taken during 2011. 
Within time phases of $\pm 10$\,ns the difference between the iterative
and non-iterative approaches with the parabolic correction applied is less than 1\%.    

The difference between the energies reconstructed using the non-iterative (with the parabolic correction applied) and iterative OF technique as a function of energy can be seen in Figure~\ref{fig:signalRecoParabolicCorrVsE} for high $\pT$ ($> 20$\,GeV) isolated muons taken from the 2010 $\sqrt{s}=7$\,TeV collision data.
For channel energies between 200\,MeV and 400\,MeV the mean difference between the two methods is smaller than 10\,MeV. For channel energies larger than 600\,MeV, the mean reconstructed energy is the same for the two methods.

\subsection{Channel time calibration and corrections}\label{subsec:timingCalib}

\begin{figure}
  \centering
  \includegraphics[width=0.49\textwidth]{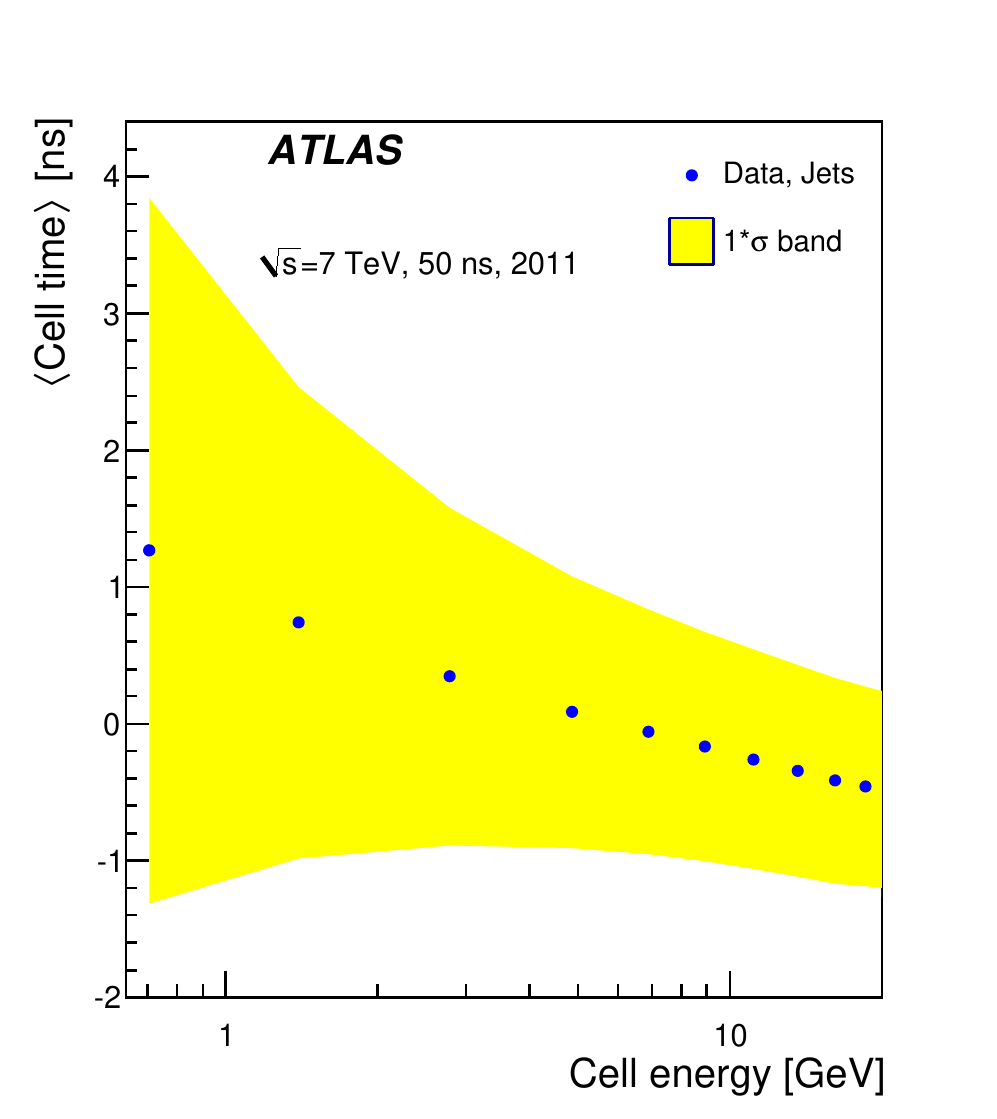}
  \includegraphics[width=0.49\textwidth]{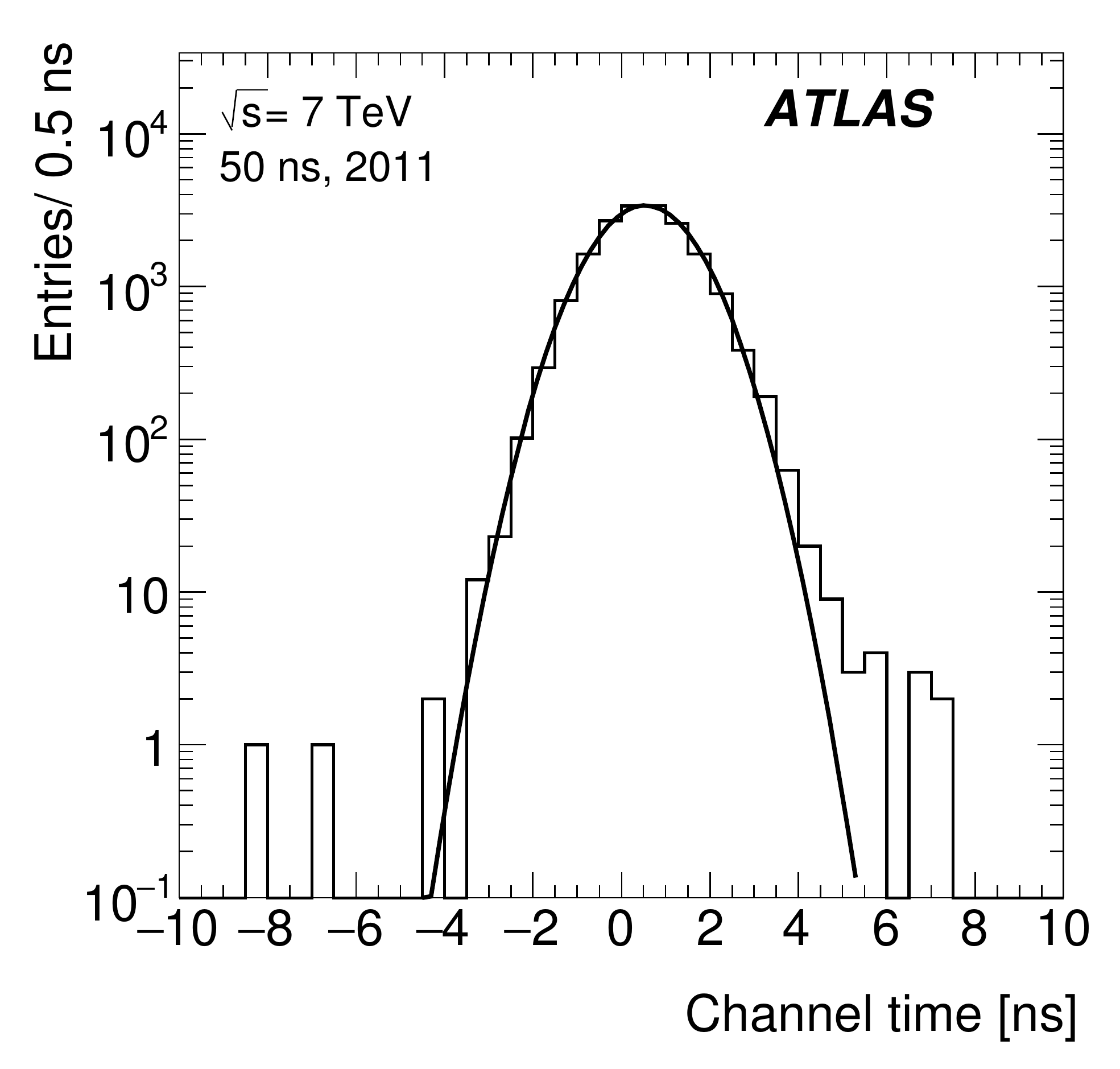}  
  \caption{Left: the mean cell reconstructed time (average of the
    times in the two channels associated with the given cell) as
    measured with jet events. The mean cell time decreases with the increase 
    of the cell energy due to the reduction of the energy fraction of 
    the slow hadronic component of hadronic showers~\cite{bib:wigmans, bib:emscaleTB}.  Right:
    example of the channel reconstructed time in jet events in 2011 data, 
    with the channel energy between 2 and 4\,GeV. The solid line represents 
    the Gaussian fit to the data.} 
  \label{fig:time_calibration}
\end{figure}

Correct channel time is essential for energy reconstruction, object selection, and for time-of-flight analyses searching for hypothetical long-lived particles entering the calorimeter. 
Initial channel time calibrations are performed with laser and cosmic-ray muon events, and are later refined using beam-splash events from a single LHC beam~\cite{bib:tileReadiness}. 
A laser calibration system pulses laser light directly into each
PMT\@. The system is used to calibrate the time of all channels in one
super-drawer such that the laser signal is sampled
simultaneously. These time calibrations are used to account for time
delays due to the physical location of the electronics.
Finally, the time calibration is set with collision data, considering
in each event only channels that belong to a reconstructed jet. This
approach mitigates the bias from pile-up noise
(Section~\ref{sec:pileupNoise}) and 
non-collision background. Since the reconstructed time slightly
depends on the energy deposited by the jet in a cell
(Figure~\ref{fig:time_calibration} left), the
channel energy is further required to be in a certain range (2--4\,GeV)
for the time calibration. An example of the reconstructed time
spectrum in a channel satisfying these conditions is shown in
Figure~\ref{fig:time_calibration} (right). The distribution shows a
clear Gaussian core (the Gaussian mean determines the time calibration constant)
with a small fraction of events at both high- and low-time tails. The higher-time
tails are more evident for low-energy bins and are mostly due to the
slow hadronic component of the shower development. Symmetric tails are
due to out-of-time pile-up (see Section~\ref{sec:pileupNoise}) and are
not seen in 2010 data where pile-up is negligible.
The overall time resolution is evaluated with jets and muons from
collision data, and is described in Section~\ref{subsec:timePerform}.

During Run~1 a problem was identified in which a digitiser could suddenly lose its time calibration settings.
This problem, referred to as a ``timing jump'', was later traced to the TTCRx chip in the digitiser board, which received clock configuration commands responsible for aligning that digitiser sampling clock with the LHC clock. During operation these settings are sent to all digitisers during configuration of the super-drawers, so a timing jump manifests itself at the beginning of a run or after a hardware failure requiring reconfiguration during a run. All attempts to avoid this feature at the hardware or configuration level failed, hence the detection and correction of faulty time settings became an important issue. Less than 15\% of all digitisers were affected by these timing jumps, and were randomly distributed throughout the TileCal. All channels belonging to a given digitiser exhibit the same jump, and the magnitude of the shift for one digitiser is the same for every jump. 

Laser and collision events are used to detect and correct for the timing jumps.
Laser events are recorded in parallel to physics data in empty bunch
crossings. The reconstructed laser times are studied for each channel
as a function of luminosity block. As the reconstructed time phase is
expected to be close to zero the monitoring algorithm searches for
differences ($> 3$\,ns) from this baseline. Identified cases are
classified as potential timing jumps, and are automatically
reported to a team of experts for manual inspection. The timing
differences are saved in the database and applied as a correction in
the offline data reconstruction.  

Reconstructed jets from collision data are used as a secondary tool to verify timing jumps, but require completion of the full data reconstruction chain and constitute a smaller sample as a function of luminosity block.
These jets are used to verify any timing jumps detected by the laser analysis, or used by default in cases where the laser is not operational. For the latter, problematic channels are identified after the full reconstruction, but are corrected in data reprocessing campaigns. 

\begin{figure}[tp]                                                                                      
\centering
\includegraphics[width=0.45\textwidth]{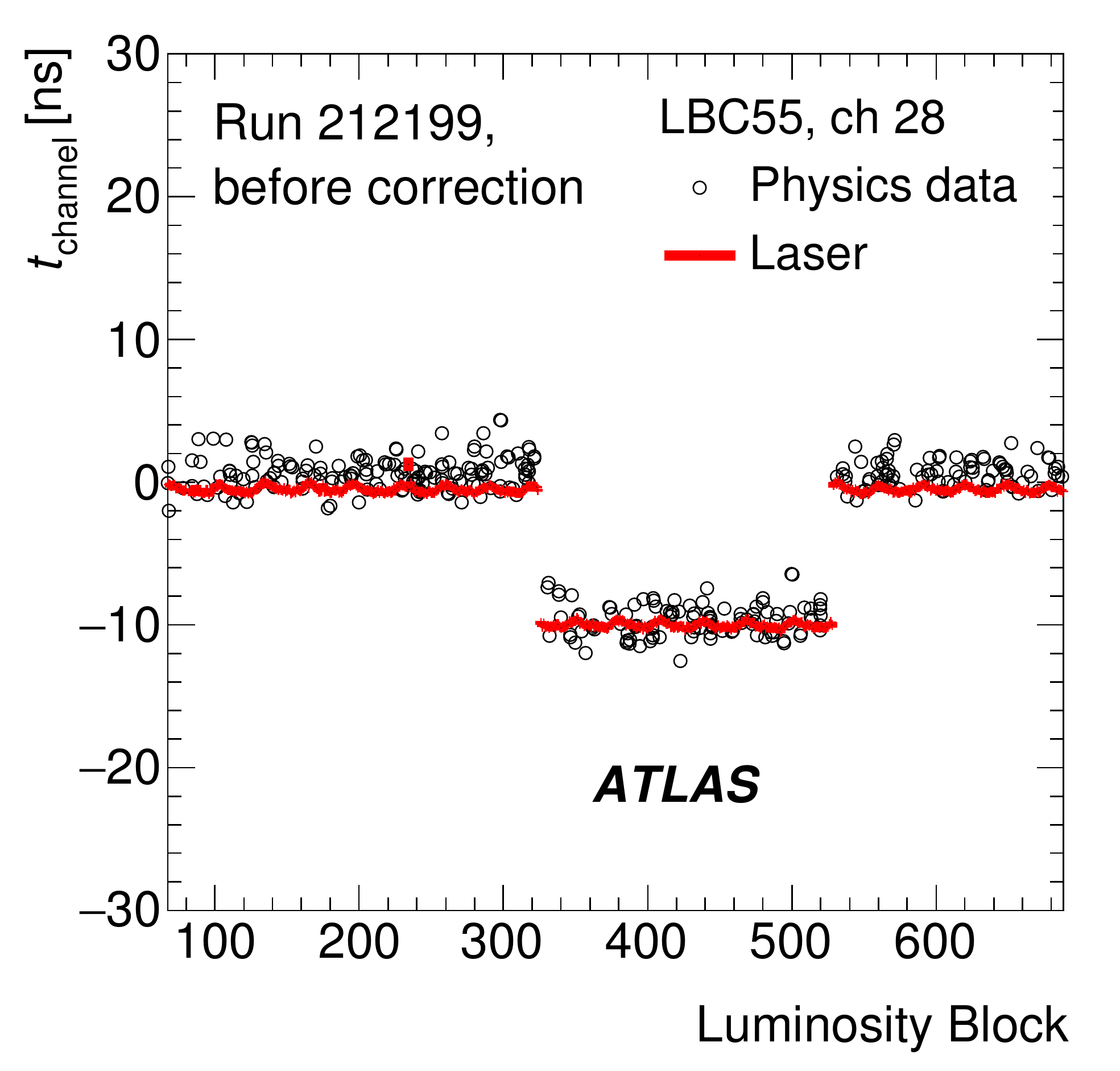}
\includegraphics[width=0.45\textwidth]{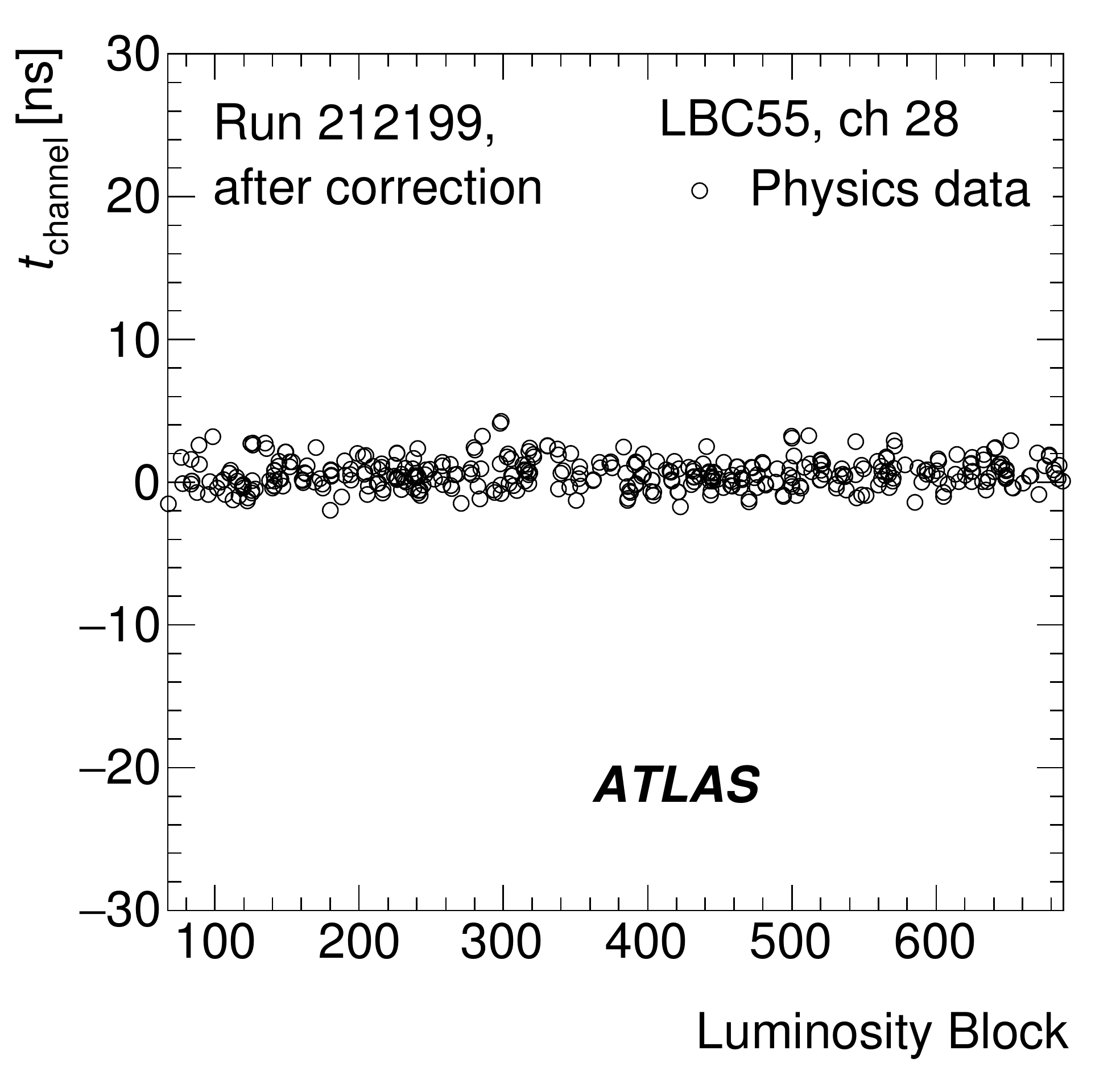}
\caption{An example of timing jumps detected using the laser (full red
  circles) and physics (open black circles) events (left) before and
  (right) after the correction. The small offset of about 2\,ns in
  collision data is caused by the energy dependence of 
  the reconstructed time in jet events (see
  Figure~\protect\ref{fig:time_calibration}, left). In these plots,
  events with any energy are accepted to accumulate enough statistics.}
\label{fig:exampleTimeJumps} 
\end{figure}

A typical case of a timing jump is shown in
Figure~\ref{fig:exampleTimeJumps} before (left) and after (right) the time correction. Before the correction the time step is clearly visible and demonstrates good agreement between the times measured by the laser and physics collision data. 

\begin{figure}[tp]                         
\centering
\includegraphics[width=0.45\textwidth]{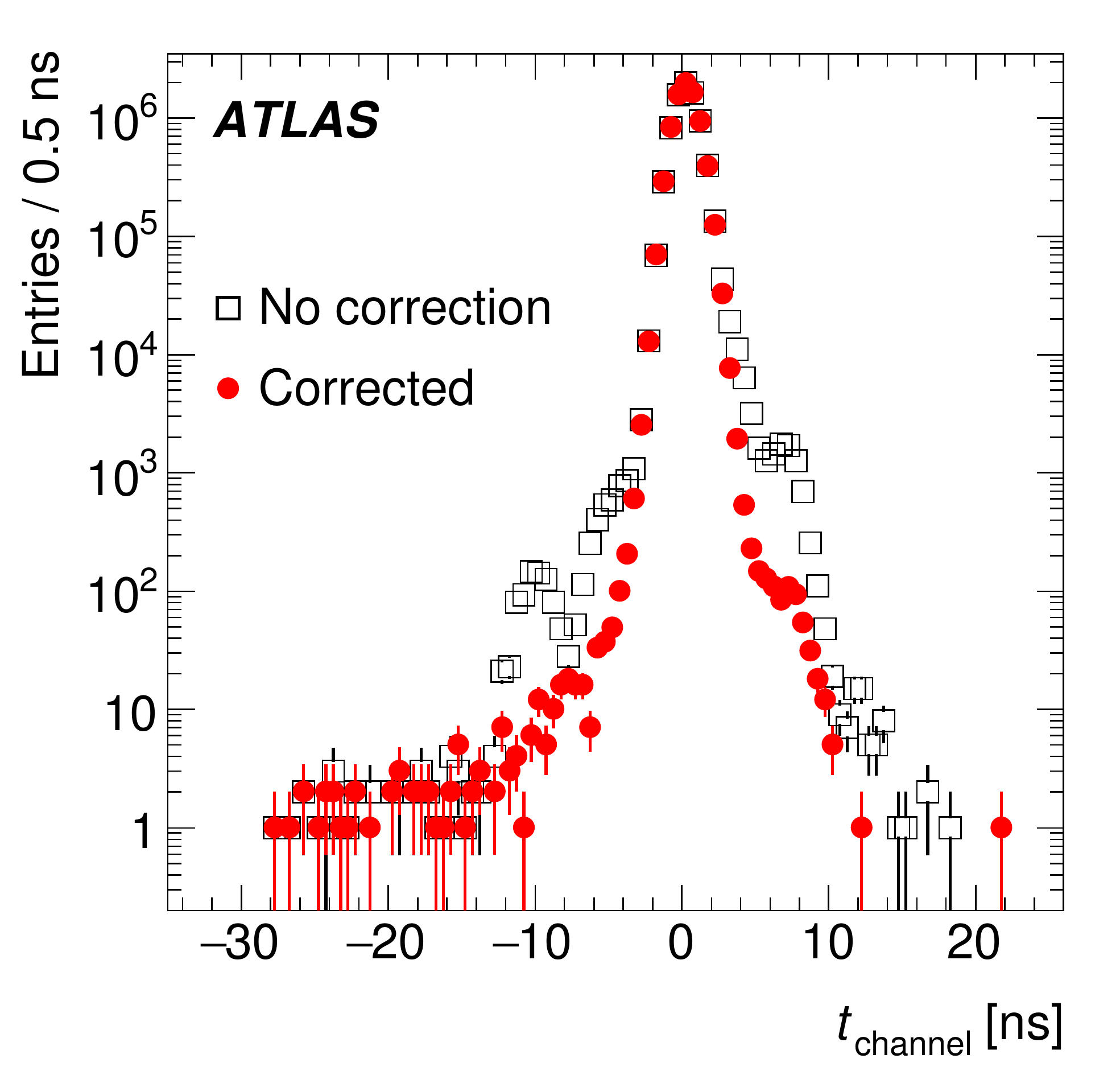}
\caption{Impact of the timing jump corrections on the reconstructed channel time in jets from collision data. Shown are all high-gain channels with $E_{\mathrm{ch}} > 4$ GeV associated with a reconstructed jet. The plot represents $1.3\,\mathrm{fb}^{-1}$ of $pp$ collision data acquired in 2012.\label{fig:impactTimingJump}}
\end{figure}

The overall impact of the timing jump corrections on the reconstructed time is studied with jets using 1.3\,fb$^{-1}$ of collision data taken in 2012. To reduce the impact of the time dependence on the reconstructed energy, the channel energy is required to be $E > 4$\,GeV, and read out in high-gain mode. The results are shown in Figure~\ref{fig:impactTimingJump}, where the reconstructed time is shown for all calorimeter channels with and without the timing jump correction. While the Gaussian core, corresponding to channels not affected by timing jumps, remains basically unchanged, the timing jump correction significantly reduces the number of events in the tails. 
The 95\% quantile range around the peak position shrinks by 12\% (from
3.3\,ns to 2.9\,ns) and the overall RMS improves by 9\% (from 0.90\,ns
to 0.82\,ns) after the corrections are applied. 
In preparation for Run~2, problematic digitisers were replaced and repaired.
The new power supplies, discussed in the next section, also contribute to the significant reduction in the number of the timing jumps since the trips almost ceased (Section~\ref{sec:overall_DQ}) and thus the module reconfigurations during the run are eliminated in Run~2.

\subsection{Electronic noise}\label{sec:electronicNoise}
The total noise per cell is calculated taking into account two components, electronic noise and a contribution from pile-up interactions (so-called pile-up noise). These two contributions are added in quadrature to estimate the total noise. Since the cell noise is directly used as input to the topological clustering algorithm~\cite{bib:topoclustering} (see Section~\ref{sec:InSitu}), it is very important to estimate the noise level per cell with good precision.

The electronic noise in the TileCal, measured by fluctuations of the pedestal, is largely independent of external LHC beam conditions.
Electronic noise is studied using large samples of high- and low-gain
pedestal calibration data, which are taken in dedicated runs without beam in the ATLAS detector. 
Noise reconstruction of pedestal data mirrors that of the data-taking period, using the OF
technique with iterations for 2010 data and the non-iterative version from 2011 onward.  

The electronic noise per channel is calculated as a standard deviation (RMS) of the energy distributions in pedestal events. 
The fluctuation of the digital noise as a function of time is studied with the complete 2011 dataset. It fluctuates by an average of 1.2\% for high gain and 1.8\% for low gain across all channels, indicating stable electronic noise constants.

\begin{figure}[tp]
\centering
\includegraphics[width=0.70\textwidth]{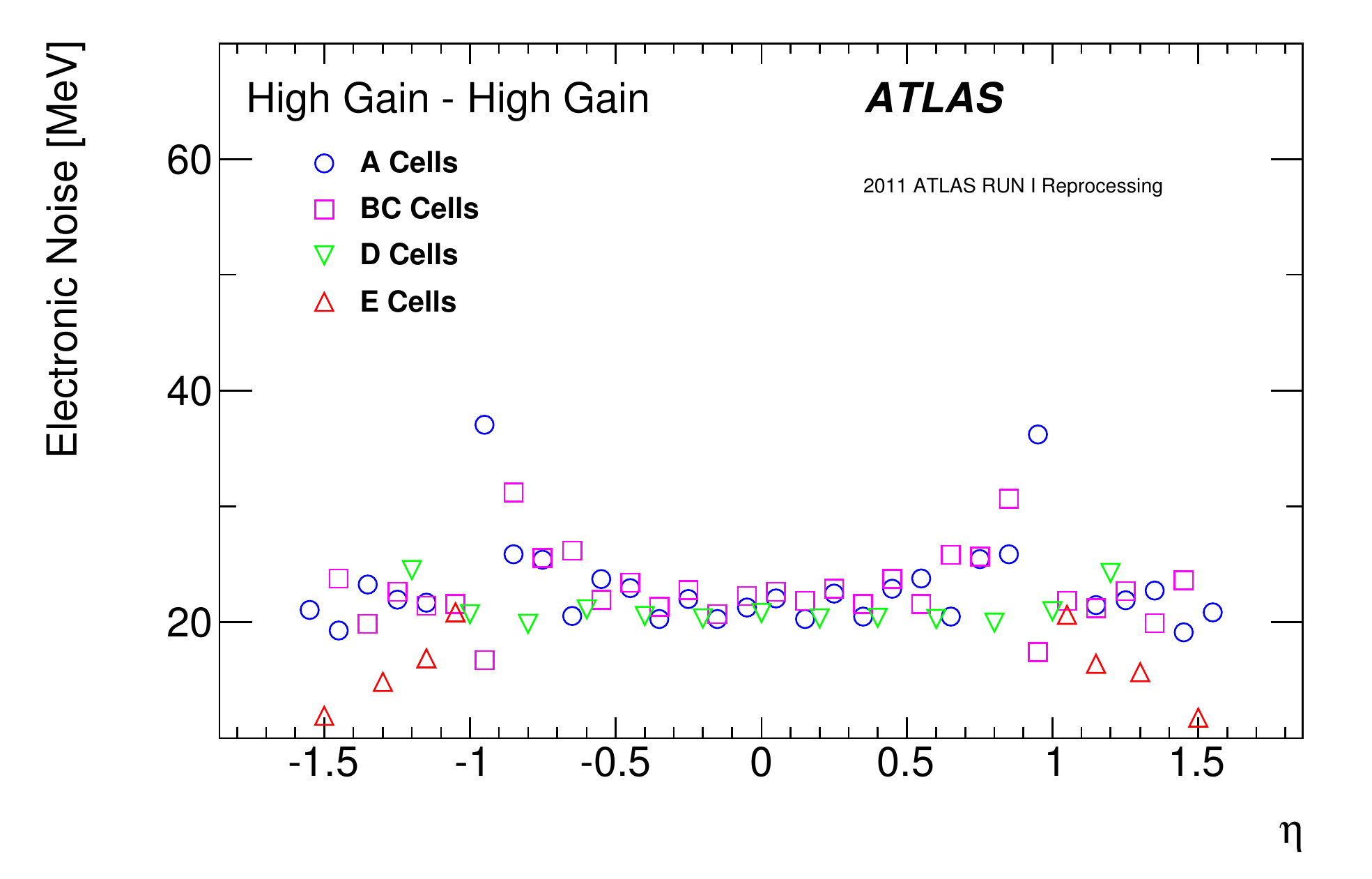}
\caption{The $\phi$-averaged electronic noise (RMS) as a function of $\eta$
  of the cell, with both contributing read-out channels in high-gain
  mode. For each cell the average value over all modules is
  taken. The statistical uncertainties are smaller than the marker size. Values are extracted using all the calibration runs
  used for the 2011 data reprocessing. The different cell types are shown separately for each layer: A, BC, D, and E (gap/crack). The transition between the long and extended barrels can be seen in the range $0.7 < |\eta| < 1.0$.\label{fig:avgNoiseCellHGHG}}
\end{figure}

As already mentioned in Section~\ref{sec:Tilecal}, a typical cell is read out by two channels. Therefore, the cell noise constants are derived for the four combinations of the two possible gains from the two input channels (high--high, high--low, low--high, and low--low). Figure~\ref{fig:avgNoiseCellHGHG} shows the mean cell noise (RMS) for all cells as a function of $\eta$ for the high--high gain combinations. The figure also shows the variations with cell type, reflecting the variation with the cell size. 
The average cell noise is approximately 23.5\,MeV. However, cells located in the highest $|\eta|$ ranges show noise values closer to 40\,MeV. These cells are formed by channels physically located near the LVPS\@. The influence of the LVPS on the noise distribution is discussed below.  
A typical electronic noise values for other combinations of gains are 400--700\,MeV for high--low/low--high gain combinations and 600--1200\,MeV for low--low gain case.
Cells using two channels with high gain are relevant when the
deposited energy in the cell is below about 15\,GeV, above that both
channels are often in low-gain mode, and if they fall somewhere in the
middle range of energies (10--20\,GeV) one channel is usually in high
gain and the other in low gain. 

During Run~1 the electronic noise of a cell is best described by a
double Gaussian function, with a narrow central single Gaussian core and a second
central wider Gaussian function to describe the tails~\cite{bib:tileReadiness}. A normalised
double Gaussian template with three parameters ($\sigma_1$,
$\sigma_2$, and the relative normalisation of the two Gaussian functions $R$)
is used to fit the energy distribution:

\begin{equation*}
f_{\mathrm{pdf}} = \frac{1}{1+R} \left(\frac{1}{\sqrt{2\pi}\sigma_1} \mathrm{e}^{-\frac{x^2}{2\sigma_1^2}}   + \frac{R}{\sqrt{2\pi}\sigma_2} \mathrm{e}^{-\frac{x^2}{2\sigma_2^2}}\right)
\end{equation*}

The means of the two Gaussian functions are set to $\mu_1 = \mu_2 = 0$, which is a good approximation for the cell noise. 
As input to the topological clustering algorithm an equivalent $\sigma_{\mathrm{eq}}(E)$ is introduced to measure the significance ($S = |E| /\sigma_{\mathrm{eq}}(E)$) of the double Gaussian probability distribution function in units of standard deviations of a normal distribution.\footnote{The $\sigma_{\mathrm{eq}}(E)$ defines the region where the significance for the double Gaussian $f_{\mathrm{pdf}}$ is the same as in the $1\sigma$ region of a standard Gaussian distribution function, i.e. $1\sigma_{\mathrm{eq}}(E)$ is defined as $\int_{-\sigma_{\mathrm{eq}}} ^{\sigma_{\mathrm{eq}}} f_{\mathrm{pdf}}\mathrm{d}x = 0.68$, $2\sigma_{\mathrm{eq}}(E)$ as $\int_{-2\sigma_{\mathrm{eq}}} ^{2\sigma_{\mathrm{eq}}} f_{\mathrm{pdf}}\mathrm{d}x = 0.954$, etc.}

\begin{figure}[tp]
\centering
\includegraphics[width=0.70\textwidth]{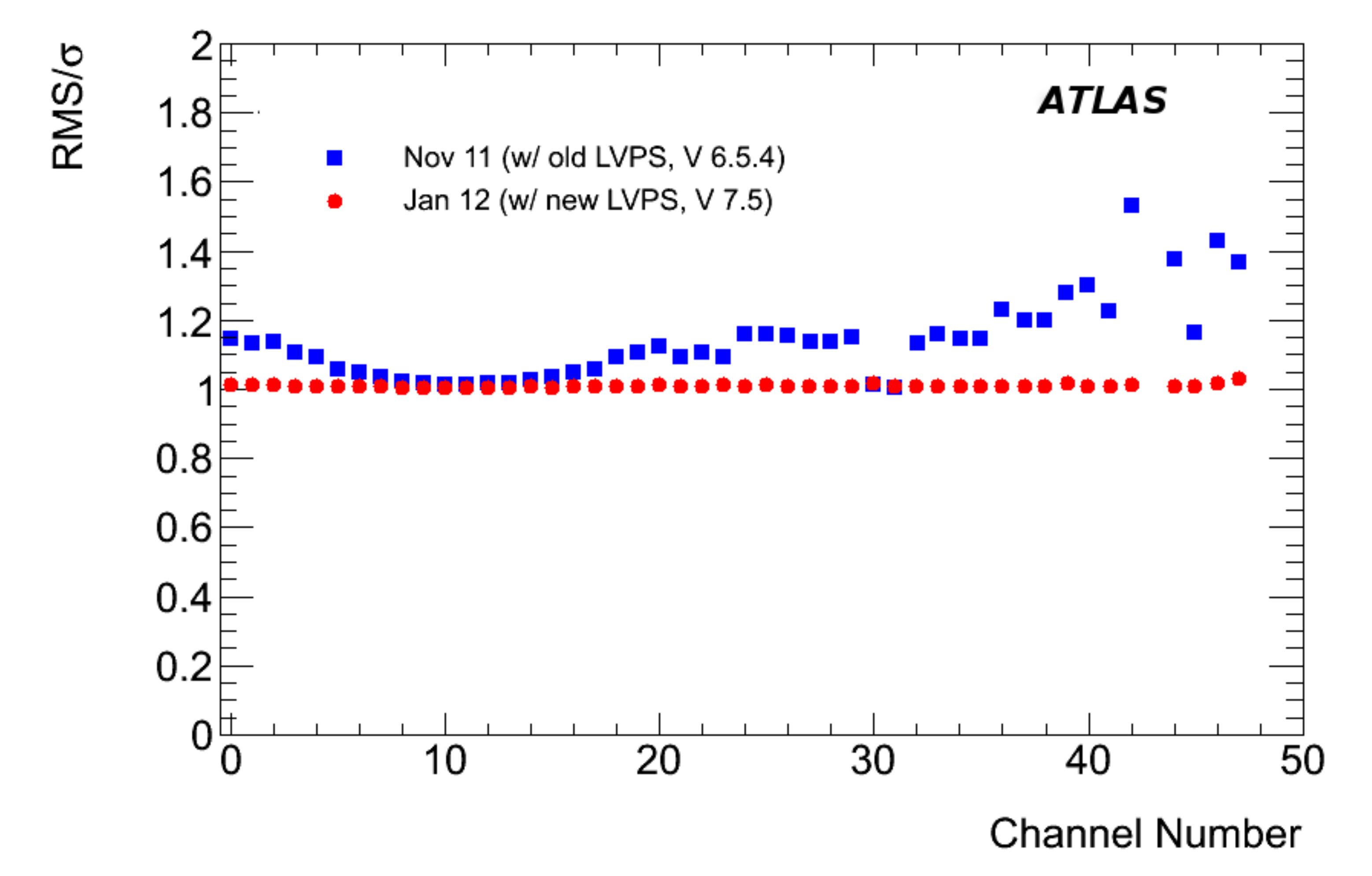}
\caption{Ratio of the RMS to the width ($\sigma$) of a single Gaussian fit to the electronic noise distribution for all channels averaged over 40 TileCal modules before (squares) and after (circles) the replacement of the LVPS\@. Higher-number channels are closer to the LVPS. \label{fig:newVsOldLVPSNoise}}
\end{figure}

The double Gaussian behaviour of the electronic noise is believed to originate from the LVPS used during Run~1, as the electronic noise in test beam data followed a single Gaussian distribution, and this configuration used temporary power supplies located far from the detector. During December 2010, five original LVPS sources were replaced by new versions of the LVPS\@. During operation in 2011 these LVPSs proved to be more reliable by suffering virtually no trips, and resulted in lower and more single-Gaussian-like behaviour of channel electronic noise. With this success, 40 more new LVPS sources (corresponding to 16\% of all LVPSs) were installed during the 2011--2012 LHC winter shutdown. Figure~\ref{fig:newVsOldLVPSNoise} shows the ratio of the RMS to the width of a single Gaussian fit to the electronic noise distribution for all channels averaged over the 40 modules before and after the replacement of the LVPS\@. It can be seen that the new LVPS have values of RMS$/\sigma$ closer to unity, implying a shape similar to a single Gaussian function, across all channels. The average cell noise in the high--high gain case decreases to 20.6\,MeV with the new LVPS\@.

The coherent component of the electronic noise was also investigated. A considerable level of correlation was only found among channels belonging to the same motherboard,\footnote{Each motherboard accommodates 12 consecutive channels in a super-drawer. One of its roles is to distribute the low voltages to the electronic components of individual channels~\cite{bib:CISRef}.} for other pairs of channels the correlations are negligible. Methods to mitigate the coherent noise were developed;\footnote{The first method estimated the coherent component of the noise as an average over all channels in the same motherboard with signals less than $3\times$ the electronic noise variation; this average value was then used to correct the individual channel energies provided at least 60\% of channels contributed to the calculation. The second method~\cite{bib:Miquel} is based on the $\chi^2$ minimisation.} they reduce the correlations from ($-40\%$, $+70\%$) to ($-20\%$, $+10\%$) and also decrease the fraction of events in the tails of the double Gaussian noise distribution.

\vspace{0.2cm}
\noindent\textbf{Electronic noise in the Monte Carlo simulations} 

\begin{figure}[tp]
\centering
\includegraphics[width=0.70\textwidth]{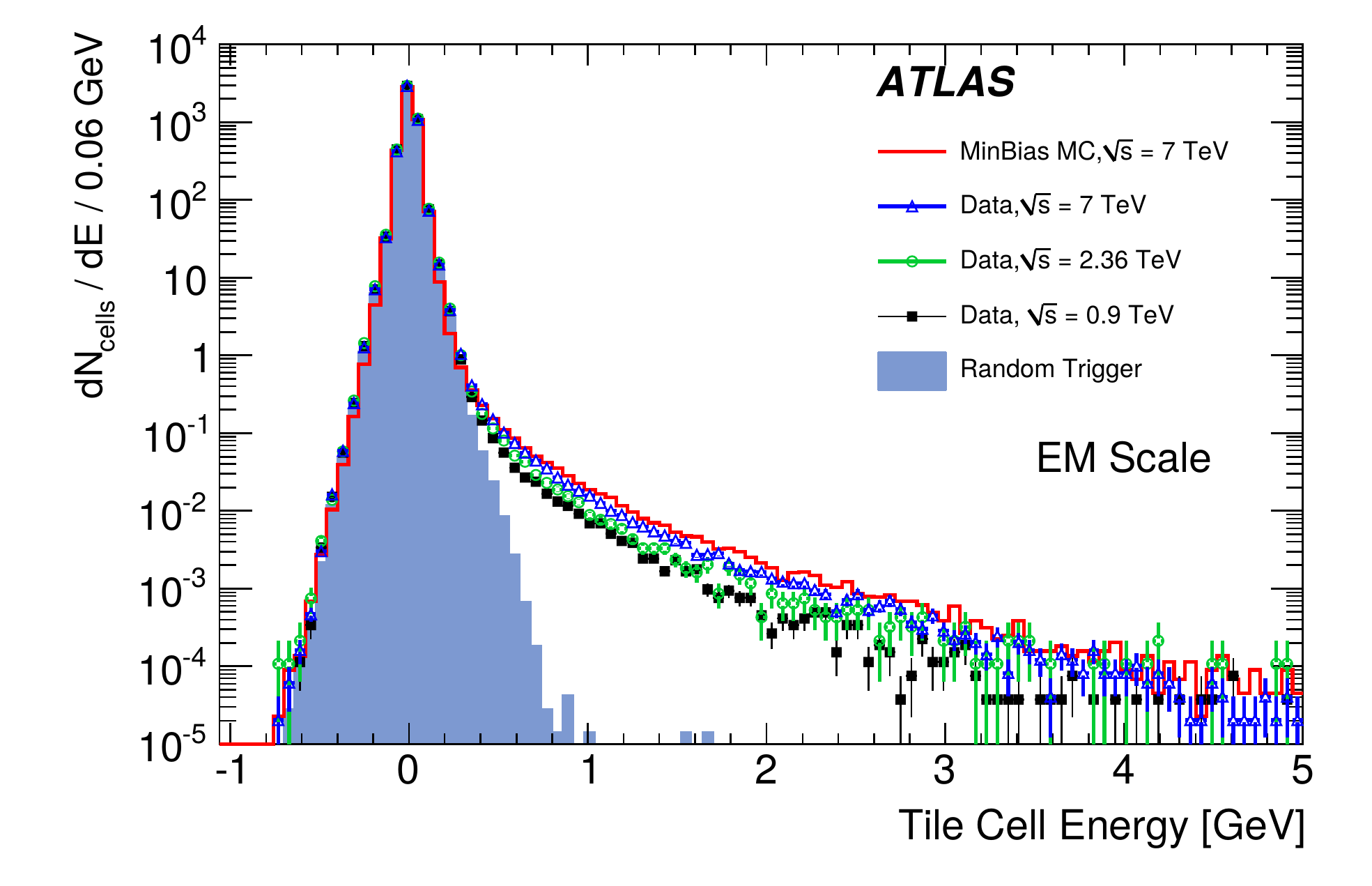}
\caption{The TileCal cell energy spectrum at the electromagnetic (EM) scale measured in 2010 data. The distributions from collision data at 7\,TeV, 2.36\,TeV, and 0.9\,TeV are superimposed with \PYTHIA minimum-bias Monte Carlo and randomly triggered events. \label{fig:cellEnergy}}
\end{figure}

The emulation of the electronic noise, specific to each individual calorimeter cell, is implemented in the digitisation of the Monte Carlo signals. It is assumed that it is possible to convert the measured cell noise to an ADC noise in the digitisation step, as the noise is added to the individual samples in the MC simulation. The correlations between the two channels in the cell are not considered. As a consequence, the constants of the double Gaussian function, used to generate the electronic noise in the MC simulation, are derived from the cell-level constants used in the real data.
As a closure test, after reconstruction of the cell energies in the
MC simulation the cell noise constants are calculated using the same procedure as
for real data. The reconstructed cell noise in the MC reconstruction is found to be in agreement with the original cell noise used as input from the real data. Good agreement between data and MC simulation of the energy of the TileCal cells, also for the low and negative amplitudes, is found (see Figure~\ref{fig:cellEnergy}). The measurement is performed using 2010 data where the pile-up contribution is negligible. The noise contribution can be compared with data collected using a random trigger.

\subsection{Pile-up noise}
\label{sec:pileupNoise}

The pile-up effects consist of two contributions, in-time pile-up
and out-of-time pile-up. The in-time pile-up originates from multiple
interactions in the same bunch crossing. In contrast, the out-of-time
pile-up comes from minimum-bias events from previous or subsequent bunch crossings. The out-of-time pile-up is present if the width of the electrical pulse (Figure~\ref{fig:pulseShapes}) is longer than the bunch spacing, which is the case in Run~1 where the bunch spacing in runs used for physics analyses is 50\,ns. These results are discussed in the following paragraphs.

The pile-up in the TileCal is studied as a function of the detector geometry and the mean number of inelastic $pp$ interactions per bunch crossing $\left<\mu\right>$ (averaged over all bunch crossings within a luminosity block and depending on the actual instantaneous luminosity and number of colliding bunches). 
The data are selected using a zero-bias trigger. This trigger unconditionally accepts events from collisions occurring a fixed number of LHC bunch crossings after a high-energy electron or photon is accepted by the L1 trigger, whose rate scales linearly with luminosity.
This triggering provides a data sample which is not biased by any residual signal in the calorimeter system. 
Minimum-bias MC samples for pile-up noise studies were generated
using \PYTHIAV{8} and \PYTHIAV{6} for 2012 and 2011 simulations, respectively.
The noise described in this section contains contributions from both electronic noise and pile-up, and is computed as the standard deviation (RMS) of the energy deposited in a given cell. 

\begin{figure}[tp]
\centering
\includegraphics[width=0.95\textwidth]{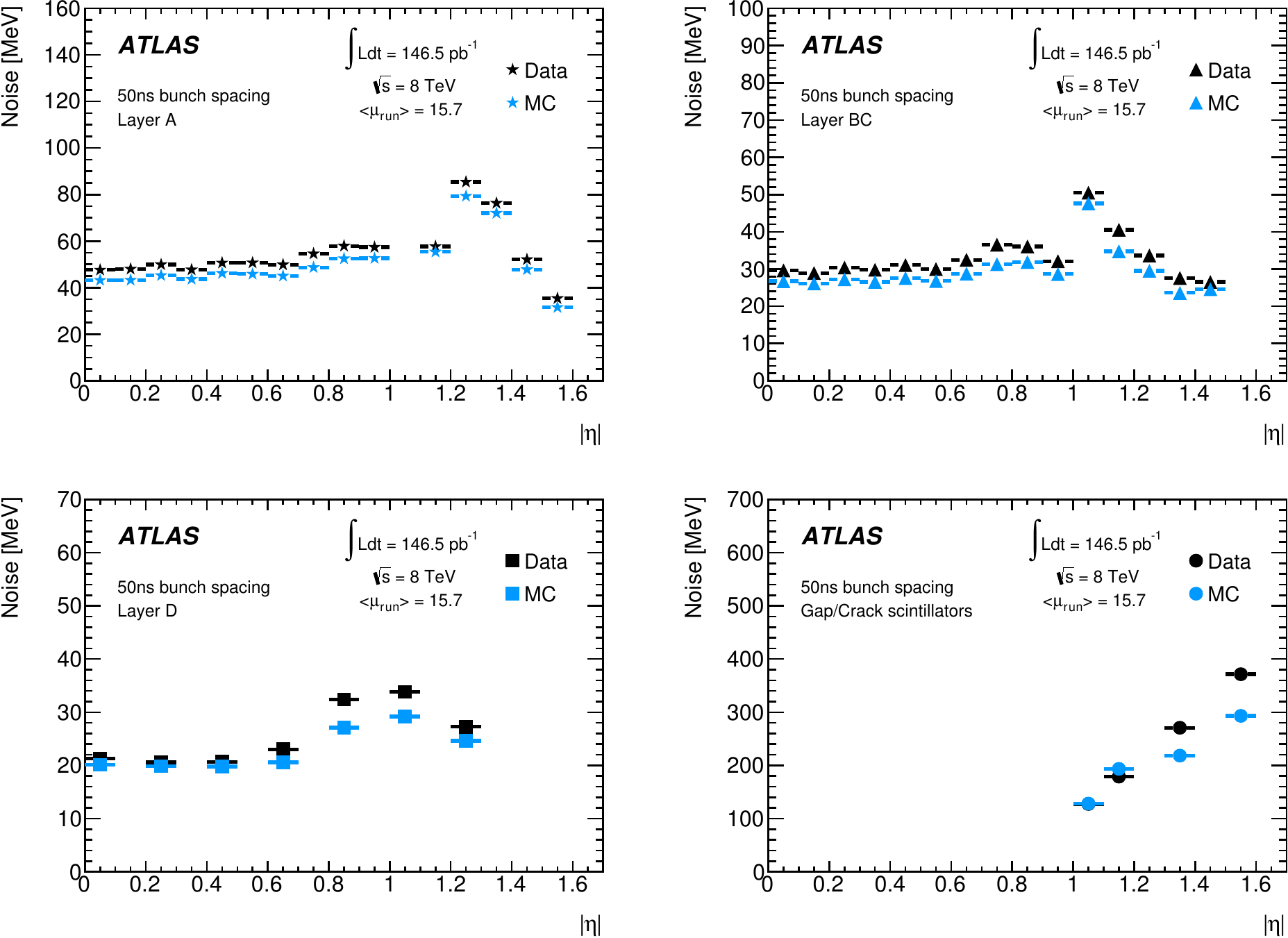}
\caption{The total noise per cell as a function of $|\eta|$ for
  $\left<\mu_{\mathrm{run}}\right>=15.7$, for the high--high gain combination.
  The data from a 2012 run, with 
  a bunch spacing of 50\,ns, are shown in black while the simulation is 
  shown in blue. 
  Four layers are displayed: layer A (top left), layer BC 
  (top right), layer D (bottom left), and the special gap and crack cells 
  (bottom right). The electronic noise component is shown in 
  Figure~\ref{fig:avgNoiseCellHGHG}.}
\label{fig:PileupAverage2012Layers}
\end{figure}

The total noise (electronic noise and contribution from pile-up) in
different radial layers as a function of $|\eta|$ for a medium pile-up
run (average number of interactions per bunch crossing over the whole run $\left<\mu_{\mathrm{run}}\right>=15.7$) taken in 2012 is shown in Figure~\ref{fig:PileupAverage2012Layers}. 
The plots make use of the $\eta$ symmetry of the detector and use cells from both $\eta$ sides in the calculation.
In the EB standard cells (all except E-cells), where the electronic noise is almost flat (see Figure~\ref{fig:avgNoiseCellHGHG}), the amount of upstream material as a function of $|\eta|$ increases~\cite{bib:detectorPaper}, causing the contribution of pile-up to the total noise to visibly decrease. 
The special cells (E1--E4), representing the gap and crack scintillators, experience the highest particle flux, and have the highest amount of pile-up noise, with cell E4 ($|\eta| = 1.55$) exhibiting about 380\,MeV of noise at $\left<\mu_{\mathrm{run}}\right>=15.7$ (of which about 5\,MeV is attributed to electronic noise). 
In general, the trends seen in the data for all layers as a function of $|\eta|$ are reproduced by the MC simulation. The total noise observed in data exceeds that in the simulation, the differences are up to 20\%.

\begin{figure}
\centering
\includegraphics[width=0.49\textwidth]{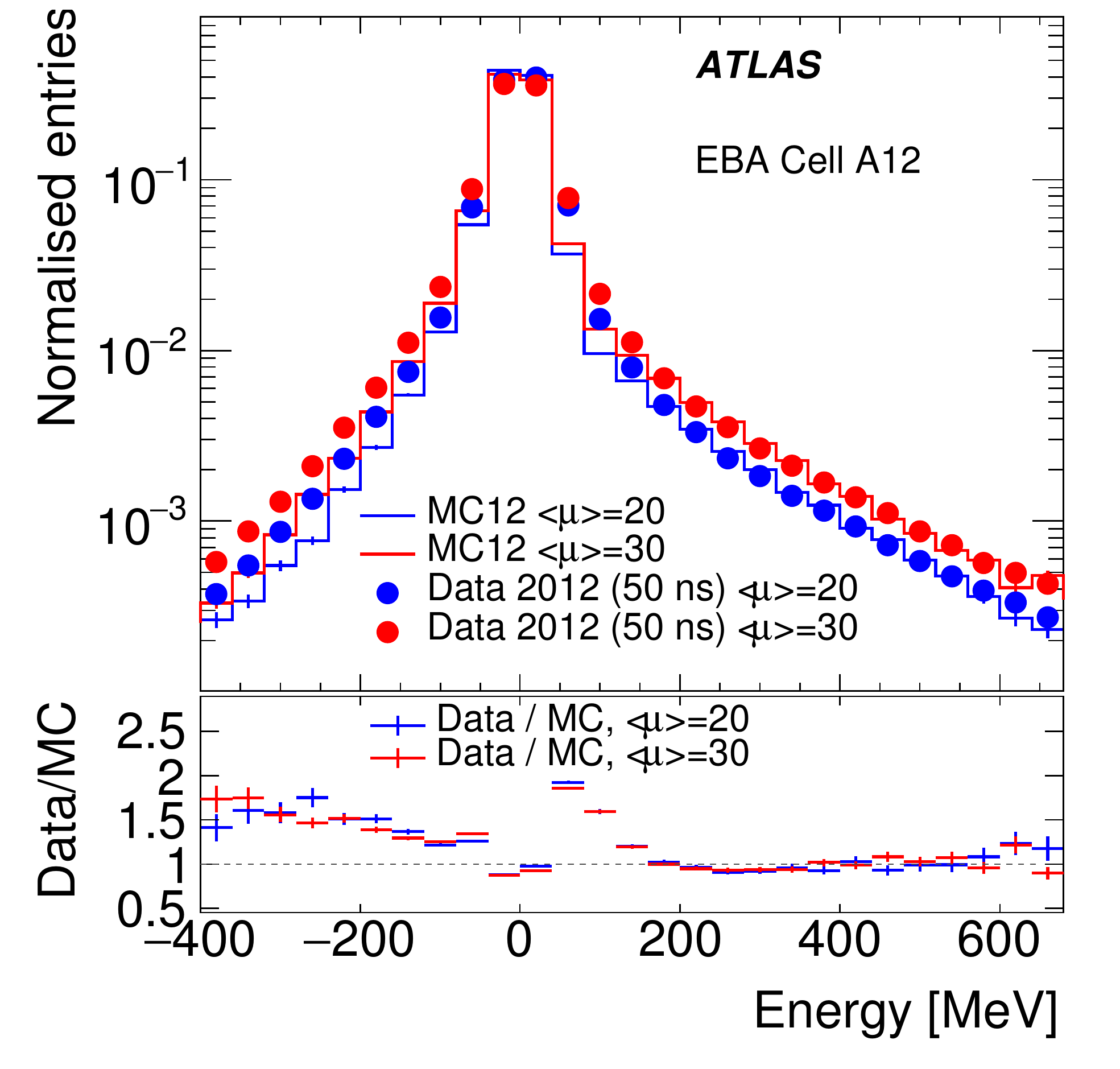}
\includegraphics[width=0.49\textwidth]{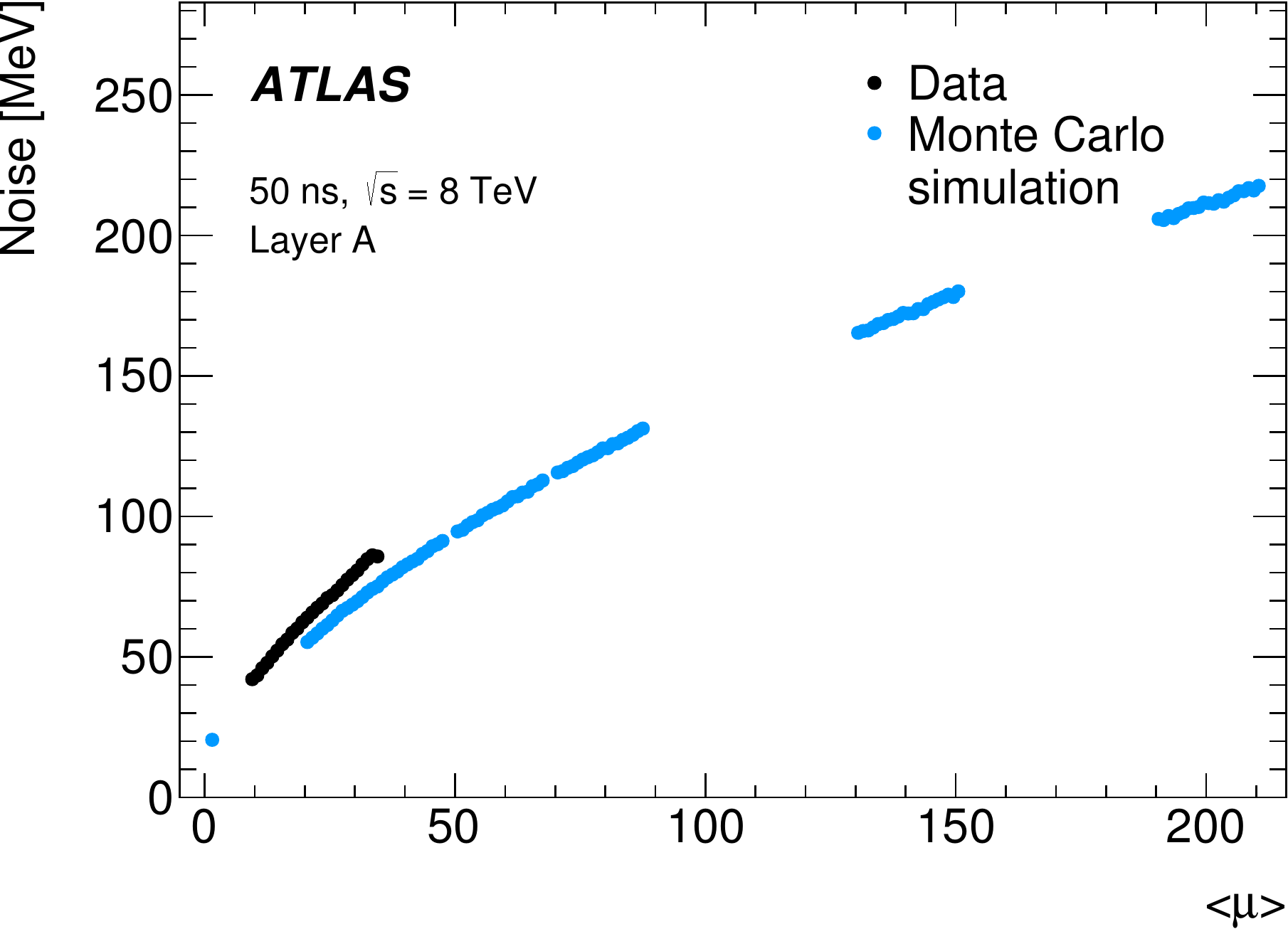}
\caption{The area-normalised energy spectra in cells A12 over all TileCal modules for two different pile-up conditions $\left<\mu\right>=20,\ 30$ (left) and the total noise, computed as the standard deviation of the energy distribution in all A-layer cells, as a function of $\left<\mu\right>$ (right) for data and minimum-bias MC simulation in 2012.}
\label{fig:PileupNoiseVsMu}
\end{figure}

The energy spectrum in the cell A12 is shown in
Figure~\ref{fig:PileupNoiseVsMu} (left) for two 
different pile-up conditions with $\left<\mu\right>=20$ and
$\left<\mu\right>=30$. 
The mean energy reconstructed in TileCal cells is centred around zero in 
minimum-bias events. Increasing pile-up widens the energy 
distribution both in data and MC simulation. Reasonable agreement between data and 
simulation is found above approximately 200\,MeV. However, below this energy, the 
simulated energy distribution is narrower than in data. 
This results in lower total noise in simulation compared with that in experimental 
data as already shown in Figure~\ref{fig:PileupAverage2012Layers}.
Figure~\ref{fig:PileupNoiseVsMu} (right) displays the average
noise for all cells in the A-layer as a function of $\left<\mu\right>$. Since this layer is the closest to the
beam pipe among LB and EB layers, it exhibits the largest increase in noise with increasing
$\left<\mu\right>$. When extrapolating $\left<\mu\right>$ to zero, the noise values are consistent with the electronic noise. 

\section{Calibration systems}\label{sec:calib}
 
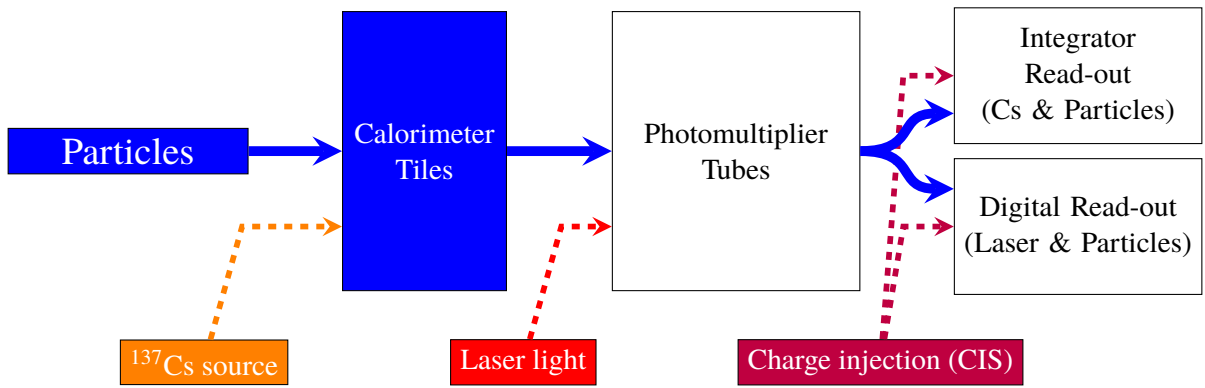
\begin{figure}
\centering

\begin{tikzpicture}[node distance=3cm,on grid,text centered]
  
  \newcommand{\CsSep}{1.4cm}
  \newcommand{\LaserSep}{1.0cm}

  \node[draw,fill=orange,text=white] (Cs) at (0.5,-2.8) {$^{137}$Cs source};

  \node[draw,fill=blue,text=white,text width=1.9cm,minimum height=3.7cm] (TileCal) at (3.4,0) {Calorimeter Tiles};
  \node[draw,text width=3cm,minimum height=3.7cm] (PMT) at (7.5,0) {Photomultiplier Tubes};
  \node[draw,text width=3.0cm,minimum height=1.80cm] (Integrator) at (12.0,1.0) {Integrator Read-out \\  (Cs \& Particles)};
  
  \draw[->,dashed,color=orange,line width=\CalibLineSize] (Cs) to (1.0,-\LaserSep) to ([yshift=-\LaserSep]TileCal.west);
  
  \newcommand{\CISSep}{1.4cm}
  \newcommand{\ShaperSep}{0.8cm}

  \node[draw,fill=purple,text=white] (CIS) at (9.4,-2.8) {Charge injection (CIS)};
  \node[draw,text width=3.0cm,minimum height=1.80cm,below=\ShaperSep of Integrator] (Shaper) at (12.0,-0.2) {Digital Read-out\\ (Laser \& Particles)};
  
  \draw[->,dashed,color=purple,line width=\CalibLineSize] (CIS) to (9.7cm,-\LaserSep) to ([yshift=-\LaserSep+\ShaperSep+0.2cm]Shaper.west);
  \draw[->,dashed,color=purple,line width=\CalibLineSize] (CIS) to (9.7cm,1.0cm) to ([yshift=-0.0cm]Integrator.west);
  
  \newcommand{\PhysicsSep}{0.0cm}
  
  \node[draw,fill=blue,text width=2.9cm,text=white,below=\PhysicsSep of Cs] (Physics) at (-0.5,0) {\Large Particles};
  
  \draw[->,color=blue,line width=\PhysLineSize] (Physics) to ([yshift=-\PhysicsSep]TileCal.west);  
  \draw[->,color=blue,line width=\PhysLineSize] ([yshift=-\PhysicsSep]TileCal.east) to ([yshift=-\PhysicsSep]PMT.west);
  \draw[->,color=blue,line width=\PhysLineSize] (PMT) edge[out=0,in=180, looseness = 1.5] ([yshift=-0.3cm+\ShaperSep]Shaper.west);
  \draw[->,color=blue,line width=\PhysLineSize] (PMT) edge[out=0,in=180, looseness = 1.5] ([yshift=-0.5cm]Integrator.west);
     
  \node[draw,fill=red,text=white] (Laser) at (4.7,-2.8) {Laser light};
  
  \draw[->,dashed,color=red,line width=\CalibLineSize] (Laser) to (5.2cm,-\LaserSep) to ([yshift=-\LaserSep]PMT.west);  
 
\end{tikzpicture}

\caption{The signal paths for each of the three calibration systems
  used by the TileCal. The physics signal is denoted by the
  thick solid line and the path taken by each of the calibration
  systems is shown with dashed lines.\label{fig:calibrationFig}} 
\end{figure}

Three calibration systems are used to maintain a time-independent
electromagnetic (EM) energy scale\footnote{The corresponding calibration constant converts the calorimeter signals, measured as electric charge in pC, to energy deposited by electrons that would produce these signals.} in the TileCal, and account for changes in the hardware and electronics due to irradiation, ageing, and
faults. 
The caesium (Cs) system calibrates the scintillator cells and PMTs but not the
front-end electronics used for collision data. The laser calibration
system monitors both the PMT and the same front-end electronics used
for physics. Finally, the charge injection system (CIS) calibrates and
monitors the front-end electronics. Figure~\ref{fig:calibrationFig}
shows a flow diagram that summarises the components of the read-out
tested by the different calibration systems. These three complementary 
calibration systems also aid in identifying the source of problematic 
channels. Problems originating strictly in the read-out electronics are seen
by both laser and CIS, while problems related solely 
to the PMT are not detected by the charge injection system. 

The signal amplitude $A$ is reconstructed in units of ADC counts using
the OF algorithm defined in Eq.~(\ref{eq:OF}). The conversion to channel
energy, $E_{\mathrm{channel}}$, is performed with the following formula:

\begin{equation}
E_{\mathrm{channel}} = A \cdot C_{\mathrm{Cs}} \cdot
C_{\mathrm{laser}} \cdot  C_{{\mathrm{ADC}\rightarrow\mathrm{pC}},\mathrm{CIS}} / C_{\mathrm{TB}} 
\label{eq:channelEnergy}
\end{equation}

where each $C_i$ represents a calibration constant or correction factor, which are described in the following paragraphs.

The overall EM scale $C_{\mathrm{TB}}$  was determined
in dedicated beam tests with electrons incident on 11\% of the
production modules~\cite{bib:tileReadiness,bib:emscaleTB}. It amounts to $1.050\pm0.003$\,pC/GeV with an RMS
spread of $(2.4 \pm 0.1)$\% in layer~A, with additional corrections applied to the other
layers as described in Section~\ref{ref:subsecCs}. The remaining
calibration constants in Eq.~(\ref{eq:channelEnergy}) are used to
correct for both inherent differences and time-varying optical and electrical
read-out differences between individual channels. They are calculated
using three dedicated calibration systems (caesium, laser, charge
injection) that are described in more
detail in the following subsections. Each calibration system determines
their respective constants to a precision better than 1\%.

\subsection{Caesium calibration}\label{ref:subsecCs} 
The TileCal exploits a radioactive $^{137}$Cs source to maintain the
global EM scale and to monitor the optical and electrical
response of each PMT in the ATLAS environment~\cite{bib:cesiumRef}.
A hydraulic system moves this Cs source through the calorimeter using
a network of stainless steel tubes inserted into small holes in each
tile scintillator.\footnote{The E3 and E4 cells are not part of this Cs
  mechanical system, and therefore are not calibrated using the Cs
  source.}
The beta decay of the $^{137}$Cs source produces 0.665\,MeV photons at
a rate of $\sim10^6$\,Hz, generating scintillation light in each
tile.\footnote{Although the Cs signal can be measured from each single tile~\cite{bib:tileOpticalInstrumentation}, the total Cs signal averaged over all tiles associated to the given cell is considered for the Cs constant evaluation.} 
In order to collect a sufficient signal, the electrical read-out
of the Cs calibration is performed using the integrator read-out path;
therefore the response is a measure of the integrated current in a PMT\@. 
As is described in Section~\ref{subsec:CIS}, dedicated
calibration runs of the integrator system show that the stability of
individual channels was better than 0.05\% throughout Run~1. 

In June 2009 the high voltage (HV) of each PMT was modified so that the Cs source response in the same PMTs was equal to that observed in the test beam.
Corrections are applied to account for differences between these two environments, namely the activity of the different sources and half-life of $^{137}$Cs.

Three Cs sources are used to calibrate the three physical
TileCal partitions in the ATLAS detector, one in the LB and one in
each EB\@. A fourth source was used for beam tests and another is used
in a surface research laboratory at CERN\@.
The response to each of the five sources was measured in April
2009~\cite{bib:tileReadiness} and again in March 2013 at the end of
Run~1 using a test module for both the LB and EB\@. The relative
response to each source measured on these two dates agrees to within
0.2\% and confirms the expected $^{137}$Cs activity during Run~1.

A full Cs calibration scan through all tiles takes approximately six
hours and was performed roughly once per month during Run~1.  
The precision of the Cs calibration in one typical cell is approximately 0.3\%. 
For cells on the extreme sides of a partition the precision is 0.5\%
due to larger uncertainties associated with the source position.
Similarly, the precision for the narrow C10 and D4 ITC cells is 3\% and
$\sim$1\%, respectively, due to the absence of an iron end-plate between the
tile and Cs pipe. It makes more challenging the distinction between the
desired response when the Cs source is inside that particular tile of
interest versus a signal detected when the source moves towards a neighbouring tile row. 

\begin{figure}[tbp]
\centering
\includegraphics[width=0.49\textwidth]{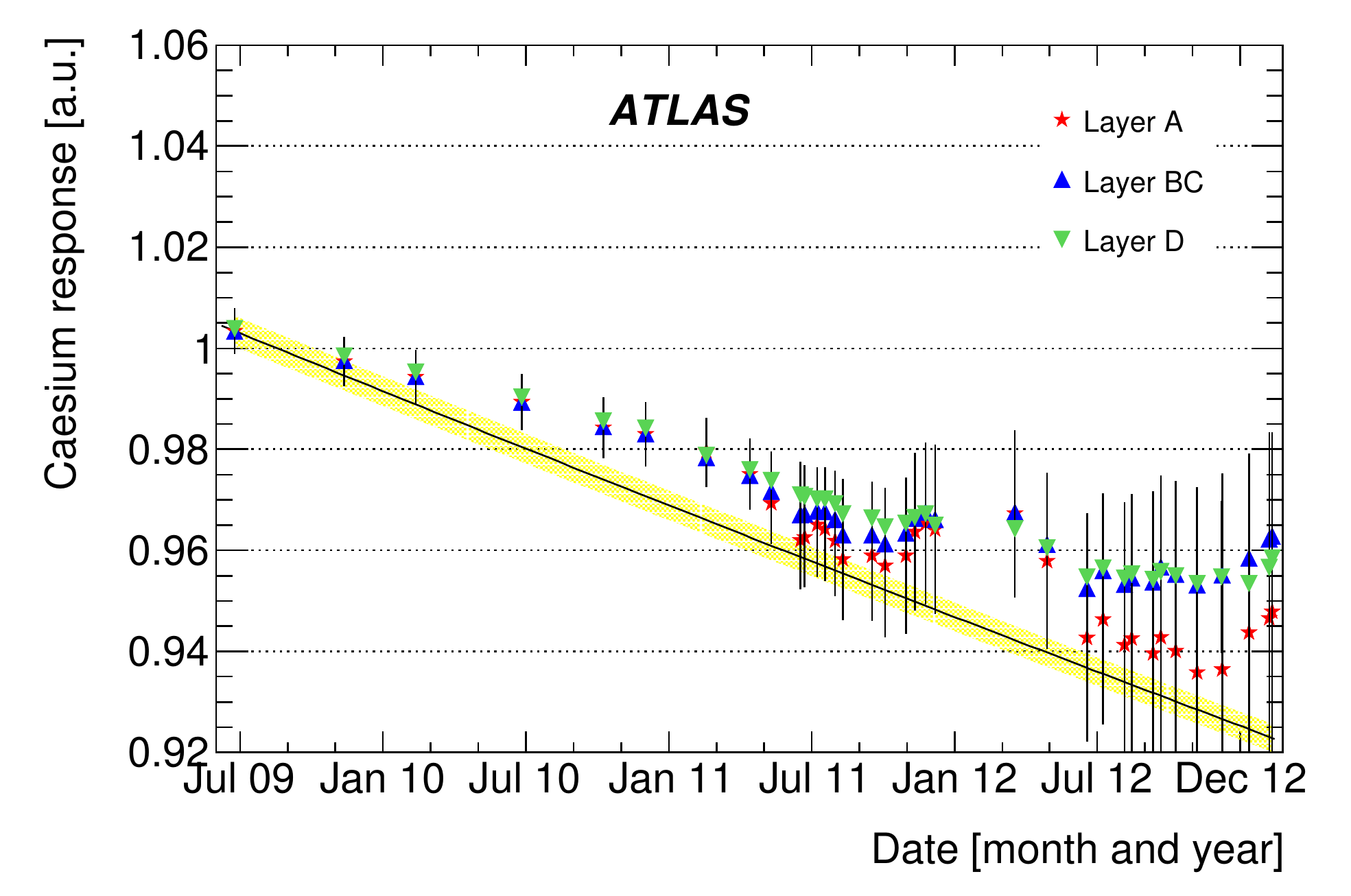}
\includegraphics[width=0.49\textwidth]{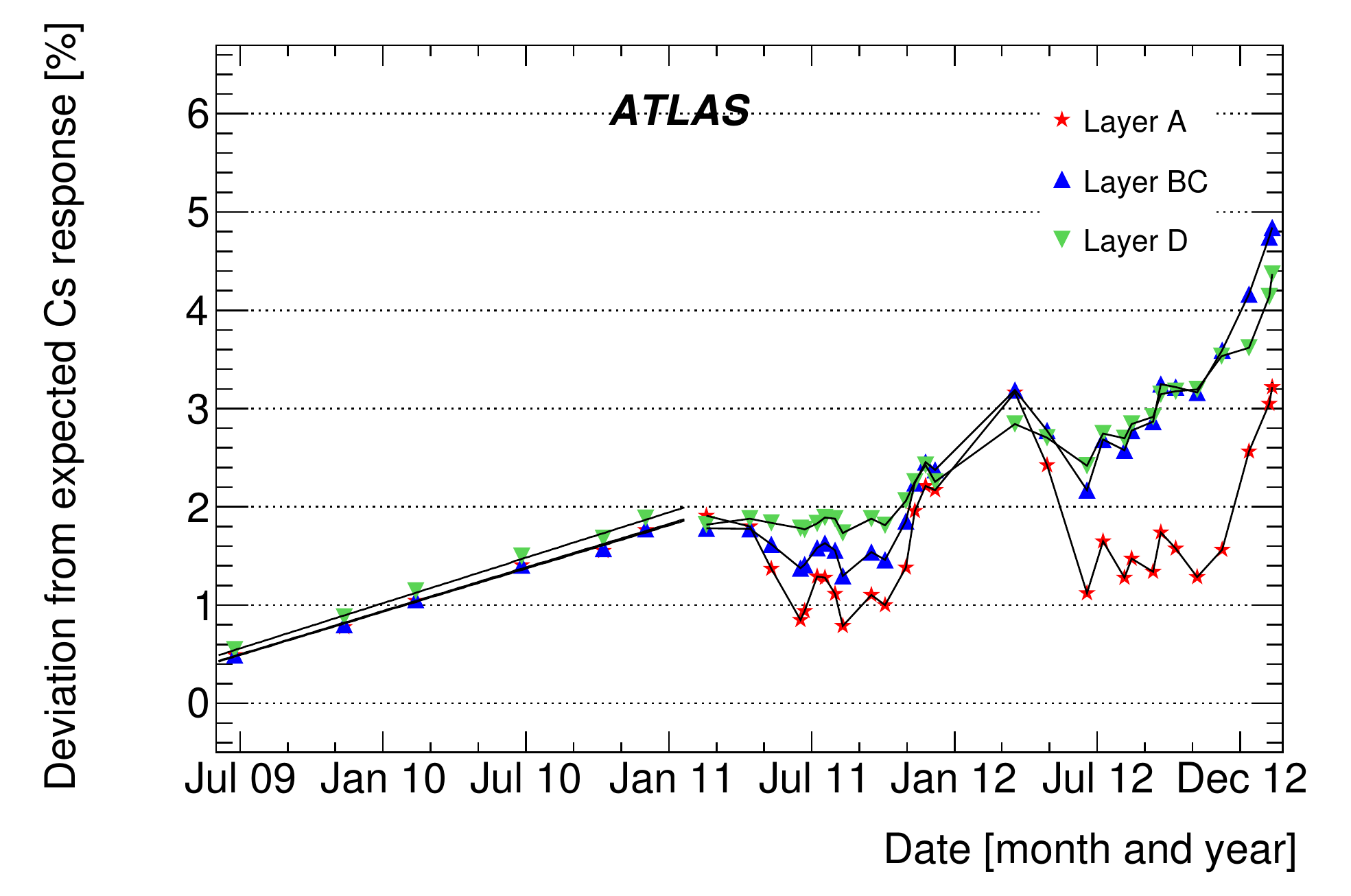}
\caption{The plot on the left shows the average response (in arbitrary units, a.u.) from all cells within a given 
  layer to the $^{137}$Cs source as a function of time from July 2009 to December 2012. 
  The solid line represents the expected response, where
  the Cs source activity decreases in time by $-2.3\%$/year. The coloured band shows the declared precision of the Cs calibration ($\pm 0.3$\%). The plot on the right shows the 
  percentage difference of the response from the expectation as a
  function of time averaged over all cells in all partitions. Both plots display only 
  the measurements performed with the magnetic field at its nominal value. The first
  points in the plot on the right deviate from zero, as the initial HV equalisation was done in
  June 2009 using Cs calibration data taken without the magnetic
  field (not shown in the plot). The increasing Cs response in the last three measurements
  corresponds to the period without collisions after the Run~1 data-taking finished.}
\label{fig:CsResponse}
\end{figure}

The Cs response as a function of time is shown in
Figure~\ref{fig:CsResponse} (left) averaged over all cells of a given
radial layer. The solid line, enveloped by an uncertainty
band, represents the expected response due to the reduced activity of
the three Cs sources in the ATLAS detector ($-2.3\%$/year). The error
bars on each point represent the RMS spread of the response in all
cells within a layer.  
There is a clear deviation from this expectation line, with the
relative difference between the measured and expected values shown in
Figure~\ref{fig:CsResponse} (right). 
The average up-drift of the response relative to the expectation
was about 0.8\%/year in 2009--2010.
From 2010 when the LHC began operation, the upward and downward trends
are correlated with beam conditions--the downward trends
correspond to the presence of colliding beams, while the
upward trends are evident in the absence of collisions.
This effect is pronounced in the innermost layer~A, 
while for layer~D there is negligible change in response.
This effect is even more evident when looking at pseudorapidity-dependent responses in individual layers. While in most LB-A cells a deviation of approximately 2.0\% is seen (March 2012 to December 2012), in EB-A cells the deviation ranges from 3.5\% (cell A13) to 0\% (outermost cell A16).
These results indicate the total effect, as seen by the Cs system,
is due to the scintillator irradiation and PMT gain changes (see
Section~\ref{subsec:comboCalib} for more details).

The Cs calibration constants are derived using Cs calibration data taken
with the full ATLAS magnetic field system on, as in the nominal
physics configuration. The magnetic field effectively increases the
light yield in scintillators approximately by {0.7\%} in the LB and
{0.3\%} in the EB\@.

Since the response to the Cs source varies across the surface of each
tile, additional layer-dependent weights are applied to maintain the
EM scale across the entire
calorimeter~\cite{bib:emscaleTB}. These weights 
reflect the different radial tile sizes in individual layers and the
fact that the Cs source passes through tiles at their outer edge.

The total systematic uncertainty in applying the EM scale from the test beam environment to ATLAS was found to be 0.7\%, with the largest contributions from variations in the response to the Cs sources in the presence of a magnetic field (0.5\%) and the layer weights (0.3\%)~\cite{bib:emscaleTB}.

\subsection{Laser calibration} 
A laser calibration system is used to monitor and correct for PMT response variations between Cs scans and to monitor channel timing during periods of collision data-taking~\cite{bib:laserRef, bib:laserRun1}.

This laser calibration system consists of a single laser source,
located off detector, able to produce short light pulses that
are simultaneously distributed by optical fibres to all 9852 PMTs. 
The intrinsic stability of the laser light was found to be 2\%, so to
measure the PMT gain variations to a precision of better than 0.5\% using the laser source, the response of the PMTs is normalised to the signal measured by a dedicated photodiode. The stability of this photodiode is monitored by an $\alpha$-source and, 
throughout 2012, its stability was shown to be 0.1\%, and the linearity of the associated electronics response within 0.2\%.

The calibration constants, $C_{\mathrm{laser}}$ in
Eq.~(\ref{eq:channelEnergy}), are calculated for each channel relative to a reference run taken just after a Cs scan, after new Cs
calibration constants are extracted and applied. 
Laser calibration runs are taken for both gains 
approximately twice per week.

\begin{figure}[tbp]
\centering
\includegraphics[width=1.0\textwidth]{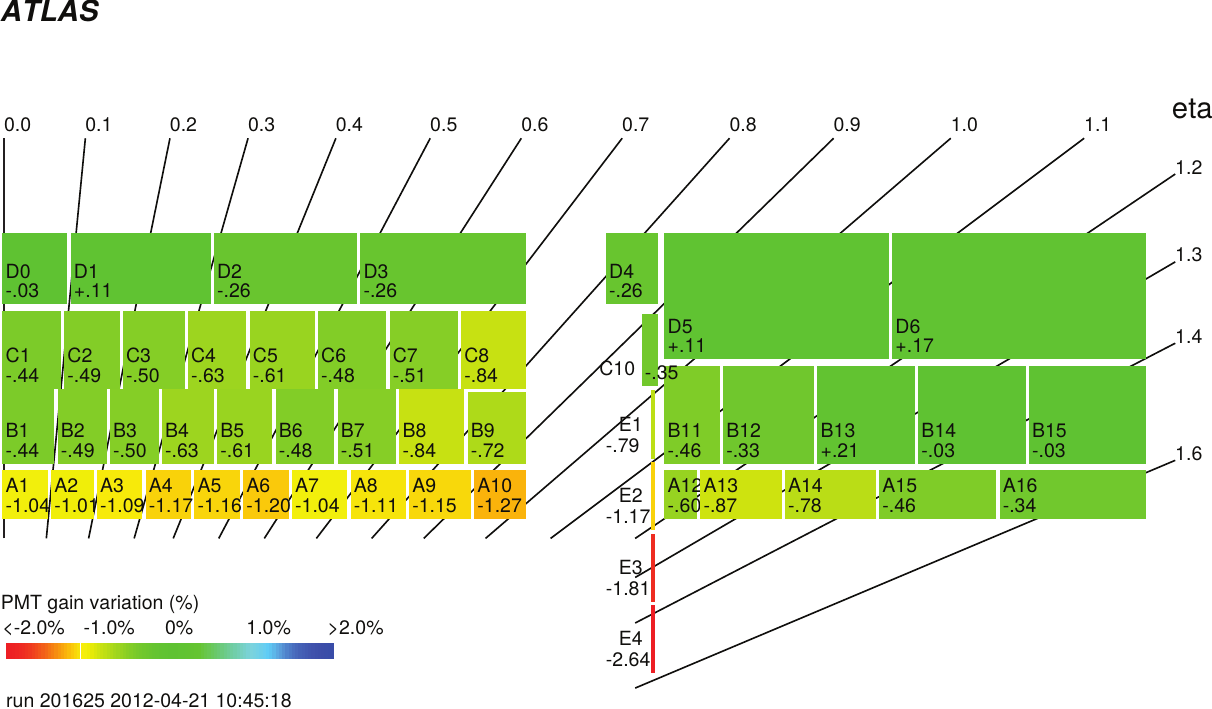}
\caption{The mean gain variation in the PMTs for each cell type averaged
  over $\phi$ between a stand-alone laser calibration run taken on 21
  April 2012 and a laser run taken before the collisions on 19 March
  2012. For each cell type, the gain variation was defined as the mean of a
  Gaussian fit to the gain variations in the channels associated with
  this cell type. A total of 64 modules in $\phi$ were used for each
  cell type, with the exclusion of known pathological
  channels.\label{fig:laserGainOneMonth}}
\end{figure}

For the E3 and E4 cells, where the Cs calibration is not possible, the
reference run is taken as the first laser run before data-taking of
the respective year.  
A sample of the mean gain variation in the PMTs for each cell type averaged over $\phi$ between 19 March 2012 (before the start of collisions) and 21 April 2012 is shown in Figure~\ref{fig:laserGainOneMonth}.
The observed down-drift of approximately 1\% mostly affects cells at
the inner radius with higher current draws.

The laser calibration constants were not used during 2010. For data
taken in 2011 and 2012 these constants were calculated and applied for
channels with PMT gain variations larger than 1.5\% (2\%) in the LB
(EB) as determined by the low-gain calibration run, with a consistent drift as measured in the equivalent high-gain run. 
In 2012 up to 5\% of the channels were corrected using the laser
calibration system. The laser calibration constants for E3 and E4 cells were 
applied starting in the summer of 2012, and were retroactively applied
after the ATLAS data were reprocessed with updated detector
conditions. 
The total statistical and systematic errors of the laser calibration
constants are 0.4\% for the LB and 0.6\% for the EBs, where the EBs
experience larger current draws due to higher exposure.

\subsection{Charge injection calibration}
\label{subsec:CIS}
The charge injection system is used to calculate the constant
$C_{{\mathrm{ADC}\rightarrow\mathrm{pC}},\mathrm{CIS}}$ in
Eq.~(\ref{eq:channelEnergy}) and applied for physics signals and
laser calibration data. A part of this system is also used to
calibrate the gain conversion constant for the slow integrator read-out.

All 19704 ADC channels in the fast front-end electronics are
calibrated by injecting a known charge from the 3-in-1 cards,
repeated for a wide range of charge values (approximately 0--800\,pC
in low-gain and 0--12\,pC in high-gain). 
A linear fit to the mean reconstructed signal (in ADC counts) yields the constant $C_{{\mathrm{ADC}\rightarrow\mathrm{pC}},\mathrm{CIS}}$.
During Run~1 the precision of the system was better than 0.7\% for each
ADC channel.   

\begin{figure}[tp]
\centering
\includegraphics[width=0.49\textwidth]{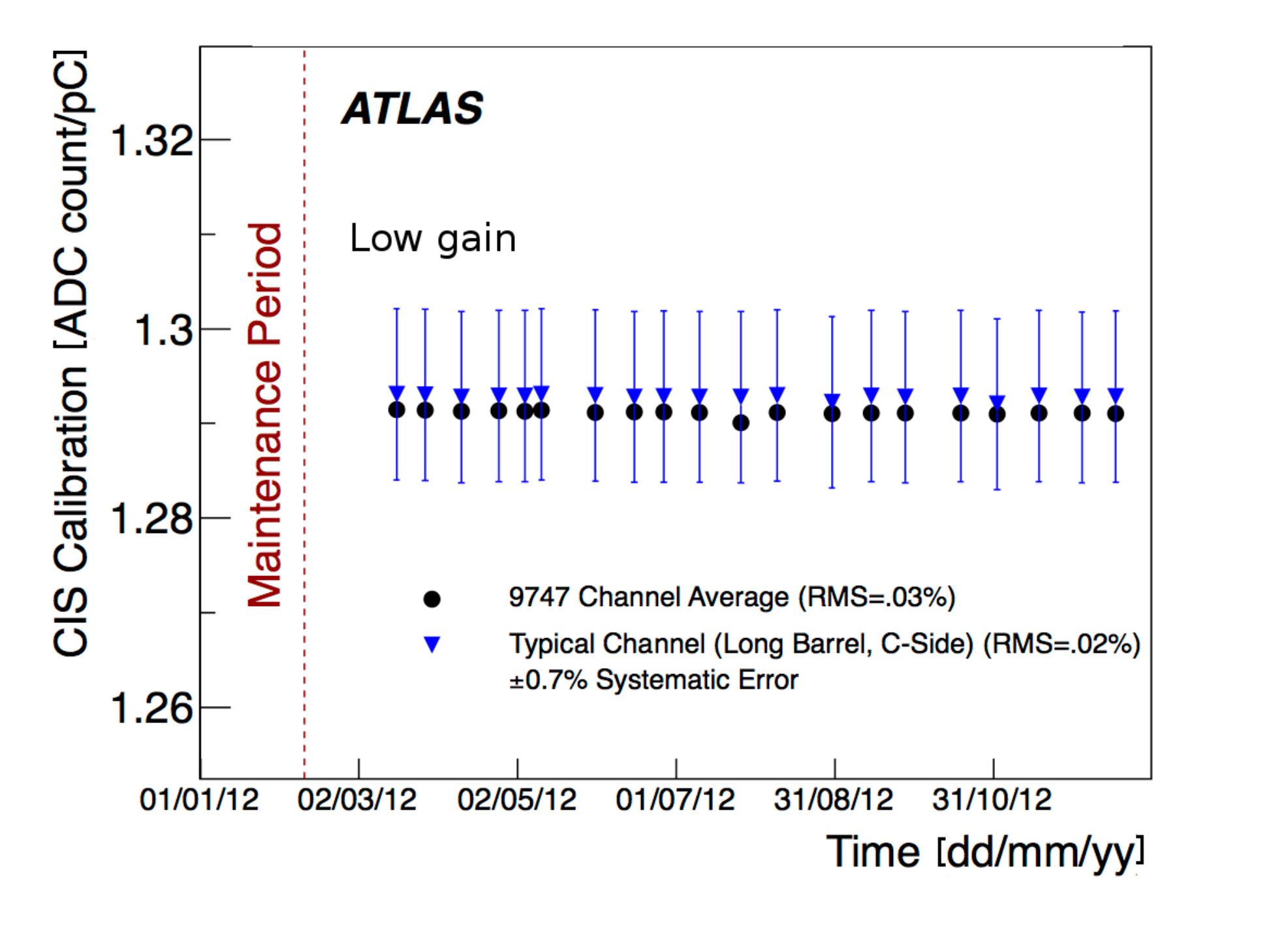}
\includegraphics[width=0.49\textwidth]{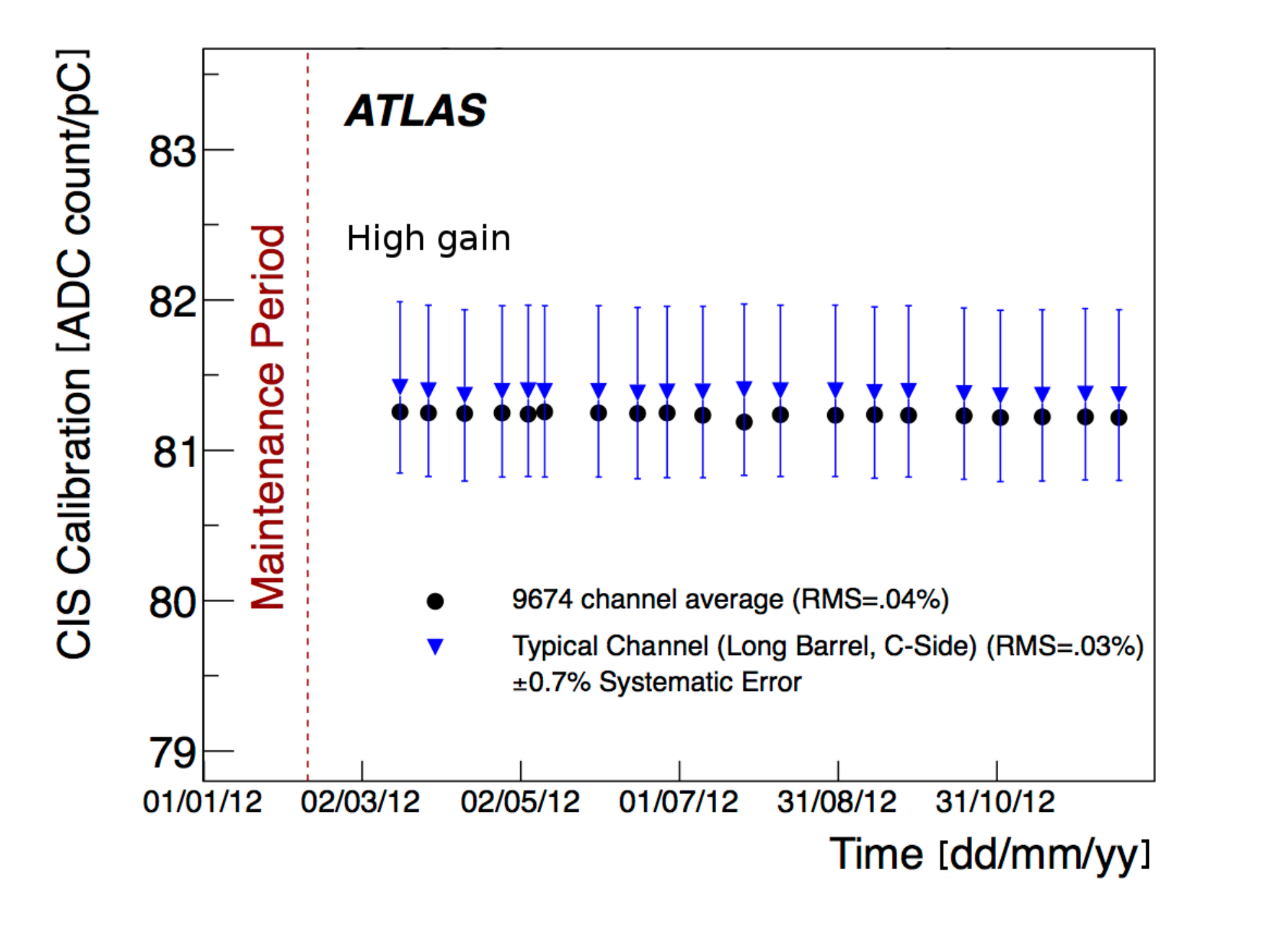}
\caption{Stability of the charge injection system constants for the low-gain ADCs (left) and high-gain ADCs (right) as a function of time in 2012. Values for the average over all channels and for one typical channel with the 0.7\% systematic uncertainty are shown. Only good channels not suffering from damaged components relevant to the charge injection calibration are included in this figure. \label{fig:CISStability}}
\end{figure}

Charge injection calibration data are typically taken twice per week
in the absence of colliding beams. 
For channels where the calibration constant varies by more than 1.0\%
the constant is updated for the energy reconstruction.
Figure~\ref{fig:CISStability} shows the stability of the charge
injection constants as a function of time in 2012 for the high-gain
and low-gain ADC channels. Similar stability was seen throughout 2010
and 2011. At the end of Run~1 approximately 1\% of all ADC channels
were unable to be calibrated using the CIS mostly due to hardware problems 
evolving in time, so default
$C_{{\mathrm{ADC}\rightarrow\mathrm{pC}},\mathrm{CIS}}$ constants
are used in such channels.

The slow integrator read-out is used to measure the PMT current over
$\sim\!\!10$\,ms. Dedicated runs are periodically taken to calculate the integrator gain conversion constant for each of the six gain settings, by fitting the linear relationship between the injected current and measured voltage response.
The stability of individual channels is better than 0.05\%, the
average stability is better than 0.01\%.

\begin{figure}[tp]
\centering
\includegraphics[width=0.57\textwidth]{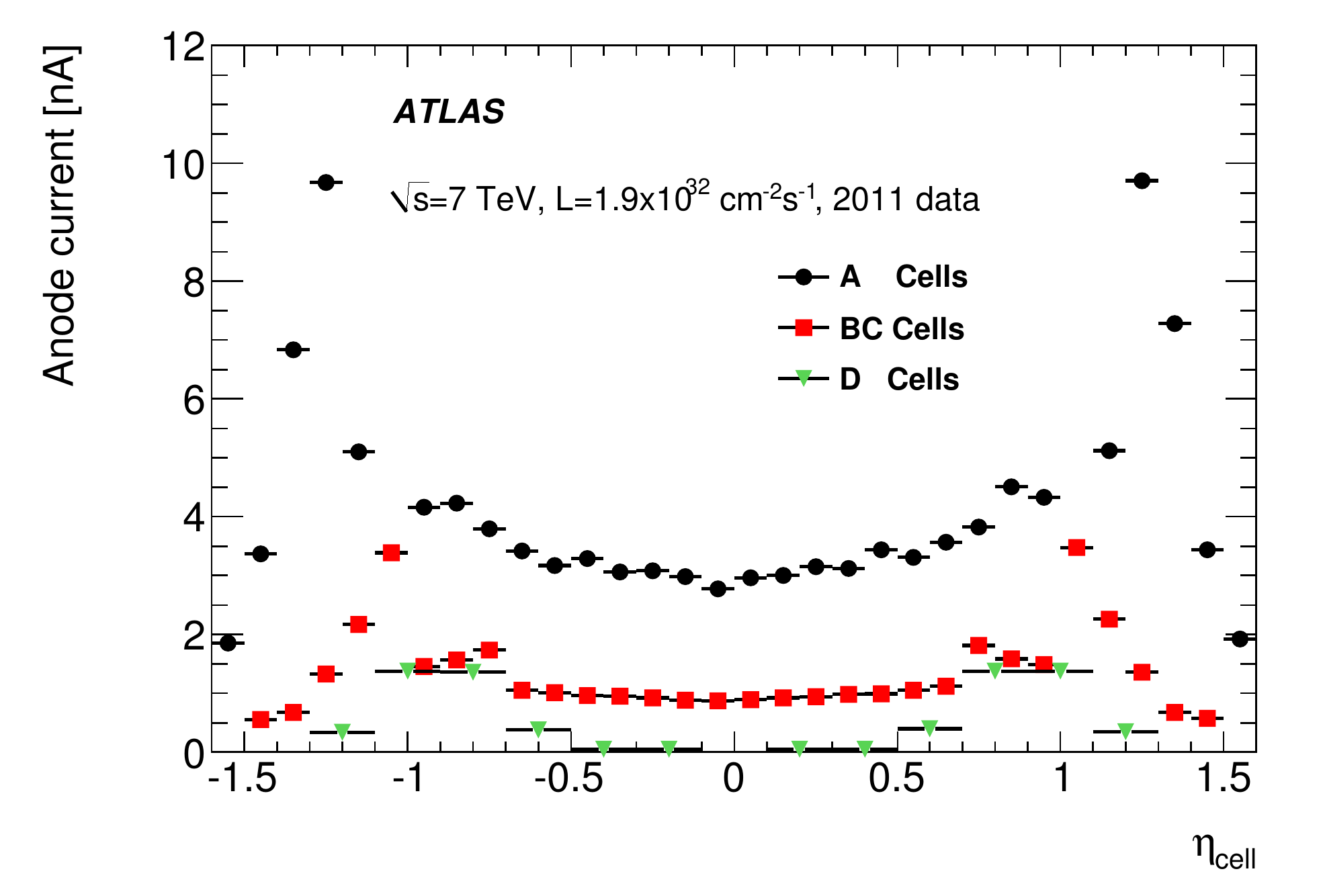}
\caption{The PMT current as measured by the slow integrator read-out as
  a function of cell $\eta$ and averaged over all modules for the three layers in the LB and EB,
  using minimum-bias data collected in 2011 at a fixed instantaneous
  luminosity ($1.9\times10^{32}$\,cm$^{-2}$s$^{-1}$).}
\label{fig:intCurrentVsEta}
\end{figure}

\subsection{Minimum-bias currents}\label{subsec:minBias}

Minimum-bias (MB) inelastic proton--proton interactions at the LHC
produce signals in all PMTs, which are used
to monitor the variations of the calorimeter response over time using the integrator read-out (as used by the Cs calibration system).\footnote{The usage of the integrators allows for a high rate of minimum-bias events, much higher than could be achieved with the fast read-out.}
The MB rate is proportional to the instantaneous luminosity, and produces signals in all subdetectors, which are uniformly distributed around the interaction point. In the integrator circuit of the Tile Calorimeter this signal is seen as an increased PMT current $I$ calculated from the ADC voltage measurement as:

\begin{equation*}
I [\mathrm{nA}] = \frac{\mathrm{ADC} [\mathrm{mV}]   - \mathrm{ped}
  [\mathrm{mV}] }{\mathrm{Int.\ gain} [\mathrm{M}\Omega]}\ ,
\end{equation*}

where the integrator gain constant (Int.\ gain) is calculated using the CIS calibration, and the pedestal (ped) from physics runs before collisions but with circulating beams (to account for beam background sources such as beam halo and beam--gas interactions).
Studies found the integrator has a linear response (non-linearity
$<1\%$) for instantaneous luminosities between $1\times10^{30}$ and $3\times10^{34}$\,cm$^{-2}$s$^{-1}$.

Due to the distribution of upstream material and the distance of cells from the interaction point the MB signal seen in the TileCal is not expected to be uniform. Figure~\ref{fig:intCurrentVsEta} shows the measured PMT current versus cell $\eta$ (averaged over all modules) for a fixed instantaneous luminosity. 
As expected, the largest signal is seen for the A-layer cells which are closer to the interaction point, with cell A13 ($|\eta| = 1.3$) located in the EB and (with minimal upstream material)  exhibiting the highest currents.  

The currents induced in the PMTs due to MB activity are used to
validate response changes observed by the Cs calibration system as
well as for response monitoring during the physics 
runs. Moreover, they probe the response in the E3 and E4 cells,
which are not calibrated by Cs.

\subsection{Combination of calibration methods}
\label{subsec:comboCalib}

\begin{figure}[tbp]
\centering
\includegraphics[width=0.75\textwidth]{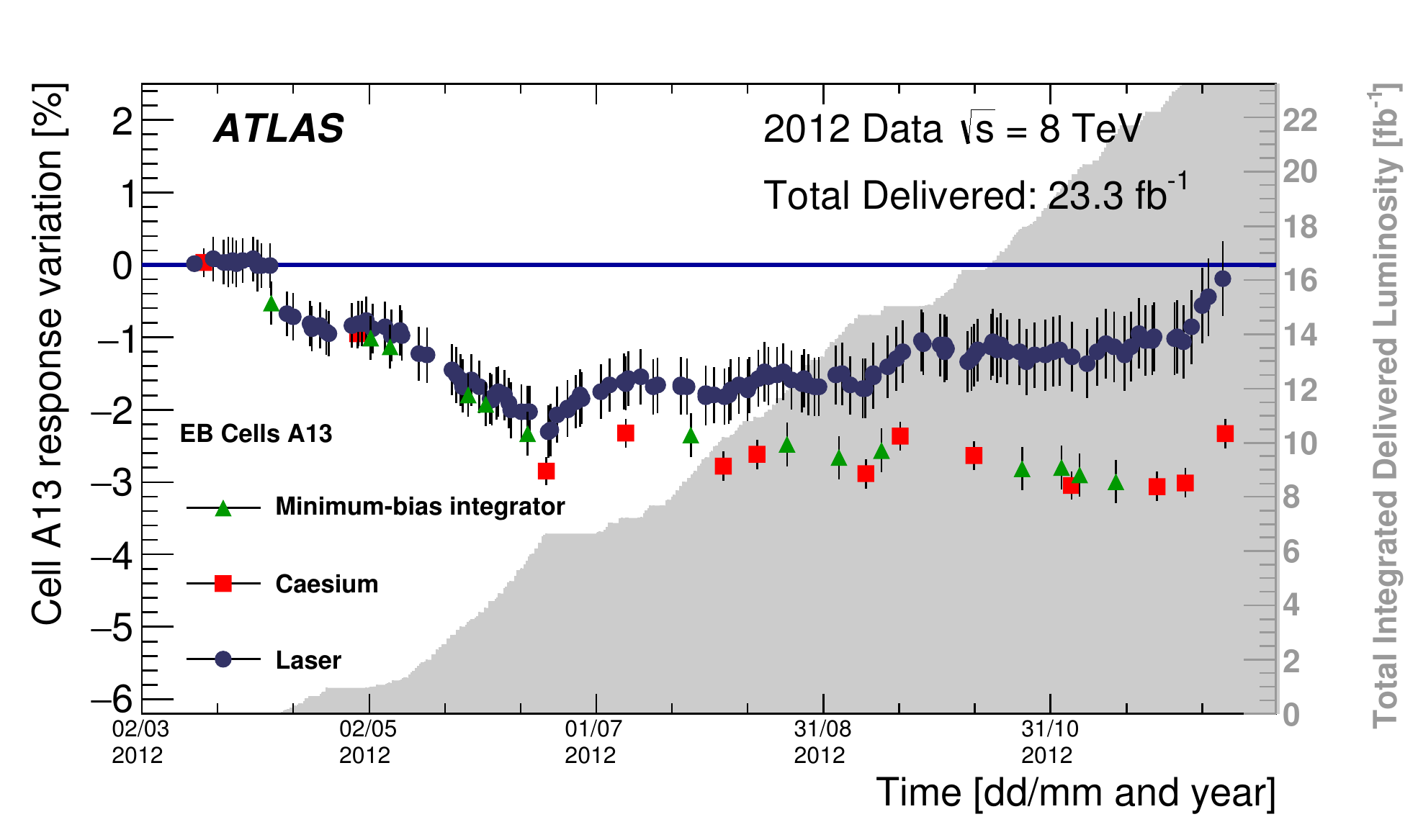}
\caption{The change of response seen in cell A13 by the minimum-bias, caesium, and laser systems throughout 2012. Minimum-bias data cover the period from the beginning of April to the beginning of December 2012. The Cs and laser results cover the period from mid-March to mid-December. The variation versus time for the response of the three systems was normalised to the first Cs scan (mid-March, before the start of collisions data-taking). The integrated luminosity is the total delivered during the proton--proton collision period of 2012. The down-drifts of the PMT gains (seen by the laser system) coincide with the collision periods, while up-drifts are observed during machine development periods. The drop in the response variation during the data-taking periods tends to decrease as the exposure of the PMTs increases. The variations observed by the minimum-bias and Cs systems are similar, both measurements being sensitive to PMT drift and scintillator irradiation.\label{fig:combinedCalib}}
\end{figure}

The TileCal response is expected to vary over time, with particular
sensitivity to changing LHC luminosity conditions.
Figure~\ref{fig:combinedCalib} shows the variation of the response to MB, Cs, and laser calibration systems for cell A13 as a function of the time in 2012. 
Cell A13 is located in the EB, and due to the smaller amount of
upstream material, it is exposed to one of the highest radiation doses
of all cells as also seen in Figure~\ref{fig:intCurrentVsEta}. 
To disentangle the effects of PMT and scintillator changes one can study the laser versus MB (or Cs) responses.

The PMT gain, as monitored with the laser, is known to decrease with
increasing light exposure due to lower secondary emissions from the
dynode surfaces~\cite{bib:Gupta}.\footnote{The decrease in the gain
 depends on several factors, including temperature, intensity and
 duration of the light exposure, and previous history of the PMT.}
When a PMT is initially exposed to light 
after a long period of `rest', its gain decreases rapidly and then a
slow stabilisation occurs~\cite{bib:Hillert}. This behaviour is
demonstrated in Figure~\ref{fig:combinedCalib}--the data-taking in
2012 started after four months of 
inactivity, followed by the gain stabilisation after several weeks of LHC
operation. The same trends were also observed in 2011.  The periods
of recovery, where the laser response tends towards initial
conditions, coincide with times when LHC is not colliding
protons. This is consistent with the known behaviour of `fatigued'
PMTs that gradually return towards original operating condition after
the exposure is removed~\cite{bib:Weitkamp}.
A global PMT gain increase of 0.9\% per year is observed even
without any exposure (e.g.\ between 2003 and 2009). This is
consistent with Figure~\ref{fig:CsResponse} (right)--after 3.5 years
the total gain increase corresponds to approximately 3.5\%.
Throughout Run~1 the maximum loss of the PMT gain in A13 is approximately 3\%,
but at the end of 2012 after periods of inactivity the gain essentially recovered from this loss.

The responses to the Cs and MB systems, which are sensitive to both the PMT gain changes and scintillator irradiation show consistent behaviour.  
The difference between MB (or Cs) and laser response variations is interpreted as an effect of the scintillators' irradiation. 
The transparency of scintillator tiles is reduced after radiation exposure~\cite{bib:scintDamage}; in the TileCal this is evident in the continued downward response to MB events (and Cs) with increasing integrated luminosity of the collisions, despite the eventual slow recovery of the PMTs as described above. 
In the absence of the radiation source the annealing process is
believed to slowly restore the scintillator material, hence improving the collected light yield. 
The rate and amount of scintillator damage and recovery are complicated combinations of factors, such as particle energies, temperatures, exposure rates and duration, and are difficult to quantify. 

The overlap between the different calibration systems allows calibration and monitoring 
of the complete hardware and read-out chain of
the TileCal, and correct for response changes with fine granularity for
effects such as changing luminosity conditions.
These methods enable the identification of sources of response
variations, and during data-taking, the correction of these variations
to maintain the global EM scale throughout Run~1.
When possible, problematic components are repaired or replaced during maintenance periods.

\section{Data quality analysis and operation}\label{sec:DQ}

A suite of tools is available to continuously monitor detector hardware and data acquisition systems during their operation.
The work-flow is optimised to address problems that arise in real time (online) and afterwards (offline). For cases of irreparable problems, data quality flags are assigned to fractions of the affected data, indicating whether those data are usable for physics analyses with care (depending on the analysis) or must be discarded entirely. 

\subsection{ATLAS detector control system}\label{subsec:HV}
An ATLAS-wide Detector Control System (DCS)~\cite{bib:DCSTDR, bib:DCS} provides a common framework to continuously monitor, control, and archive the status of all hardware and infrastructure components for each subsystem.
The status and availability of each hardware component is visually
displayed in real time on a web interface. This web interface also
provides a detailed history of conditions over time to enable tracking
of the stability.
The DCS infrastructure stores information about individual device properties in databases.

The TileCal DCS is responsible for tracking the low voltage, high voltage, front-end electronics cooling systems, and back-end crates.
The DCS monitoring data are used by automatic scripts to generate alarms if the actual values are outside the expected operating conditions. 
Actions to address alarm states can be taken manually by experts, or subject to certain criteria the DCS system can automatically execute actions. 

The TileCal DCS system monitors the temperature of the front-end electronics with seven probes at various locations in the super-drawer. A temperature variation of $1\,^{\circ}$C would induce a PMT gain variation of 0.2\%~\cite{bib:tileReadiness}. Analyses done over several data periods within Run~1 indicated the temperature is maintained within $0.2\,^{\circ}$C.

One key parameter monitored by the Tile DCS is the HV applied to each
PMT; typical values are 650--700\,V. Since the HV changes alter the PMT gain, an 
  update of the calibration constants is required to account for the
  response change.
The relative PMT gain variation $\Delta G$ between a reference time
$t_\mathrm{r}$ and a time of interest $t$ depends on the HV variation over the
same period according to: 

\begin{equation}
  \frac{\Delta G}{G} = \frac{ \mathrm{HV}^{\beta} (t)}{\mathrm{HV}^\beta (t_\mathrm{r}) } - 1
  \label{eq:gainVariationWithHV}
\end{equation}

where the parameter $\beta$ is extracted experimentally for each
PMT\@. Its mean value is $\beta = 7.0$ with an RMS of 0.2 across 97\% of the measured
PMTs; hence a variation of 1\,V corresponds to a gain variation of 1\%
(for $\beta = 7$).

The TileCal high-voltage system is based on remote HV bulk power
supplies providing a single high voltage to each super-drawer. Each
drawer is equipped with a regulator system (HVopto card) that provides fine adjustment of the voltage for each PMT\@. 
One controller (HVmicro card) manages two HVopto cards of the
super-drawer. The HVmicro card reports actual HV values to the DCS
through a CANbus network every few seconds.

Several studies were performed to quantify the stability of the HV of the PMTs and to identify unstable PMTs. 
One study compares the value of the measured HV with the expected HV for each PMT over the 2012 period. 
The difference between the measured and set high voltage ($\Delta \mathrm{HV}$) for each PMT is fitted  with a Gaussian distribution, and the mean value is plotted for all good channels in a given partition. Good channels are all channels except those in modules that were turned off or in the so-called emergency state (described later). For each partition the mean value is approximately 0\,V with an RMS spread of 0.44\,V, showing good agreement.
Another study investigates the time evolution of $\Delta\mathrm{HV}$ for a given partition. The variation of the mean values versus time is lower than 0.05\,V, demonstrating the stability of the HV system over the full period of the 2012 collision run.

\begin{figure}[tp]
\centering
\includegraphics[width=0.85\textwidth]{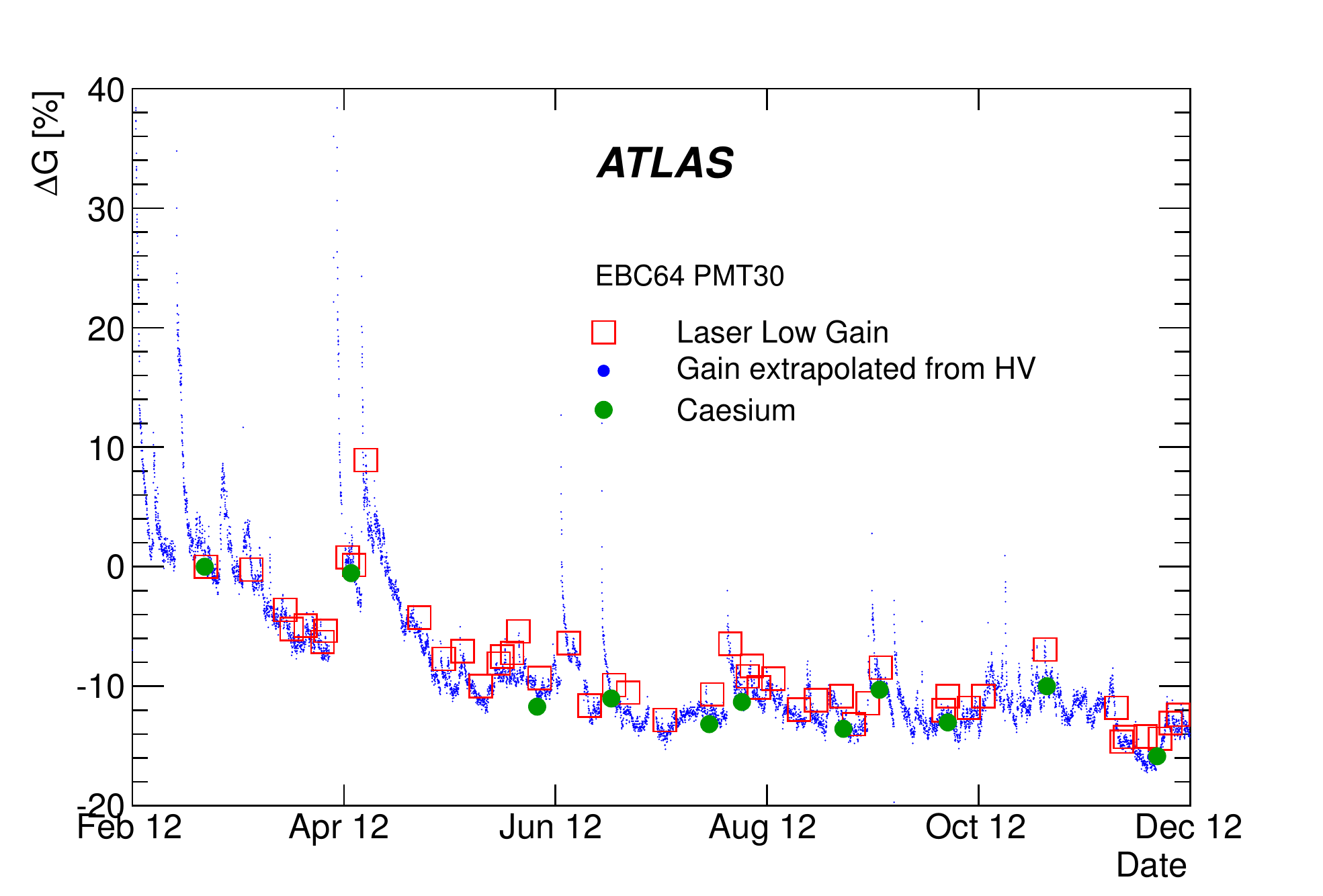}
\caption{One PMT of the EBC64 module with the largest gain
  variation. This plot presents a comparison between the gain
  expected from the HV instability (tiny dots), the one measured by
  the laser (open squares) and Cs (full circles) systems during
  the whole 2012 run. One HV point represents the averaged gain
  variation over one hour. The vertical structures are due to power
  cycles. There is very good agreement between the three methods,
  meaning that even large variations can be detected and handled by
  the TileCal monitoring and calibration systems.}
\label{fig:DGEBC64}
\end{figure}

In order to identify PMTs with unstable HV over time, $\Delta \mathrm{HV}$ is computed every hour over the course of one day for each PMT\@. Plots showing the daily variation in HV over periods of several months are made. PMTs with $\Delta \mathrm{HV} > 0.5$\,V are classified as unstable. The gain variation for these unstable channels is calculated using Eq.~(\ref{eq:gainVariationWithHV}) (with knowledge of the $\beta$ value for that particular PMT), and compared with the gain variation as seen by the laser and Cs calibration systems. These calibration systems are insensitive to electrical failures associated with reading back the measured HV and provide a cross-check of apparent instabilities. 
Figure~\ref{fig:DGEBC64} shows the gain variation for one PMT that suffered from large instabilities in 2012, as measured by the HV and calibration systems. The gain variations agree between the three methods used. Only those channels that demonstrate instabilities in both the HV and calibration systems are classified as unstable. During 2012, a total of only 15 PMTs (0.15\% of the total number of PMTs) were found to be unstable. 

\subsection{Online data quality assessment and monitoring}
During periods of physics collisions, the Tile Calorimeter has experts in the ATLAS control room 24 hours per day and a handful of remote experts available on call to assist in advanced interventions. The primary goal is to quickly identify and possibly correct any problem that cannot be fixed later in software, and that can result in overall data loss. 
The ATLAS data quality framework is designed to perform automatic checks of the data and to alert experts to potential problems that warrant further investigation~\cite{bib:ATLASDQMF}. 

Common problems identified by TileCal experts during the online shifts include hardware failures that do not automatically recover, or software configuration problems that might present themselves as data corruption flags from the ROD data integrity checks.  
The trigger efficiency and data acquisition, as well as higher-level reconstruction data quality, might be influenced by such problems.

\subsection{Offline data quality review}
Shortly after the data are taken, a small fraction is quickly
reconstructed using the Tier-0 computing farm within the ATLAS Athena
software framework~\cite{bib:ComputingTDR}. Reconstructed data are
then used by the offline data quality experts with more complex tools
to evaluate the quality of the data. The experts are given 48~hours to
identify, and, where possible, to correct problems, before the bulk
reconstruction of the entire run is made. The TileCal offline experts
can update the conditions database, where information such as the
calibration constants and status of each channel is stored.  
Channels that suffer from high levels of noise have calibration constants in the database updated accordingly.
For channels that suffer from intermittent data corruption problems, data quality flags are assigned to the affected data to exclude the channels in the full reconstruction during that period.
This 48-hour period is also used to identify cases of digitiser timing
jumps and to add the additional time phases to the time constants of the digitiser affected to account for the magnitude of the time jump.

Luminosity blocks can be flagged as defective to identify periods of
time when the TileCal is not operating in its nominal
configuration. These defects can either be tolerable whereby
corrections are applied but additional caution should be taken while
analysing these data, or intolerable in which case the data are not
deemed suitable for physics analyses. Defects are entered into the
ATLAS Data Quality Defect database~\cite{bib:ATLASDefects} with the
information propagating to analyses as well as to integrated
luminosity calculations.

One luminosity block nominally spans one minute, and removing all data within that time can accumulate to a significant data loss. For rare situations where only a single event is affected by the data corruption, an additional error-state flag is introduced into the reconstruction data. This flag is used to remove such events from the analysis.

Once all offline teams review the run, it is sent to the Tier-0 computing farm for bulk reconstruction, where the entire run is reconstructed using the most up-to-date conditions database. Subsequently the data can be re-reconstructed when reconstruction algorithms are improved and/or the conditions database is further refined to improve the description of the detector.\footnote{An example is the correction of time constants due to timing jumps discovered only from fully reconstructed physics data, see Section~\ref{subsec:timingCalib}.} These data reprocessing campaigns typically occur several months after the data are taken.

\subsection{Overall Tile Calorimeter operation} 
\label{sec:overall_DQ}
Overall the TileCal operation was highly successful in Run~1, with an extremely high fraction of data acceptable for a physics analysis. A summary of the total integrated luminosity delivered to ATLAS and approved for analysis is shown in Table~\ref{tab:DQSummary}, along with the fraction of data passing the Tile Calorimeter data quality reviews.

\begin{figure}[tp]
\centering
\includegraphics[width=0.80\textwidth]{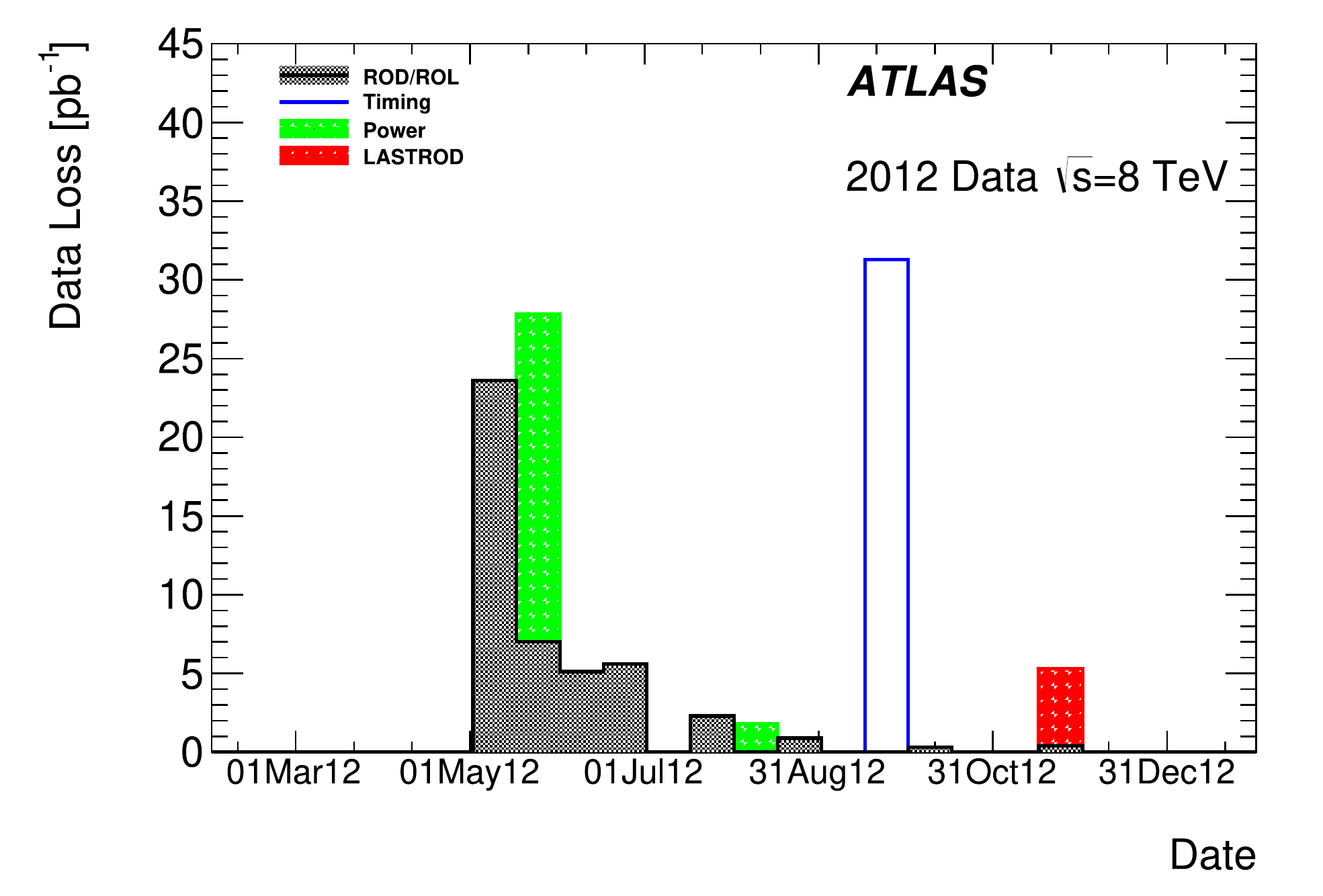}
\caption{The sources and amounts of integrated luminosity lost due to Tile Calorimeter data quality problems in 2012 as a function of time. The primary source of luminosity losses comes from the stop-less read-out link (ROL) removal in the extended barrels accounting for 45.2\,pb$^{-1}$ of this loss. Power cuts or trips of the 200V power supplies account for 22.6\,pb$^{-1}$. The last 4.9\,pb$^{-1}$ of losses stem from Laser Calibration ROD (LASTROD) busy events. The loss of 31.3\,pb$^{-1}$ due to a $-25$\,ns timing shift in EBC are recovered after the data are reprocessed with updated timing constants. Each bin in the plot represents about two weeks of data-taking.\label{fig:OnlineLoss}}
\end{figure}

In 2012, the total integrated luminosity lost after the first bulk reconstruction of the data due to TileCal data-quality-related problems was $104\,\mathrm{pb}^{-1}$ out of $21.7\,\mathrm{fb}^{-1}$, and is summarised in Figure~\ref{fig:OnlineLoss} as a function of time for various categories of intolerable defects.\footnote{The integrated luminosity values, quoted in this section and used in Figure~\ref{fig:OnlineLoss}, are estimated during the data-taking and are preliminary. These values therefore slightly differ from the final numbers listed in Table~\ref{tab:DQSummary}, obtained with the most recent offline calibration.}
The primary source of Tile luminosity losses are cases when a read-out
link (ROL), which transmits data from the ROD to the subsequent chain
in the trigger and data acquisition system, is removed from
processing. It implies no data are received from the four
corresponding modules. ROLs are disabled in situations when they are flooding the trigger with data (malfunctioning configuration or difficulty processing data), putting the trigger into a busy state where effectively no data can be read from any part of the detector. Removing a ROL during a run is done in a so-called stop-less recovery state, whereby the run is not stopped, as restarting a run can take several minutes. 
One role of the online experts is to identify these cases and to respond by correcting the source of the removal and re-enabling the ROL in the run. 
After a new run begins any ROLs that were previously removed are re-included.
Improvements for handling ROL removals include adding monitoring plots counting the number of reconstructed Tile cells, where large drops can indicate a ROL removal, and an automatic ROL recovery procedure. With the automatic recovery in place, a single ROL removal lasts less than 30 seconds, and losses due to ROL removal dramatically dropped in the second half of 2012.
As the removal of a ROL affects four consecutive modules, this defect is classified as intolerable, and it accounted for 45.2\,pb$^{-1}$ of data loss in 2012.

Power cuts or trips of the HV bulk power supply sources accounted for 22.6\,pb$^{-1}$ of lost integrated luminosity. The last 4.9\,pb$^{-1}$ of loss came from situations when the laser ROD became busy.\footnote{Laser events are recorded in parallel with physics data in empty bunch crossings.} During 2012 this was improved by prompting the online expert to disable the laser ROD.

An additional loss of 31.3\,pb$^{-1}$ was due to a 25\,ns timing
shift in a large fraction of the EBC partition which was not caught by the online or offline experts or tools. 
Improvements for large timing shifts include data quality monitoring warnings when the reconstructed time for large numbers of Tile channels differs from the expected value by a large amount.
These data are subsequently recovered in later data reprocessing
campaigns when the timing database constants are updated accordingly.

\begin{table}[tbp]
\begin{center}
\begin{tabular}{| l | c | c | c | }
\hline
 Integrated luminosity                                      & 2010    &   2011    & 2012 \\
\hline\hline
LHC delivered                             &  48.1\,pb$^{-1}$    &   5.5\,fb$^{-1}$     &  22.8\,fb$^{-1}$  \\
ATLAS recorded                         &   45.0\,pb$^{-1}$ (93.5\%)    &  5.1\,fb$^{-1}$ (92.7\%)     & 21.3\,fb$^{-1}$ (93.4\%)  \\
ATLAS analysis approved        &   45.0\,pb$^{-1}$ (100\%)    &   4.6\,fb$^{-1}$ (90.2\%)      &  20.3\,fb$^{-1}$ (95.3\%) \\ 
Tile data quality efficiency         & 100\%                    &   99.2\%                 &  99.6\%  \\

\hline
\end{tabular}
\end{center}
\caption{Summary of total integrated luminosity delivered by the LHC, recorded by ATLAS and approved for physics analyses (the data quality deemed as good simultaneously from all ATLAS subsystems). The numbers in the parentheses denote the fraction of the integrated luminosity relative to the entry on the previous line. The last row lists the fraction of the ATLAS recorded data approved as good quality by the Tile Calorimeter system.}
\label{tab:DQSummary}
\end{table}

There are several operational problems with the LVPS sources that contribute to the list of tolerable defects. In some cases the LVPS fails entirely, implying an entire module is not analysed. 
The failure rate was one LVPS per month in 2011 and 0.5 LVPS per month in 2012. The faulty LVPS sources were replaced with spares during the maintenance campaigns in the ATLAS cavern at the end of each year.

In addition to overall failures, sometimes there are problems
with the low voltage supplied to the HVopto card, which means the PMT
HV can be neither controlled nor measured. In this case the applied HV is set to the minimum value, putting the module in an emergency state. The calibration and noise constants for all channels within a module in emergency mode are updated to reflect this non-nominal state.

Finally, the LVPS suffered from frequent trips correlated with the
luminosity at a rate of 0.6 trips per 1\,pb$^{-1}$. Automatic recovery
of these modules was implemented, to recover the lost
drawer. During the maintenance period between 2011 and 2012, 40 new LVPS
sources (version 7.5) with improved design~\cite{bib:newLVPS} were
installed on the detector. In 2012 there were a total of about 14\,000 LVPS
trips from all modules, only one of which came from the new LVPS
version. After the LHC Run~1, all LVPS sources were replaced with version 7.5.  

\begin{figure}[tp]
\centering
\includegraphics[width=0.95\textwidth]{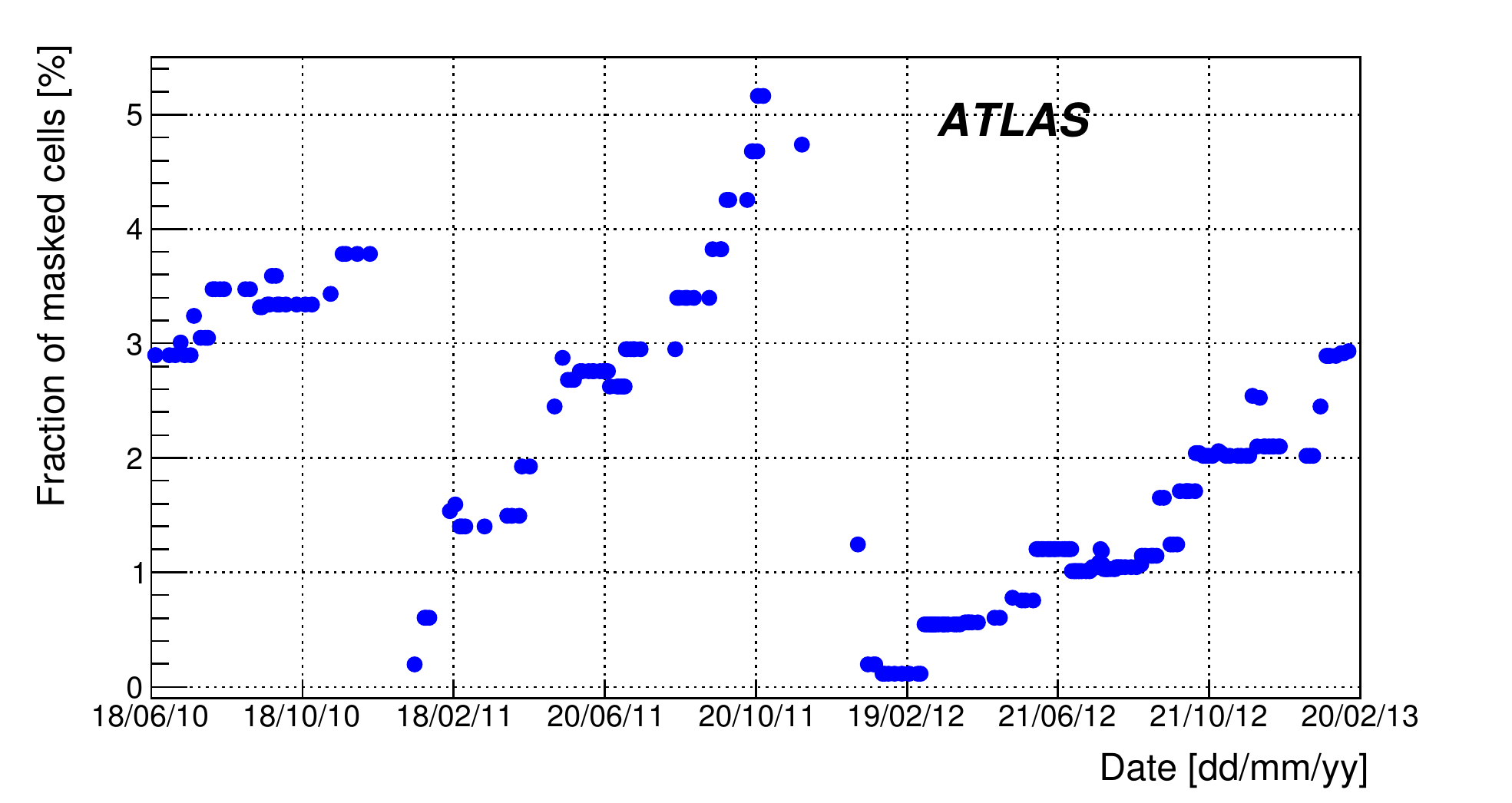}
\caption{The percentage of the TileCal cells that are masked in the
  reconstruction as a function of time starting from June 
  2010. Periods of recovery correspond to times of hardware
  maintenance when the detector is accessible due to breaks in the
  accelerator schedule. Each super-drawer LB (EB) failure corresponds
  to 0.43\% (0.35\%) of masked cells. The total number of cells
  (including gap, crack, and minimum-bias trigger scintillators) is
  5198. Approximately 2.9\% of cells were masked in February
  2013, at the end of the proton--lead data-taking period closing the
  Run~1 physics programme.}
\label{fig:deadCellsVsTime}
\end{figure}

Figure~\ref{fig:deadCellsVsTime} shows the percentage of the TileCal cells masked in the reconstruction as a function of time. These cells are located in all areas of the detector, with no one area suffering from a large number of failures.
The main reasons for masking a cell are failures of LVPS sources, evident by the steep steps in the figure. 
Other reasons are severe data corruption problems or very large noise. 
The periods of maintenance, when faulty hardware components are repaired or replaced (when possible), are visible by the reduction of the number of faulty cells to near zero.
For situations when cell energy reconstruction is not possible the energy is interpolated from neighbouring cells.
The interpolation is linear in energy density (energy per cell volume) and is done independently in each layer, using all possible neighbours of the cell (i.e.\ up to a maximum of eight).
In cases where only one of two channels defining a cell is masked the energy is taken to be twice that of the functioning channel.

\section{Performance studies}\label{sec:InSitu}

The response of each calorimeter channel is calibrated to the EM scale
using Eq.~(\ref{eq:channelEnergy}). The sum of the two channel
responses associated with the given read-out cell forms the cell energy, which
represents a basic unit in the physics object reconstruction
procedures. Cells are combined into clusters with the topological
clustering algorithm~\cite{bib:topoclustering} based on the
significance of the absolute value of the reconstructed cell energy
relative to the noise, $S = |E|/\sigma$. The noise $\sigma$ combines
the electronic (see Section~\ref{sec:electronicNoise}) and pile-up
contributions (Section~\ref{sec:pileupNoise}) in quadrature. Clusters
are then used as inputs to jet reconstruction algorithms.  

The ATLAS jet performance~\cite{bib:jet7tev, bib:jet_energy_resolution}
and measurement of the missing transverse momentum~\cite{bib:met} are documented in detail in other papers. The performance studies reported here focus
on validating the reconstruction and calibration methods, described in
previous sections, using the isolated muons, hadrons and jets entering the Tile Calorimeter.

\subsection{Energy response to single isolated muons}

Muon energy loss in matter is a well-understood process~\cite{bib:pdgPassage}, and can be used to probe the response of the Tile Calorimeter.
For high-energy muons, up to muon energies of a few hundred GeV, the dominant energy loss process is ionisation. 
Under these conditions the muon energy loss per unit distance is approximately constant.
This subsection studies the response to isolated muons from cosmic-ray sources 
and to $W\rightarrow\mu\nu$ events from $pp$ collisions.

Candidate muons are selected using the muon RPC and TGC triggering subsystems of the Muon Spectrometer. 
A muon track measured by the Pixel and SCT detectors is extrapolated through the calorimeter volume, taking into account the detector material and magnetic field~\cite{bib:trackExtropolation}.
A linear interpolation is performed to determine the exact entry and exit points of the muon in every crossed cell to compute the distance traversed by the muon in a given TileCal cell. 
The distance ($\Delta x$) together with the energy deposited in the cell ($\Delta E$) are used to compute the muon energy loss per unit distance, $\Delta E/\Delta x$.

The measured $\Delta E/\Delta x$ distribution for a cell can be described by a Landau function convolved with a Gaussian distribution, where the Landau part describes the actual energy loss and the Gaussian part accounts for resolution effects.
However, the fitted curves show a poor $\chi^2$ fit to the data, due
to high tails from rare energy loss mechanisms, such as bremsstrahlung
or energetic gamma rays. 
For this reason a truncated mean $\langle \Delta E/\Delta x \rangle_{\mathrm{t}}$ is used to define the average muon response. For each cell the truncated mean is computed by removing a small fraction (1\%) of entries with the highest  $\Delta E/\Delta x$ values.
The truncated mean exhibits a slight non-linear scaling with the path
length $\Delta x$. This non-linearity and other residual
non-uniformities, such as the differences in momentum and incident
angle spectra, are to a large extent reproduced by the MC simulation. To
compensate for these effects, a double ratio formed by the ratio of the experimental and simulated truncated means is defined for each calorimeter cell as:

\begin{equation}
R \equiv \frac{\langle \Delta E/\Delta x \rangle_{\mathrm{t}}^{\mathrm{data}}}{\langle \Delta E/\Delta x \rangle_{\mathrm{t}}^{\mathrm{MC}}}
\label{eq:doubleRatioCosmics}
\end{equation}

The double ratio $R$ is used to estimate the calorimeter response as a function of various detector geometrical quantities (layer, $\phi$, $\eta$, etc). Deviations of the double ratio from unity may indicate poor EM energy scale calibration in the experimental data.

\subsubsection{Cosmic-ray muon data}\label{subsec:cosmics}

Muons from cosmic-ray showers, called cosmic muons, are used as a
cross-check of energy reconstruction and calibration complementary to the collision data. At sea level, cosmic muons can have energies up to a TeV or more, but most of the muons are at lower energies, with the mean energy being approximately 4\,GeV~\cite{bib:pdgPassage}.  

Candidate cosmic muons are triggered during empty bunch crossings in physics runs in a dedicated data stream allocated for muon candidates identified by the muon spectrometer trigger system if at least one track is matched to the inner detector tracking system.  In total there are approximately one million such events triggered in each year studied (2008, 2009, 2010).

The energy in TileCal channels is reconstructed using the iterative OF method (see Section~\ref{sec:SignalReco}). The muon tracks, reconstructed using Pixel and SCT detectors with a dedicated algorithm, are extrapolated through the volume of the calorimeter in both upward and downward directions. This allows to study the response of the TileCal modules in top and bottom parts of the detector.
 
The event selection criteria used to select events for the cosmic muons analysis are summarised in Table~\ref{tab:cosmicSelection}. 
A candidate cosmic-muon event is required to have exactly one track associated with a reconstructed muon (Cut 1), with at least eight hits in the Pixel plus SCT detectors (Cut 2).
A cut on the maximum distance of the reconstructed track from the origin of the coordinate system in both the transverse ($d_0$) and longitudinal ($z_0$) components (Cut 3) is used to select well-reconstructed tracks that follow the projective geometry of the calorimeter. 
Muons with a trajectory close to the vertical direction are poorly measured in the TileCal due to the vertical orientation of the scintillating tiles, hence Cut 4 is used to remove the very central cells located within the vertical coverage of the inner detector. The last two requirements (Cut 3 and Cut 4) effectively remove muons at very low pseudorapidities.
The muon is required to have momentum in the range 10--30\,GeV to minimise the effects of multiple scattering at low momentum, and to reduce radiative energy losses at higher momentum, which could produce large fluctuations in the results. 
The muon path length through a cell must be larger than 200\,mm. An energy of at least 60\,MeV\footnote{The value corresponds approximately to $3\sigma$ of the typical cell noise distribution.} has to be released in that cell to remove contributions from noise.
Cut 8 is used to reduce contributions from multiple scattering, such that the track's azimuthal impact point at the inner (outer) radial point of the cell, $\phi_{\mathrm{inner}} (\phi_{\mathrm{outer}})$ is within 0.04 radians of the cell centre $\phi_{\mathrm{c}}$ coordinate (with a cell width of $\Delta\phi = 0.1$). 

\begin{table}[htp]
\begin{center}
\begin{tabular}{| c | l  l | }
\hline
      Cut      &      Variable    &       Requirement                   \\
\hline\hline           
     1       & number of muon tracks $N_{\mu}$                  & $N_{\mu} = 1$  \\
     \hline
     2       & number of track hits in Pixel + SCT                  & $\geq 8$   \\
     \hline
     3       & reconstructed track distance from origin         & $|d_0| \leq 380$\,mm (transverse), \\
               &                                                                                  & $|z_0| \leq 800$\,mm (longitudinal) \\
     \hline          
     4       & polar angle of track relative to vertical axis & $|\theta_\mu| > 0.13$ rad  \\
     \hline
     5       & muon momentum                                                 & $10\,\mathrm{GeV} < p_\mu < 30\,\mathrm{GeV}$   \\
     \hline
     6       & muon path length through cell                           & $\Delta x > 200$\,mm \\
     \hline
     7       & cell energy                                                             & $\Delta E > 60$\,MeV \\
     \hline          
     8       & track impact point at inner \& outer radial point of cell  & $|\phi_{c} - \phi_{\mathrm{inner}}| < 0.04$, \\
               &                                                                                 & $ |\phi_{c} - \phi_{\mathrm{outer}}| < 0.04$ \\
\hline
\end{tabular}
\end{center}
\caption{Selection criteria applied to the event, track, and muon used in the cosmic muons analysis. A description and motivation of each cut can be found in the text.
\label{tab:cosmicSelection}} 
\end{table}

\begin{figure}[tp]
\centering
\includegraphics[width=0.49\textwidth]{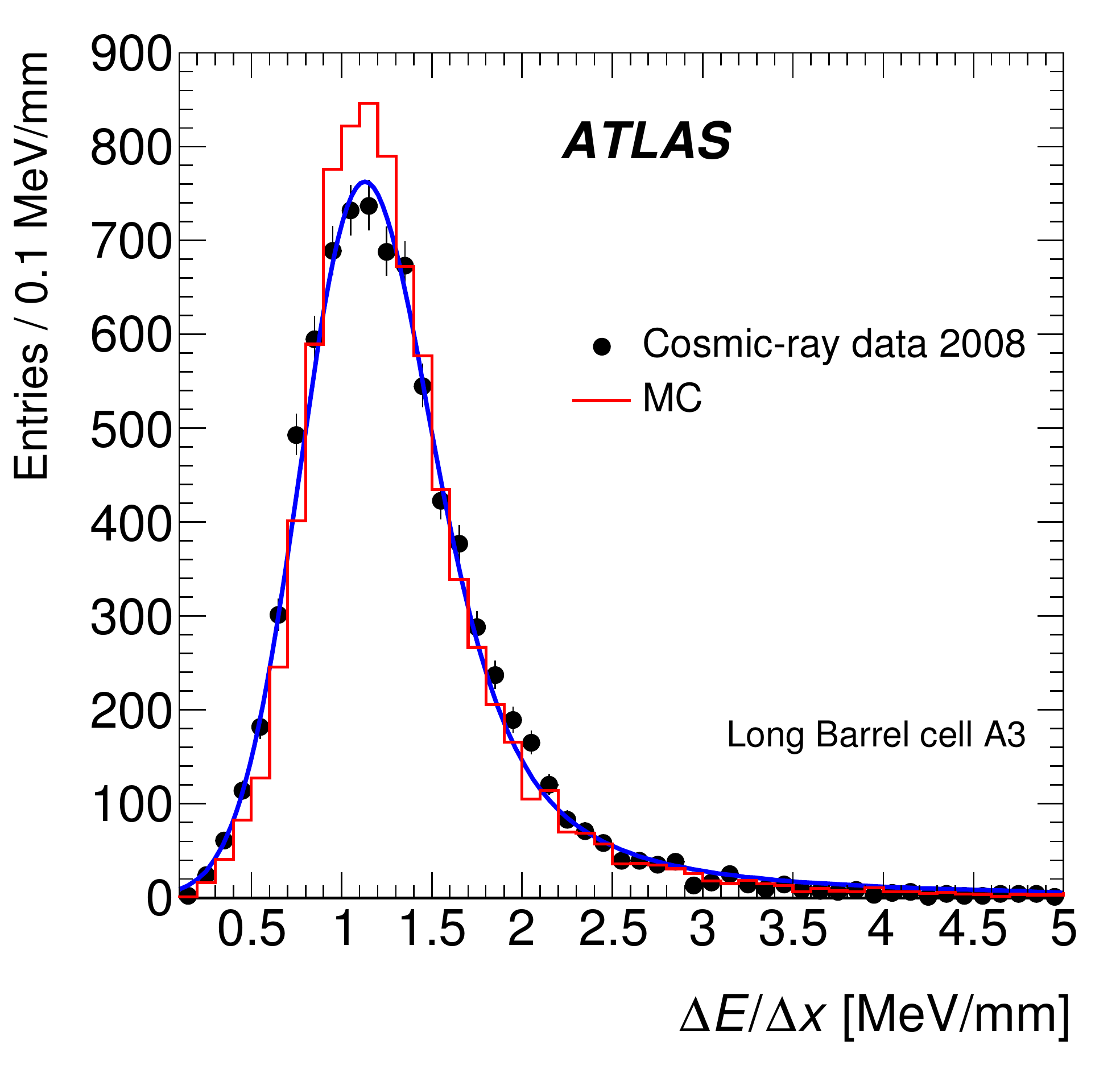}
\includegraphics[width=0.49\textwidth]{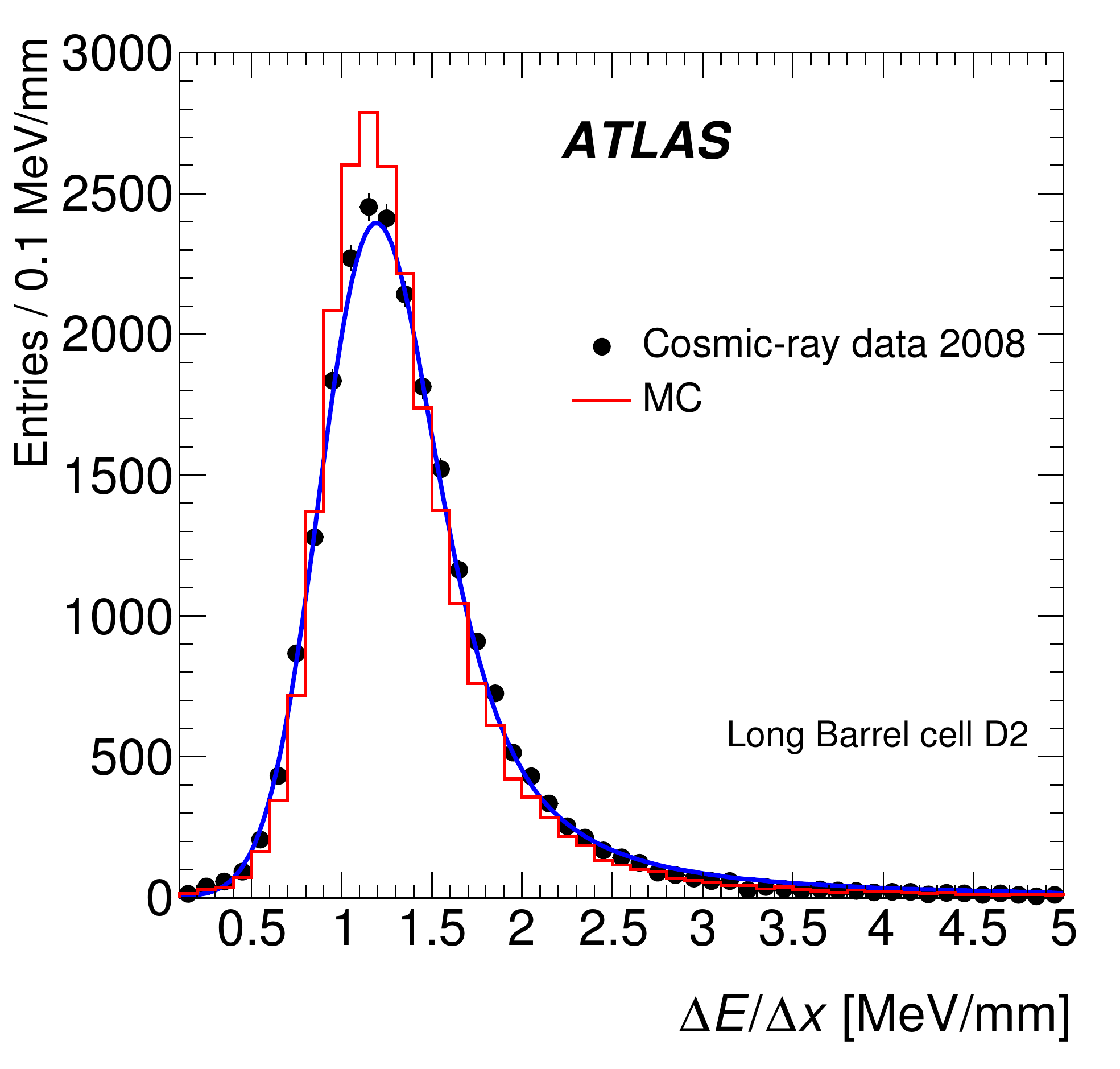}
\caption{Distributions of the energy deposited by cosmic muons per unit of path length, $\Delta E/\Delta x$, in the two cells in the long barrel,  A3 (left) and D2 (right) obtained using 2008 experimental (full points) and simulated (solid lines) data. The A3 (D2) cell in the long barrel covers the region $0.2 < |\eta| < 0.3$ ($0.3 < |\eta| < 0.5$) and is located in the innermost (outermost) calorimeter layer. The function curve overlaid on top of the experimental data is a Landau distribution convolved with a Gaussian distribution.  \label{fig:cosmicDEDXforA3D2}}
\end{figure}

The response to cosmic muons in the calorimeter is also studied using MC simulated data. 
The cosmic-muon energy and flux spectra as measured at sea-level~\cite{bib:cosmicMuonTheory} are used as input into the simulation.
The material between the surface and the ATLAS cavern is simulated, including the cavern volumes and detector access shafts. 
Air showers are not simulated but have negligible impact due to the selection requirements for single-track events. 
The $\Delta E/\Delta x$ distributions for the 2008 cosmic-muon data and MC simulation
are shown in Figure~\ref{fig:cosmicDEDXforA3D2} for cells A3 (left) and D2 (right) in the long barrel. The A3 (D2) cell covers the region $0.2 < |\eta| < 0.3$ ($0.3 < |\eta| < 0.5$) and is located in the innermost (outermost) calorimeter layer. Differences between the experimental and simulated data are discussed in the following paragraphs.

\bigskip
\noindent\textbf{Verification of the radial layer intercalibration}
\smallskip

The calibration between cells within the same layer is investigated
using the double ratio formed by the ratio of the experimental and
simulated truncated means, as shown in Eq.~(\ref{eq:doubleRatioCosmics}). The typical non-uniformity of all cells in a given layer is found to be approximately 2\% for all layers every year. This can be explained by the variations in the optical and electrical components of the calorimeter. 

Several sources of systematic uncertainty, summarised in Table~\ref{tab:cosmicUncert}, are considered in the studies of cosmic muons. The systematic variations 1--5 are related to the selection criteria. The results are assumed to be stable for different values used in the selection criteria. This assumption is checked by varying the values in the specified range and repeating the analysis for every variation, both for data and MC simulations. The resulting differences contribute to the total systematic uncertainty. 
Differences in the response along the muon path through the detector and due to signal evaluation method should be well described by MC. Two systematic variations, applied both to data and MC, are introduced to verify this assumption. Variation 6 compares the response in the upper part of the detector ($\phi_{\mathrm{c}} > 0$), where these muons enter the detector, and in the lower part ($\phi_{\mathrm{c}} < 0$), after the muons pass through a large fraction of the detector. The uncertainty of the method used to evaluate the detector’s response to cosmic muons is considered as source 7. Source 8 reflects the different spread of the experimental and simulated $\Delta E/ \Delta x$ distributions. The effect on the determination of the truncated mean is estimated to be 0.3\% using a toy MC simulation. 
The final classes of uncertainties, 9 and 10, concern the signal calibration procedures performed in the test beam and in situ in ATLAS (already discussed in Section~\ref{sec:calib}). The corresponding variations are only applied to the MC. The parameters of each source of systematic uncertainty are considered as random variables and their values are selected according to the distributions reported in Table~\ref{tab:cosmicUncert}.
In the case of sources 3, 8 and 9 the errors are treated as uncorrelated and a different value is considered for each layer. To evaluate the total systematic uncertainty, 2500 working points are generated in the parameter phase-space and for each of them the analysis is performed and the double ratio calculated. A Gaussian distribution is observed for each layer and the standard deviation, $\sigma$, is taken as the associated systematic uncertainty. The contribution of statistical errors is negligible. 
  
\begin{table}[htp]
\begin{center}
\begin{tabular}{| c | l | l | }
\hline
 Source   &   Systematic uncertainty              &  Parameter distribution and variation   \\
\hline\hline              
 1   &  $|\theta_\mu|$ (Cut 4 Table~\ref{tab:cosmicSelection})                            & Uniform [0.10, 0.15] \\
 \hline
     &                                                     &  Uniform [$5\,\mathrm{GeV} < p_{\mu} < 10\,\mathrm{GeV}$, \\
  2  & $p_{\mu}$   (Cut 5 Table~\ref{tab:cosmicSelection}) & $10\,\mathrm{GeV} < p_{\mu} < 30\,\mathrm{GeV}$, \\
     &                                                     & $30\,\mathrm{GeV} < p_{\mu} < 50\,\mathrm{GeV}$] \\
 \hline
 3   &  $\Delta x$   (Cut 6 Table~\ref{tab:cosmicSelection})                               & Uniform [$\Delta x_{\mathrm{min}}$, $\Delta x_{\mathrm{max}}$] \\
     &  with $\Delta x_{\mathrm{min}} = \mathrm{MaxPath}/2 - 100$\,mm                &  \\
       & $\Delta x_{\mathrm{max}} = \Delta x_{\mathrm{min}} + (\mathrm{MaxPath})/2 $   &  \\
 \hline                                                  
 4  &  $\Delta E$ (Cut 7 Table~\ref{tab:cosmicSelection})                                  & Uniform [30\,MeV, 90\,MeV] \\
 \hline                                                                          5 & $|\phi_{\mathrm{c}} - \phi_{\mathrm{inner}}|, |\phi_{\mathrm{c}} - \phi_{\mathrm{outer}}|$ (Cut 8 Table~\ref{tab:cosmicSelection})  & Uniform [0.03, 0.05] \\
  \hline                                                          
 6   & $\phi_{\mathrm{c}}$                                                             & Uniform $\phi_{\mathrm{c}} > 0, \phi_{\mathrm{c}} < 0$ \\
 \hline
 7   & $\Delta E/\Delta x$ truncation                        & Uniform 0\%, 1\%, 2\%             \\   
 \hline                 
8   & Smearing of simulated $\Delta E/\Delta x$   & Gaussian $\mu = 0, \sigma=0.3\%$ \\
\hline
9   &  Uncertainty in radial calibration correction  & Gaussian $\mu = 0, \sigma=0.3\%$ \\
\hline
10   &   Uncertainty in up-drift and magnetic field effects & Gaussian $\mu = 0, \sigma=1.0\%$ (LB), 0.6\% (EB) \\
\hline
\end{tabular}
\end{center}
\caption{Different sources of systematic uncertainty considered in the analysis of the cosmic muons. Sources 1--5 are associated with the event selection procedure, other sources are relevant for the double ratio responses. The distributions of the parameters used in the analysis are reported. In the case of source~3 for each track and each cell the value of the maximal path length, MaxPath, is determined by the dimensions of the cell. 
\label{tab:cosmicUncert}} 
\end{table}

Table~\ref{tab:cosmicLayers} shows the double ratio and its total
uncertainty per layer for all three years under study. These results
can be used to validate the calibration procedure including all corrections as mentioned in Section~\ref{sec:calib}. A method based on Bayes' theorem is used to establish the uniformity of the layer response in each year~\cite{ATL-TILECAL-PUB-2011-001}. The probability function that the six measured double ratios $\vec{R}=(R_{\mathrm{LB-A}},\ldots,R_{\mathrm{EB-D}})$ correspond to layer responses $\vec{\mu} = (\mu_{\mathrm{LB-A}},\ldots,\mu_{\mathrm{EB-D}})$ is proportional to the likelihood $\mathcal{L}(\vec{\mu}|\vec{R})$, as uniform prior probabilities are assumed. Since the distribution of the double ratio is found to be Gaussian in each layer, the likelihood is constructed as six-dimensional Gaussian function 

\begin{equation}
\label{eq:log_likelihood_6dim}
\mathcal{L}(\mu|R) \propto \mathrm{exp}\left(-0.5\cdot(\vec{\mu}-\vec{R})^T V^{-1} (\vec{\mu}-\vec{R})\right)
\end{equation}

where $V$ is the error matrix obtained from the analyses over 2500 working points described above. For each pair of layers $l, l^\prime$, the posterior probability $f(\mu_l,\mu_{l^\prime}|\vec{R})$ is evaluated by integrating Eq.~(\ref{eq:log_likelihood_6dim}) over the remaining layers. It is found that the response of layer D in the long barrel differs from that of layers A and BC by $4\sigma$ and $3\sigma$, respectively, for all years (see Section~\ref{sec:performance_summary} for more details). The response for all other layer pairs is found to be consistent.
The total error in the EM energy scale for all cells in a fixed layer is found to be approximately 2\%, including uncertainties of the cosmic muons analysis, uncertainties in the determination of the EM scale at test beams and subsequent application in ATLAS, and the uncertainty in the simulation of the TileCal response to muons.

\begin{table}[htp]
\begin{center}
\begin{tabular}{| l | c | c | c | }
\hline
                     &     $R_{2008}$                & $R_{2009}$                  & $R_{2010}$ \\
\hline\hline              
    LB-A        &  0.966 $\pm$ 0.012   & 0.972 $\pm$ 0.015  &  0.971 $\pm$ 0.011 \\
    LB-BC     &  0.976 $\pm$ 0.015   & 0.981 $\pm$ 0.019  & 0.981 $\pm$ 0.015 \\
    LB-D        & 1.005 $\pm$ 0.014      & 1.013 $\pm$ 0.014  & 1.010 $\pm$ 0.013 \\
    EB-A        & 0.964 $\pm$ 0.043    & 0.965 $\pm$ 0.032  & 0.996 $\pm$ 0.037 \\
    EB-B        & 0.977 $\pm$ 0.018   & 0.966 $\pm$ 0.016   & 0.988 $\pm$ 0.014 \\
    EB-D       & 0.986 $\pm$ 0.012    & 0.975 $\pm$ 0.012   & 0.982 $\pm$ 0.014 \\ 
\hline
\end{tabular}
\end{center}
\caption{Double ratio given in Eq.~(\ref{eq:doubleRatioCosmics}) by
  the ratio of the experimental and simulated $\Delta E/\Delta x$ truncated means for different layers in the long barrel (LB) and extended barrel (EB), for the three data periods using cosmic-muon data. The sources of uncertainty are described in the text. Larger uncertainties in the EB-A layer reflect that fewer cosmic muons satisfy the selection criteria. A maximum difference of 4\% is observed between the layer calibrations. 
\label{tab:cosmicLayers}} 
\end{table}

A maximum-likelihood fit is used to estimate the mean calorimeter
response ($\hat{\mu}_y$) over all layers for a given year ($y$),
taking into account the uncertainties and correlations. The ratios
$\hat{\mu}_y/\hat{\mu}_{y'}$ for $y \neq y'$ and $ y, y' \in \{2008,
2009, 2010\}$ are then computed, and are found to be consistent with unity. 
Within uncertainties the response of the calorimeter layers to cosmic-muon data is found to be stable, confirming the calibration systems are able to follow the variations of the PMT gains and to compensate for the drift of response per year to better than 1\% in the long barrel and better than 3\% in the extended barrel. 

The double ratios listed in Table~\ref{tab:cosmicLayers} are approximately 0.97, except for the LB-D layer, with a quoted uncertainty of the order of 1.5\%. Nevertheless, the differences from unity are well within the TileCal EM scale uncertainty of 4\% measured in studies of isolated particles and in the beam tests~\cite{bib:tileReadiness}. Detailed discussion and the comparison with the results of the  isolated collision muons’ analysis (next section) are presented in Section~\ref{sec:performance_summary}. 

\subsubsection{Isolated collision muons}
\label{sec:collision_muons}

The calorimeter performance is also assessed with isolated muons from $W\rightarrow\mu\nu$ processes originating in proton--proton collisions, complementary to the cosmic-muon studies presented in previous subsection. Data from proton--proton collisions in 2010--2012 are analysed.
Events were collected using a L1 muon trigger which accepts events with sizeable muon $p_{\mathrm{T}}$ originating from the interaction point. A total of approximately one billion events are selected for these three years. The event selection is further refined using the criteria listed in Table~\ref{tab:collisionSelection}.
Cuts 1--3 are used to select $W\rightarrow\mu\nu$ events and to suppress background from multi-jet processes. 
The transverse mass ($m_{\mathrm{T}}$), Cut 2, is defined as follows:

\begin{equation*}
 m_{\mathrm{T}}  = \sqrt{ 2 p_{\mathrm{T}}^\mu    E_{\mathrm{T}}^{\mathrm{miss}}  (1 - \cos [\Delta\phi(  \textbf{p}_{\mathrm{T}}^\mu, \textbf{p}_{\mathrm{T}}^{\mathrm{miss}}     )  ] )   },
\label{eq:mT}
\end{equation*}

where $\textbf{p}_{\mathrm{T}}^\mu$ is the vector of the muon's transverse momentum and $\textbf{p}_{\mathrm{T}}^{\mathrm{miss}}$ stands for the vector of the missing transverse momentum. The scalar variables denote the corresponding vector magnitude, $p_{\mathrm{T}}^\mu \equiv |\textbf{p}_{\mathrm{T}}^\mu|$ and $E_{\mathrm{T}}^{\mathrm{miss}} \equiv |\textbf{p}_{\mathrm{T}}^{\mathrm{miss}}|$.

An explicit cut on missing transverse momentum is made (Cut 3) by
requiring $E_{\mathrm{T}}^{\mathrm{miss}} > 25$\,GeV in order to
further reduce background from jet production.
Similar to the cosmic muons analysis, a cut on the polar angle relative to the vertical axis is applied (Cut~4) and only muons in a low momentum range [20\,GeV, 80\,GeV] are selected (Cut 5).
The contribution from nearby particles is suppressed by requiring
the selected tracks to be well isolated within a cone of size $\Delta R =
\sqrt{(\Delta\phi)^2  + (\Delta\eta)^2} = 0.4$ in the tracking detector
(Cut~6) and the response in the upstream liquid argon (LAr) calorimeter must
be compatible with a minimum-ionising particle (Cut 7).
The muon path length through a cell is required to be larger than 100\,mm (Cut 8), and the cell energy has to be greater than 60~MeV to remove residual noise contributions (Cut 9). 

\begin{table}[htp]
\begin{center}
\begin{tabular}{| c | l  l | }
\hline
      Cut      &      Variable    &       Requirement                   \\
\hline\hline           
     1       & number of muon tracks $N_{\mu}$                  & $N_{\mu} = 1$  \\
     \hline
     2      & transverse mass $m_\mathrm{T}$                    & $m_\mathrm{T} > 40$\,GeV \\
     \hline
     3      & missing transverse momentum $E_{\mathrm{T}}^{\mathrm{miss}}$   &  $E_{\mathrm{T}}^{\mathrm{miss}} > 25$\,GeV \\
     \hline
     4      & polar angle of track relative to vertical axis & $|\theta_\mu| > 0.13$ rad  \\
     \hline
     5      & muon momentum                                                 & $20\,\mathrm{GeV} < p_\mu < 80\,\mathrm{GeV}$   \\
     \hline
     6      & transverse momentum around track within $\Delta R < 0.4$: & $p_\mathrm{T}^{\mathrm{cone40}} < 1$ GeV \\
     \hline
     7     & LAr calorimeter energy around track within $\Delta R < 0.4$:  &    $E_{\mathrm{LAr}} < 3$\,GeV   \\
     \hline
     8      & muon path length through cell                           & $\Delta x > 100$\,mm \\
     \hline
     9       & cell energy                                                             & $\Delta E > 60$\,MeV \\
     \hline
\end{tabular}
\end{center}
\caption{Selection criteria applied to the events, tracks, and muons for the collision muon analysis.
\label{tab:collisionSelection}} 
\end{table}

The same selection criteria are applied to MC simulated data. The
$W\rightarrow\mu\nu$ events were generated using
 the leading-order generators \PYTHIAV{6}~\cite{bib:Pythia6} in 2010, and \SHERPA~\cite{bib:SherpaW} in
2011 and 2012. The full ATLAS digitisation and reconstruction is
performed on the simulated MC data. Unfortunately, data and MC events in
2010 were processed with different reconstruction
algorithms\footnote{Data in 2010 were reconstructed with iterative OF
  method, while in MC simulation the non-iterative approach was applied.}
that in the end biases the data/MC 
ratio. Therefore, only the results from 2011 and 2012 are reported here.

\bigskip
\noindent\textbf{Cell response uniformity}
\smallskip

The double ratios given in Eq.~(\ref{eq:doubleRatioCosmics}) by ratios
of the truncated means of the data and MC $\Delta E/\Delta x$
distributions are used to quantify the cell response uniformity in $\phi$.
The systematic uncertainty associated with the non-uniformity in the response for cells of the same type in the considered $\phi$ slice is quantified using a maximum-likelihood method. The likelihood function with mean response $\mu$ and non-uniformity $s$ is defined as follows:

\begin{equation}
\mathcal{L} = \prod_{c = 1}^{N_c} \frac{1}{\sqrt{2\pi} \cdot \sqrt{(\sigma_c^{2} + s^{2})}} \exp\left[ -\frac{1}{2} \left(    \frac{R_c - \mu}{ \sqrt{\sigma_c^{2} + s^{2}}  }    \right)^2  \right]
\label{eq:likelihoodCells}
\end{equation}

where the product runs over 64 modules in $\phi$ for each cell
$c$ of the same type. Here $R_c$ is the double ratio from
Eq.~(\ref{eq:doubleRatioCosmics}) and  $\sigma_c$ the statistical
uncertainty for the cell under consideration. The maximum is
effectively found by minimising the unbinned log-likelihood $-2
\log{\mathcal{L}}$, varying the non-uniformity $s$.

\begin{sidewaysfigure}[tbp]
\centering
\includegraphics[width=0.99\textwidth]{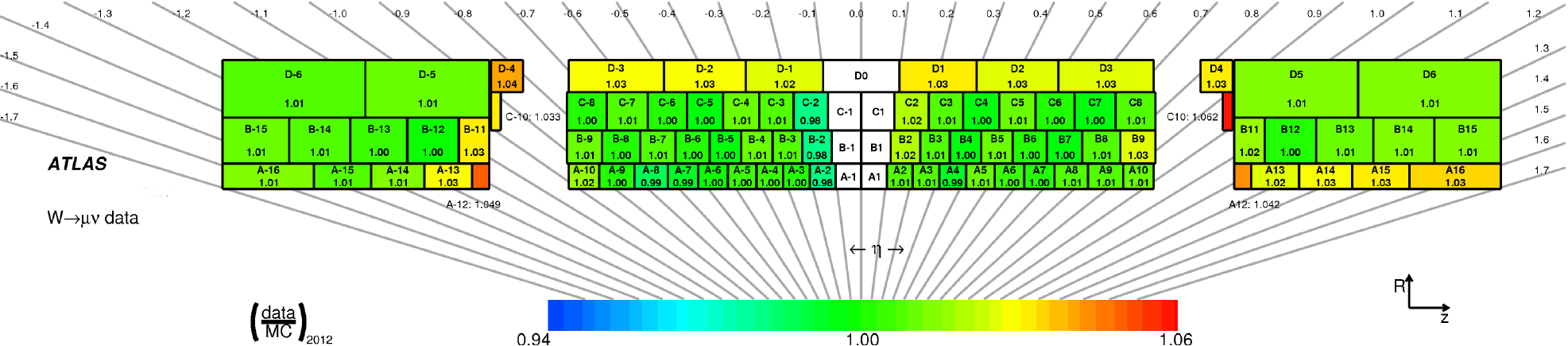}
\includegraphics[width=0.99\textwidth]{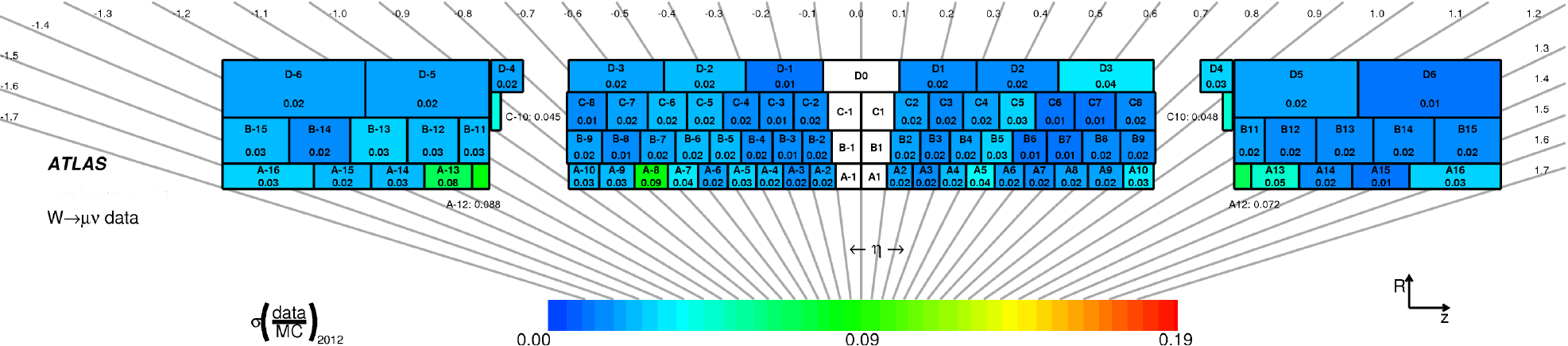}
\caption{Visualisation of the TileCal in the ($z, r$) plane showing the results for the fit parameters of Eq.~(\ref{eq:likelihoodCells}) for (top) the mean double ratio response $\hat{\mu}$ and (bottom) the non-uniformity in $\phi$ of the double ratio $\hat{s}$. Shown for all cells with $|\eta| > 0.1$, using the 2012 data and MC simulation.     
\label{fig:colFitParams2012}}
\end{sidewaysfigure}

The results of the fits are shown in Figure~\ref{fig:colFitParams2012}
for the mean response $\hat{\mu}$ (top) and non-uniformity in the
azimuthal angle $\hat{s}$ (bottom) in 2012. Cut 4 in
Table~\ref{tab:collisionSelection} reduces the number of muons
crossing the most central calorimeter cells, and there are too few
detected muons to include the cells with $|\eta| < 0.1$ in the analysis.
A similar study is done also for 2011 data. The mean double
ratio across all cells is consistent with unity. Moreover, the double ratio 
is found to be constant across $\eta$ in each layer.
Upper limits on the average non-uniformity in $\phi$, quantified by
the spread in response amongst calorimeter cells of a given cell type,
is found to be about 5\% in both 2011 and 2012 data.

The amount of energy deposited in a cell depends on the geometrical
properties, such as the amount of upstream dead material and
cell-specific calibration constants. In general there exists a
symmetry between $\eta > 0$ and $\eta < 0$. As one goes to increasing
radius (layers A$\rightarrow$BC$\rightarrow$D), the values of the truncated means
remain approximately the same. A similar trend is observed for the 2011 data.

\bigskip
\noindent\textbf{Verification of the radial layer intercalibration}
\smallskip

The double ratio of the observed and simulated response is
calculated for each radial calorimeter layer for each data-taking year
considered in the analysis. 
The systematic uncertainties associated with the event selection and the response evaluation are listed in Table~\ref{tab:collisionUncert}. These
variations are considered as random variables and their values
selected according to uniform probability distributions. 
In total, 1000 combinations in the parameter phase-space are generated
by varying each of the applied cuts. The analysis is repeated for each
combination, similarly to the cosmic muons analysis
(Section~\ref{subsec:cosmics}). The same method exploiting a six-dimensional Gaussian function is used and the mean response per layer is determined by maximum-likelihood fit for each data-taking year (2011, 2012), taking into account the correlations of the systematic uncertainties between the layers.

\begin{table}[htp]
\begin{center}
\begin{tabular}{| c | l | l | }
\hline
 Source   &   Systematic uncertainty              &  Parameter distribution and variation   \\
\hline\hline              
  1  &   $|\theta_\mu|$ (Cut 4 Table~\ref{tab:collisionSelection})                            & Uniform [0.1238, 0.1365] \\
     &                                                         &  Uniform [$20\,\mathrm{GeV} < p_{\mu} < 35\,\mathrm{GeV}$,\\
  2  & $p_{\mu}$   (Cut 5 Table~\ref{tab:collisionSelection})  &  $35\,\mathrm{GeV} < p_{\mu} < 50\,\mathrm{GeV}$,\\
     &                                                         &  $50\,\mathrm{GeV} < p_{\mu} < 80\,\mathrm{GeV}$]\\
  3   &  $\Delta x$   (Cut 8 Table~\ref{tab:collisionSelection})                                    & Uniform [95\,mm, 105\,mm] \\
  4  &   $\Delta E$ (Cut 9 Table~\ref{tab:collisionSelection})                                   & Uniform [30\,MeV, 90\,MeV] \\
  5  &   fraction of high tail excluded  & Uniform [0\%, 1\%, 2\%] \\
  & to compute truncated mean  & \\ 
\hline
\end{tabular}
\end{center}
\caption{The variations associated with the event selection procedure and response evaluation considered as the sources of systematic uncertainty in the collision muon analysis. The distributions of the parameters are reported. 
\label{tab:collisionUncert}} 
\end{table}
The double ratios together with the total uncertainties are
reported in Table~\ref{tab:collisionRatioLayer}. The results
indicate that the radial layers LB-A, LB-BC, EB-A, EB-B and EB-D
were well intercalibrated in 2011 and 2012. It was found that
the layer LB-D had higher response than the layers LB-A and LB-BC;
the difference of $+3\%$ is further discussed in Section~\ref{sec:performance_summary}.

\begin{table}[htp]
\begin{center}
\begin{tabular}{| l | c | c | }
\hline
                & $R_{2011}$                   & $R_{2012}$ \\
\hline\hline    
    LB-A        &  0.996$\pm$0.006            &  1.003$\pm$0.006 \\       
    LB-BC       & 1.001$\pm$0.004             &  1.005$\pm$0.005 \\ 
    LB-D        & 1.031$\pm$0.009             &  1.028$\pm$0.008 \\
    EB-A        & 1.007$\pm$0.013             &  1.025$\pm$0.008 \\
    EB-B        & 1.001$\pm$0.006             &  1.012$\pm$0.007 \\ 
    EB-D        & 1.008$\pm$0.010             &  1.012$\pm$0.010 \\ 
\hline
\end{tabular}
\end{center}
\caption{Double ratio given in Eq.~(\ref{eq:doubleRatioCosmics}) by
  the ratio of the experimental and simulated
  $\Delta E/\Delta x$ truncated means for different layers in the long
  barrel (LB) and extended barrel (EB), using isolated muons from
  $W\rightarrow\mu\nu$ in data and MC events in 2011 and 2012. The sources of uncertainty
  are described in the text.  
\label{tab:collisionRatioLayer}} 
\end{table}

\bigskip
\noindent\textbf{Time stability}
\smallskip

The double ratio defined in Eq.~(\ref{eq:doubleRatioCosmics}) as the ratio of the responses in experimental and simulated data, averaged over all calorimeter cells of the same type, is calculated for all cell types for each year (2011, 2012). 
The selection criteria associated with the systematic uncertainties reported in Table~\ref{tab:collisionUncert} are varied and used to generate 1000 working points. For each such point, the analysis is repeated. Similarly to the radial layer intercalibration studies, a model with a two-dimensional (2011, 2012) Gaussian function is applied. The log-likelihood is minimised to fit the mean double ratio response for each year taking into account the correlations between the years, also obtained from the varied analyses.  The relative difference of the fitted responses between two years is computed as

\begin{equation*}
  \label{eq:collision_muons_2011_2012}
  \Delta_{2011\rightarrow 2012} \equiv \frac{\hat{R}^c_{2012} -
  \hat{R}^c_{2011}}{\hat{R}^c_{2011}}
\end{equation*}

for each cell of a given type, to quantify the response change. The
average difference across all cells is found to be $\langle
\Delta_{2011\rightarrow 2012} \rangle = (0.6 \pm 0.1) \%$, indicating
good stability of the response.\footnote{The precision of
  the Cs system is approximately 0.3\% as mentioned in
  Section~\ref{ref:subsecCs}. The steeper response up-drift observed
  in second half of 2012 (see Figure~\ref{fig:CsResponse}, right)
  might also contribute, since the Cs calibration constants were
  typically updated once per month.}
The distribution of $\Delta_{2011\rightarrow 2012}$ over cell types 
shows an RMS spread of $0.96\%$.

\subsection{Energy response with hadrons}

The calorimeter response can be also tested using single hadrons and jets. Compared to muons, these objects deposit more energy in the hadronic calorimeter and therefore the response to higher energies can be probed. In addition, the MC simulations of objects interacting hadronically are compared with experimental data.

\subsubsection{Single hadrons}
\label{sec:single_hadrons}
The energy response of the TileCal is probed in situ by
studying the ratio of a charged hadron's energy ($E$), as measured
by the TileCal, to that of the hadron's momentum ($p$), as measured by
the ATLAS inner detector system~\cite{bib:detectorPaper}.
The energies of hadrons in data and MC events are calibrated to the
electromagnetic energy scale.
The data-to-MC double ratio given by $\langle
E/p \rangle_{\mathrm{data}}/\langle E/p \rangle_{\mathrm{MC}}$ should be
approximately one, with deviations from unity possibly due to
poor EM scale calibration in the data or differences in the MC description of the more complex hadron shower development (relative to the muon studies).

The datasets used in this analysis are based on the collision data taken at the LHC during 2010--2012. In 2010, 92\,nb$^{-1}$ of data were collected using the Minimum Bias Trigger Scintillators (MBTS). In 2011 and 2012 the data were triggered using fixed-rate random triggers, corresponding to effective integrated luminosities of 15.5\,nb$^{-1}$ and 129\,nb$^{-1}$, respectively.
The MC datasets were generated using \PYTHIAV{6}~\cite{bib:Pythia6}
(2010, 2011) and \PYTHIAV{8}~\cite{bib:Pythia8} (2012) to simulate minimum-bias non-diffractive events.
The MC events are weighted to reproduce the average number of interactions per bunch
crossing, $\langle \mu \rangle$, as seen in data. The MC events are also reweighted such that the spectra of the number of tracks match that of the data for 8 bins in $\eta$ and 16 bins in $p$.

The data and MC events are required to meet the selection criteria listed in
Table~\ref{tab:EoverPSelection}. First, a candidate track is required
to have transverse momentum greater than 2\,GeV in order
to reach the TileCal (Cut~1). The extrapolated tracks must have an absolute
pseudorapidity less than 1.7 to be within the TileCal geometrical acceptance (Cut~2). Only
tracks matched to non-problematic cells in the TileCal and with a
maximum energy deposit not in the gap or crack scintillators are
considered (Cut~3). In addition, the track is required to meet
isolation criteria, such that the total transverse momentum of all
other tracks in a cone of $\Delta R = 0.4$ in the $\eta$--$\phi$ plane
around the particle direction is required to be less than 15\% of the
candidate track's transverse momentum (Cut~4). The track must have at least
a minimum number of hits in the three inner detector systems (Cut~5),
and is required to have an impact point close to the primary vertex
(Cut~6). Only one track per event is considered.
Next, the energy associated with a track is defined as the sum of the energy deposited in calorimeter cells (LAr, TileCal, or LAr + TileCal) calibrated to the EM scale belonging to topological clusters with a barycentre within a cone of size $\Delta R = 0.2$ around the
projected track direction. The sum of energy deposited in the upstream electromagnetic
calorimeter is required to be compatible with that of a
minimum-ionising particle, $E_{\mathrm{LAr}}<1$\,GeV
(Cut~7).\footnote{This
  analysis uses a lower upper bound than in
  Section~\ref{sec:collision_muons} in order to suppress early
  showering particles.}
Finally, the amount of energy deposited in the TileCal must be at least 75\% of the total energy of associated calorimeter cells to reject muons (Cut 8). Only events with $\langle \mu \rangle$ between 3 and 25 are accepted in 2012 analysis to have a reasonable sample size in both data and MC simulation at the low and high edges of the $\mu$ distribution (Cut 9). Approximately 2.5\% of events survive these selections. 

\begin{table}[htp]
\begin{center}
\begin{tabular}{| c | c | }
\hline
      Cut      &      Selection Criteria            \\
\hline\hline           
     1       & track $\pT > 2$\,GeV                 \\
     2       & extrapolated tracks $|\eta_{\mathrm{track}}| < 1.7$    \\   
     3       & extrapolated tracks outside problematic regions in TileCal   \\   
     4       & $p_\mathrm{T}^{\mathrm{cone40}}$ / $\pT$(track)$< 0.15$ \\
     5       & one hit in Pixel and TRT, six hits in SCT        \\
     6       & interaction point $d_0 < 1.5$\,mm and $z_0\sin\theta<1.5$\,mm      \\ 
     7       & energy in LAr $E_\mathrm{LAr} < 1$\,GeV                          \\
     8       & fraction of energy in TileCal $> 75\%$      \\     
     9       &  $3 < \langle \mu \rangle < 25$ (2012 only)    \\
\hline
\end{tabular}
\end{center}
\caption{The selection criteria used for the $E/p$ analysis with single isolated hadrons.}
\label{tab:EoverPSelection}
\end{table}

\begin{figure}[tp]
\centering
\includegraphics[width=0.45\textwidth]{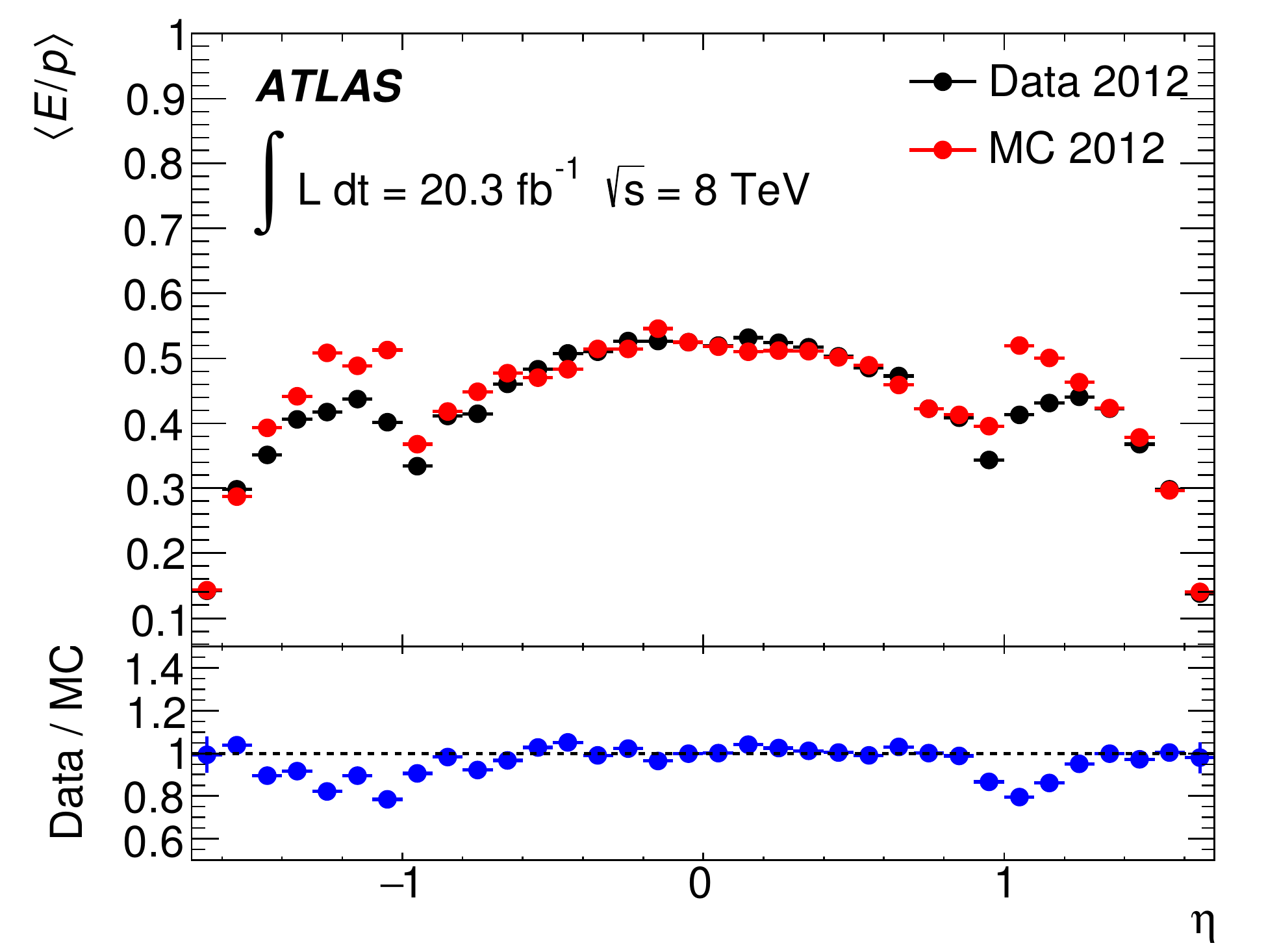}
\hfill
\includegraphics[width=0.45\textwidth]{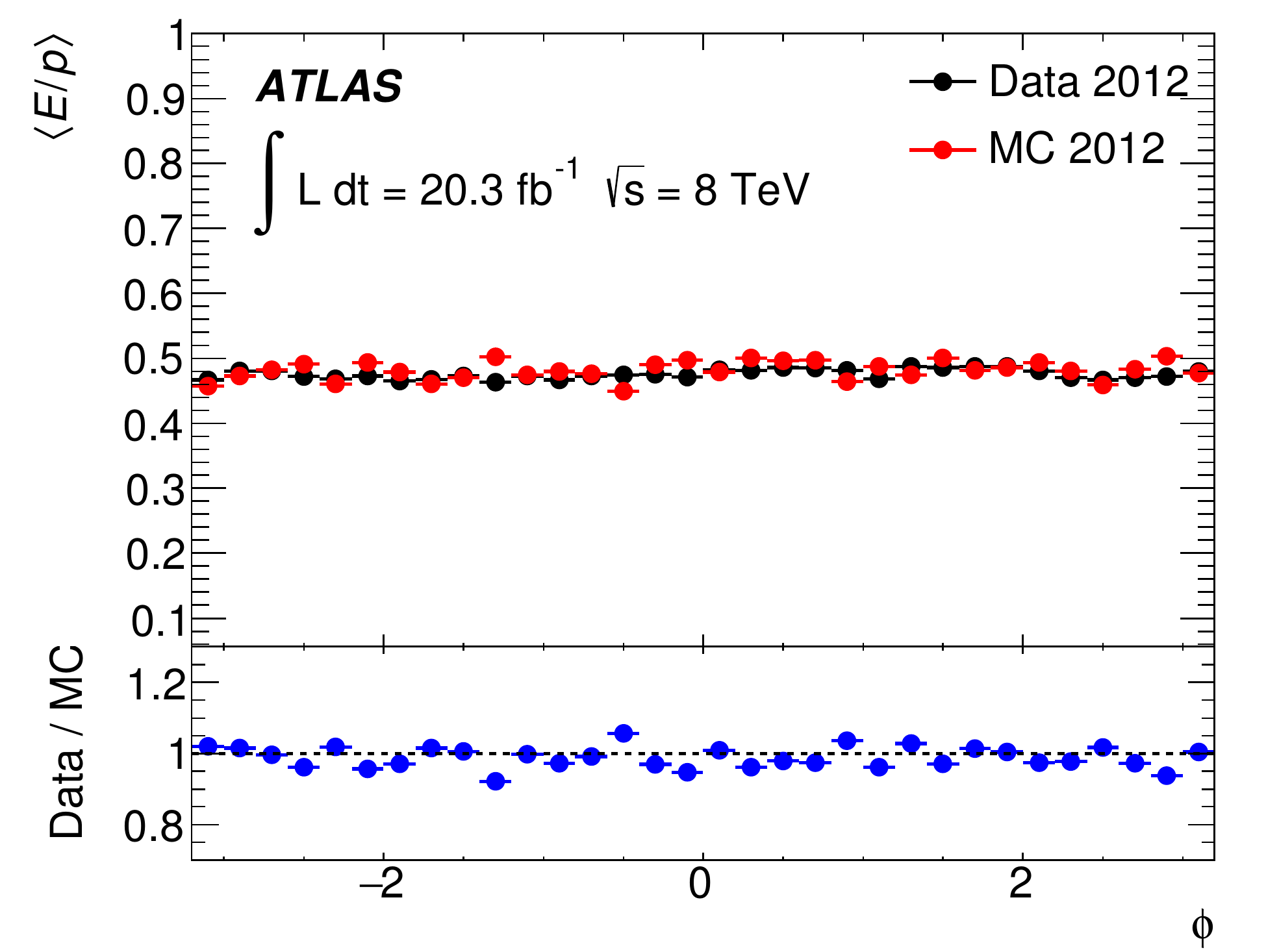}\\
\vspace{3mm}
\includegraphics[width=0.45\textwidth]{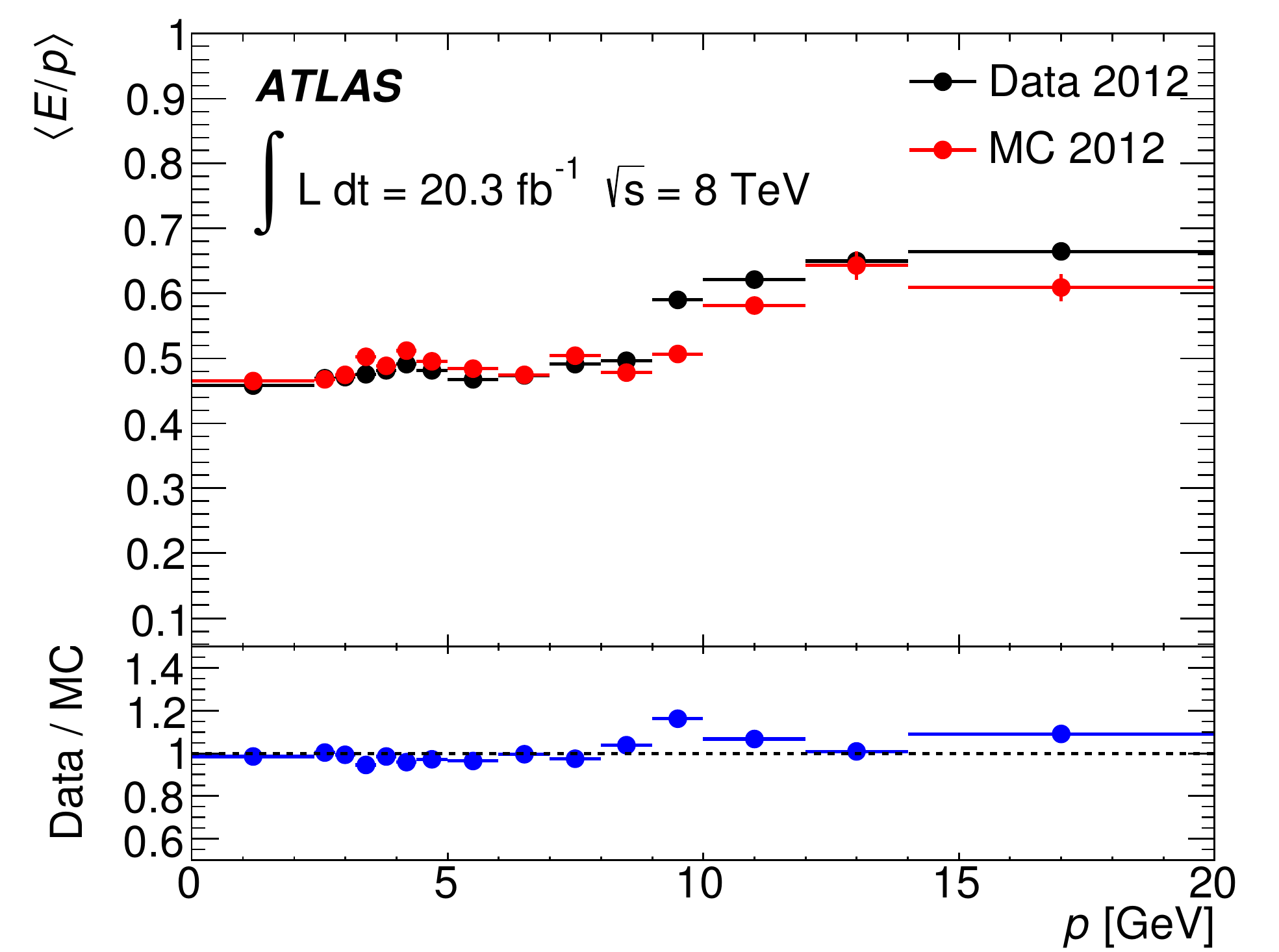}
\hfill
\includegraphics[width=0.45\textwidth]{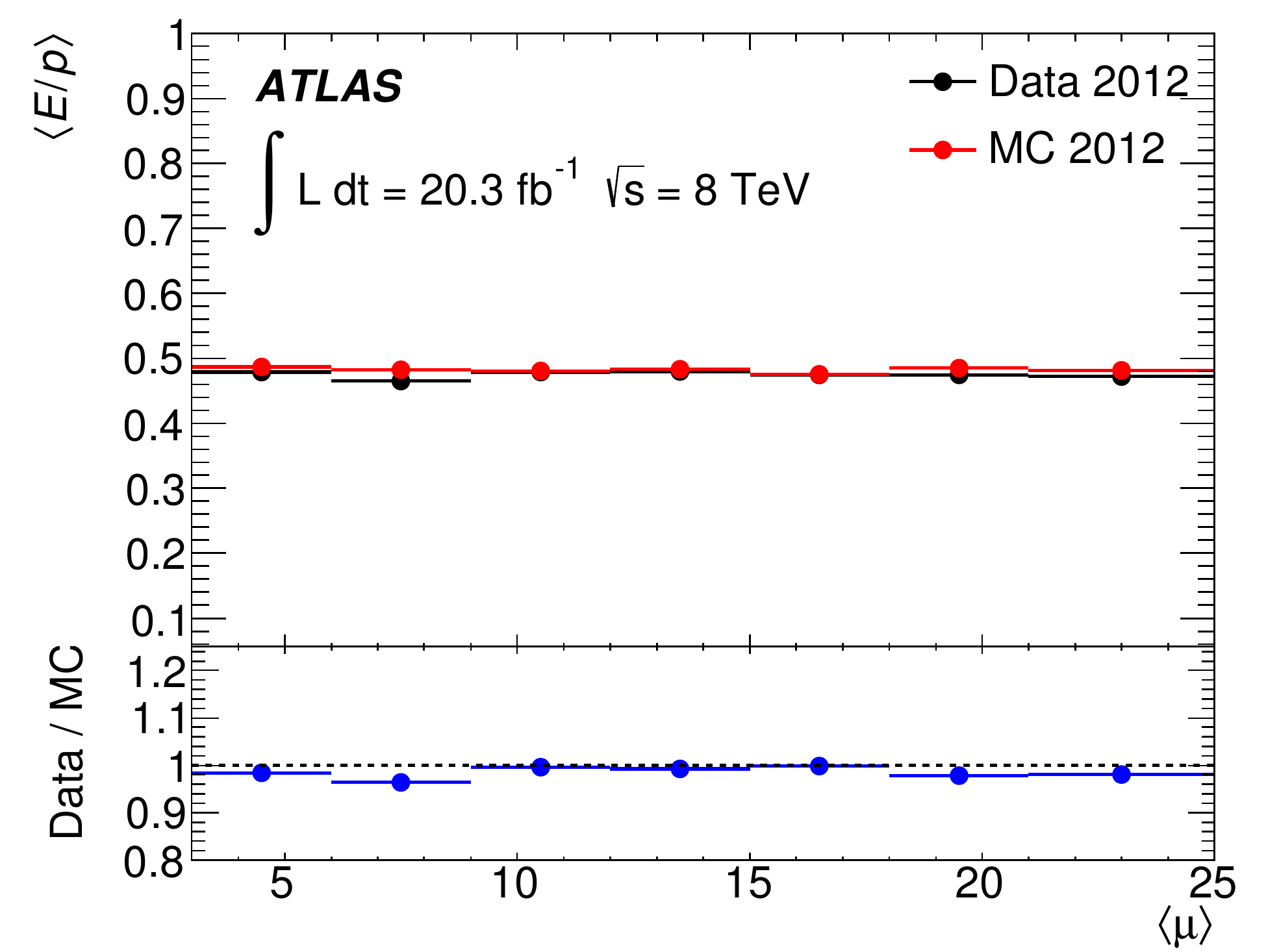}
\caption{Distributions of $\langle E/p \rangle$ as a function of
  $\eta$ (top left), $\phi$ (top right), $p$ (bottom left) and
  $\langle \mu \rangle$ (bottom right) measured in 2012. Only statistical uncertainties are shown.
  The sources of systematic uncertainty are discussed in the text.
\label{fig:EoverP2012}}
\end{figure}

Distributions of $\langle E/p \rangle$ \footnote{Here $E$ corresponds only to the sum of energy deposited in the TileCal cells calibrated to the EM scale and belonging to topological clusters with a barycentre within a cone of 0.2 around the projected track direction.} as a function of $\eta$,
$\phi$, $p$ and $\langle \mu \rangle$ are studied in all three
years. The results with statistical uncertainties as measured in 2012
are shown in Figure~\ref{fig:EoverP2012}. The agreement between data
and MC simulation is overall good in all cases. The value of $\langle E/p
\rangle$ is approximately 0.5, reflecting the non-compensating response of the
calorimeter, and very stable as a function of the
azimuthal angle $\phi$. The dependence on the pseudorapidity is well
reproduced in simulated data and the maximum disagreement is in the
region $\eta \approx \pm 1$ (crack region) in all three years. 
The distribution of material in the crack region is not known precisely and 
therefore it cannot be described accurately in the Monte Carlo simulations.
The difference between the data and the MC simulation is
reduced, even in this less well-described region, once the jets are calibrated to
the jet energy scale using in situ techniques. 
The ratio $\langle E/p \rangle$ measured in $pp$ collision data increases 
from~0.5 to~0.6 for track momenta of about 10\,GeV. This rise is not 
reproduced very
well in the MC simulation, the largest difference between
data and simulation (16\%) is observed at $p\approx 9$\,GeV.
The ratio $\langle E/p \rangle$ is found to be stable versus pile-up.

The systematic uncertainties associated with the
event selection procedure, the energy scale in the TileCal, and the MC
simulation are 
considered. The event selection systematic uncertainties are evaluated using
variations in the cuts applied in the analysis. 
The cuts on the number of hits in inner detector (Cut 5), the energy
deposition in the LAr calorimeter (Cut 7) and fraction of energy in
the TileCal (Cut 8) in Table~\ref{tab:EoverPSelection} are changed up/down
by an amount corresponding $1 \sigma$ of the
relevant distribution. The variation of the distance of the track from
the primary vertex (Cut 6) within $1 \sigma$ was found to be
negligible. No additional systematic uncertainty is assigned to the changing
pile-up conditions since no dependence on $\langle \mu \rangle$ is
observed. Other cuts are used to ensure that the hadron reaches
the TileCal and therefore are not varied. The mean value is recalculated
for all considered variations. The deviations from the nominal value
are summed in quadrature for lower and upper limits due to each source of 
systematic uncertainty.
%
The uncertainty of the EM energy scale (4\%) is fully
correlated across the momentum, pseudorapidity, azimuth and
$\langle\mu\rangle$, so the data/MC differences observed in a few momentum bins cannot be explained.
Other possible sources of systematic uncertainties are associated with the MC 
simulation, namely with the neutral particle production and modelling 
of the particle's passage through matter. The neutral particles (neutrons, 
$K^{0}_{\mathrm{L}}$), if produced close to the measured charged hadron, 
alter the calorimeter signal. While this effect plays some role in 
electromagnetic calorimeters, it is found to be negligible in the hadronic 
calorimeters~\cite{bib:single_particle_calo_response}. 
Two hadronic interaction models implemented in the {\GEANT}4 toolkit were 
compared, the difference in simulated $\langle E/p\rangle$ 
in the hadronic calorimeter was found to be well below 
5\%~\cite{bib:single_particle_calo_response}.
To conclude, the total systematic uncertainties associated with individual 
points in the $\langle E/p\rangle$ plots shown in Figure~\ref{fig:EoverP2012}
are highly correlated and are estimated to be of the order of 6\,\%. They do 
not cover some of the data/MC discrepancies, which leaves open the 
possibility of further simulation development.

Double ratios $\langle E/p \rangle_{\mathrm{data}}/\langle E/p
\rangle_{\mathrm{MC}}$ are used to validate the agreement between data
and MC simulation. The results for all three years are listed in
Table~\ref{tab:Eopresults}. An overall double ratio, averaged over all three years,
$0.986 \pm 0.003 \, \mathrm{(stat)} ^{+0.059} _{-0.018} \, \mathrm{(sys)}$ is measured. 
The double ratio in 2011, which shows the largest deviation from unity, 
agrees with this result within $1.5\sigma$ assuming an uncorrelated systematic uncertainty across different years. The results from
2010--2012 shows the cell energy is well calibrated to the
EM scale and also good agreement between experimental
data and MC predictions for the single hadrons.  

\begin{table}[htp]
\begin{center}
\begin{tabular}{| l | c | c | c | }
\hline
                     &     2010                & 2011                  & 2012 \\
\hline              
    Data-to-MC ratio of $\langle E/p \rangle$   &  $1.000\pm 0.004$  & $0.927\pm 0.007$  & $0.987\pm 0.004$ \\
\hline
\end{tabular}
\end{center}
\caption{Double ratios $\langle E/p \rangle_{\mathrm{data}}/\langle E/p \rangle_{\mathrm{MC}}$ with their statistical uncertainties for years 2010 to 2012.
\label{tab:Eopresults}}
\end{table}

\subsubsection{High transverse momentum jets}
The performance of ATLAS jet reconstruction is strongly influenced by
the quality of energy reconstruction in the TileCal, as this
calorimeter reconstructs about 30\% of the total jet energy (for jets
with energies above 140\,GeV at the electromagnetic scale). It is important that MC simulation correctly describe the complicated structure of jets as they propagate through the detector to the TileCal, since MC simulation are often used to optimise reconstruction algorithms and compute initial calibrations. The MC simulation are also heavily used by searches for new physics to quantify the statistical (dis)agreement of predictions with the observed data. 
This subsection studies how well the longitudinal shower profile of high-$\pT$ jets is described in the MC simulation by looking at the fraction of energy deposited in each TileCal layer.
The analysis uses jets that are clustered using the anti-$k_{t}$ clustering algorithm with a radius parameter of 0.4~\cite{bib:antiKt}. The inputs of the jet algorithm are the topological clusters. All jets considered here are calibrated to the EM energy scale.

The results are based on the full dataset from $pp$ collisions in 2012. Candidate events are selected such that they contain one isolated high-$\pT$ photon ($\pT > 100$\,GeV) and one jet ($\pT > 140$\,GeV). 
Photons are required to meet the tightest ATLAS definition based on shower shape quantities~\cite{bib:photons}.
Jets are selected after passing standard procedures to remove sources of mismeasured jets such as beam backgrounds and detector read-out problems~\cite{bib:jet7tev}.
Jet candidates are removed if they geometrically overlap with a photon within a cone of $\Delta R = 0.4$ centred around the jet candidate.
In addition, jets are removed if they are reconstructed adjacent to masked cells which have energies interpolated from working neighbouring cells. 
Finally, jets and photons are required to be separated by an azimuthal angle larger than 2~radians to suppress events with additional jets from radiated quarks and gluons.

The experimental data are compared with MC simulation in which a prompt photon is produced in association with a jet at parton level. These events are generated using \PYTHIAV{8} with the CTEQ6L1 PDF set with the ATLAS AU2 set of tuned parameters of \PYTHIAV{8}~\cite{bib:Pythia8}, a leading-order matrix-element MC generator.

\begin{figure}[tp]
\centering
\includegraphics[width=0.32\textwidth]{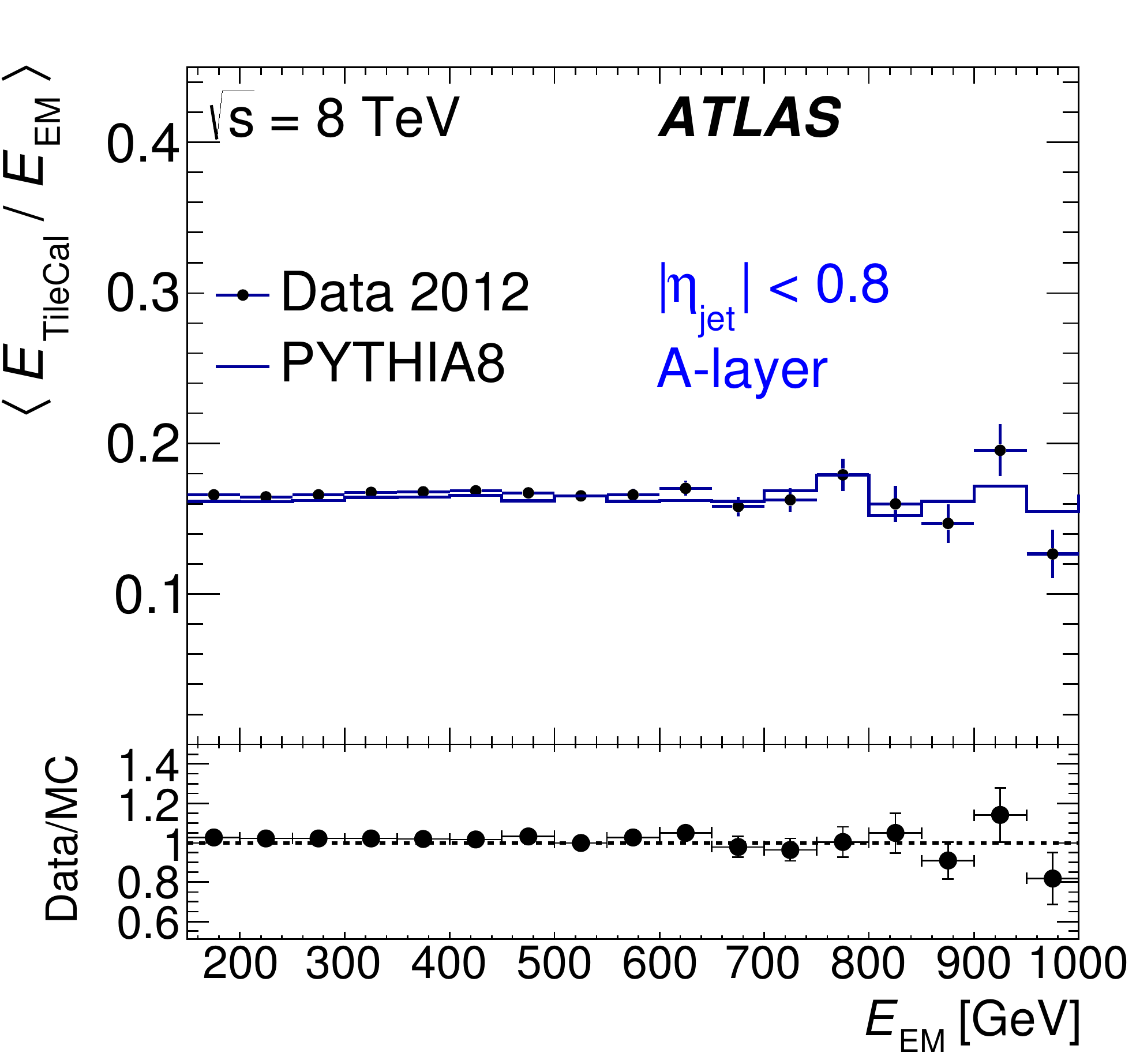}
\includegraphics[width=0.32\textwidth]{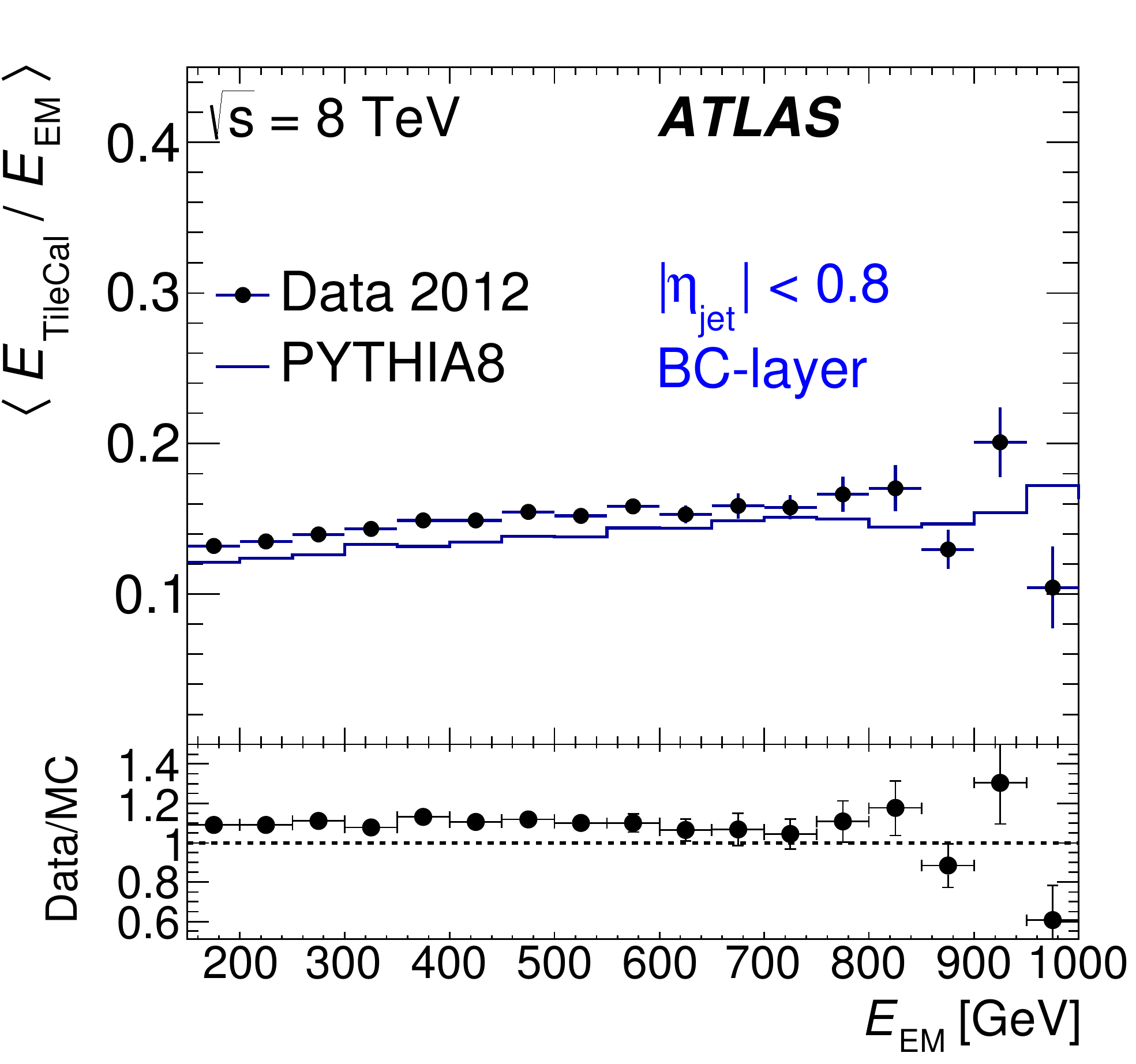}
\includegraphics[width=0.32\textwidth]{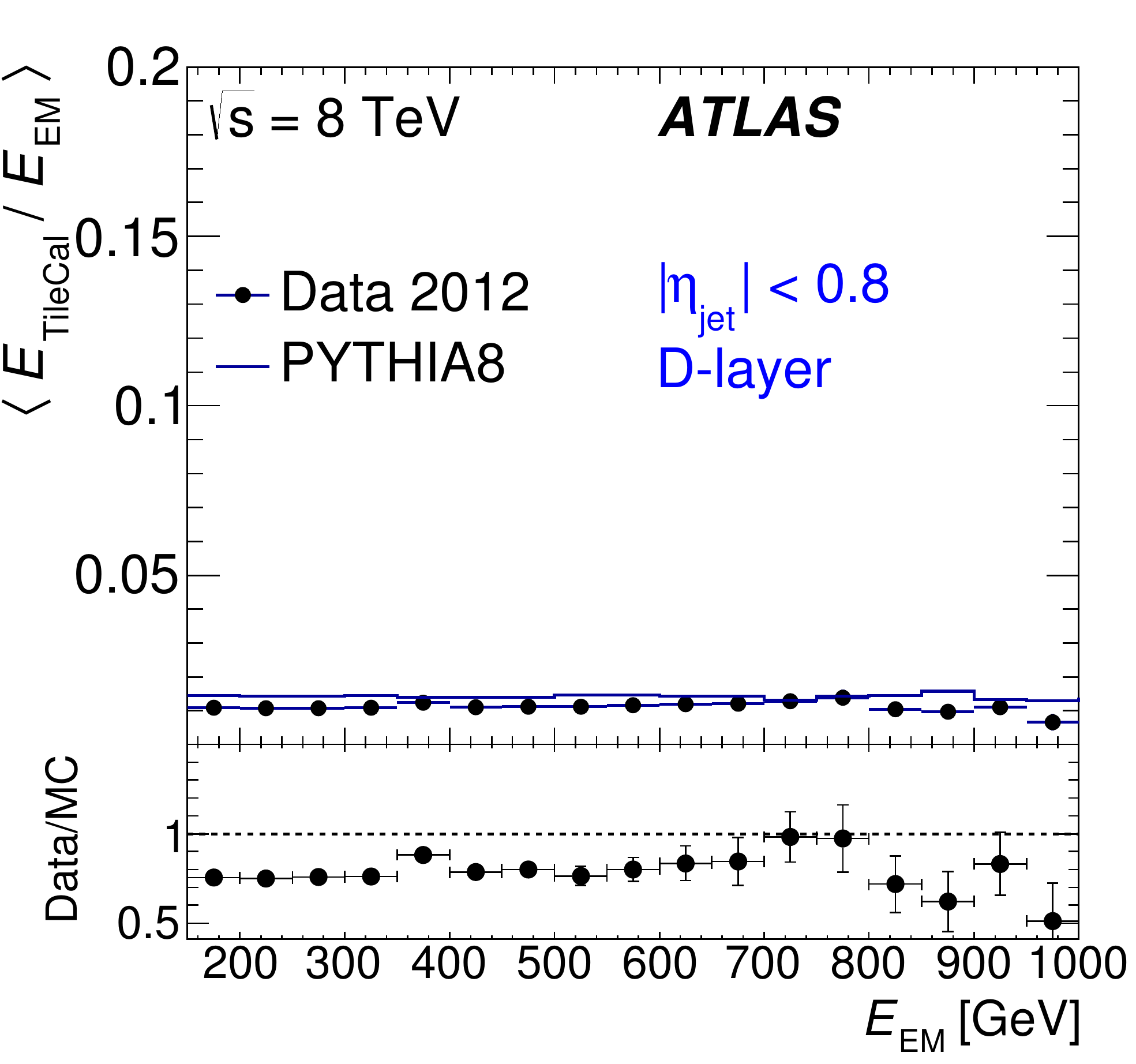}
\caption{Electromagnetic scale jet energy fraction in the TileCal for jets with $\pT>140$\,GeV in the long barrel ($|\eta_\mathrm{jet}| < 0.8$) for layer A (left), layer BC (middle), and layer D (right). Error bars represent statistical uncertainties. \label{fig:highPtJetsLB}}
\end{figure}

\begin{figure}[tp]
\centering
\includegraphics[width=0.32\textwidth]{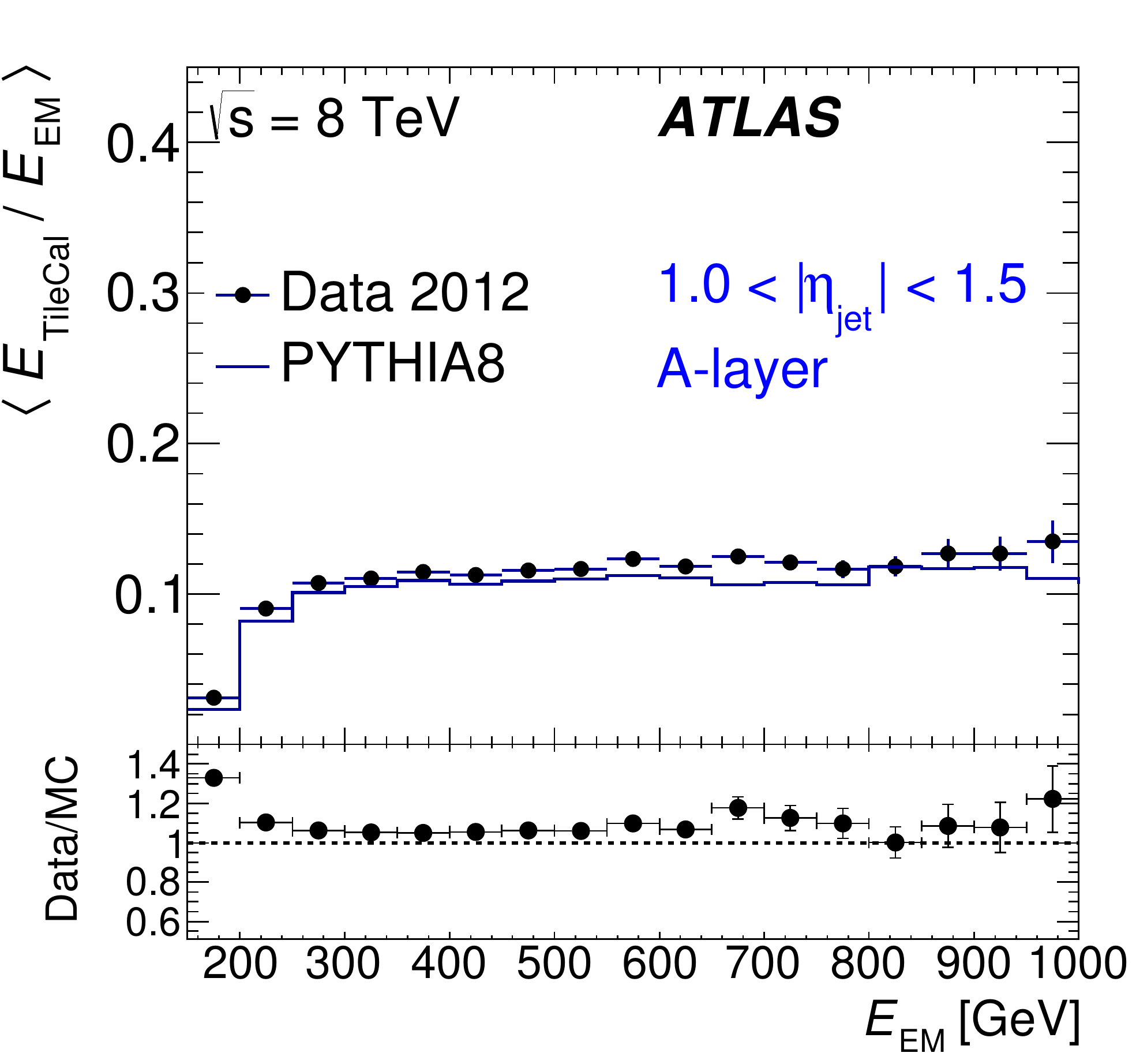}
\includegraphics[width=0.32\textwidth]{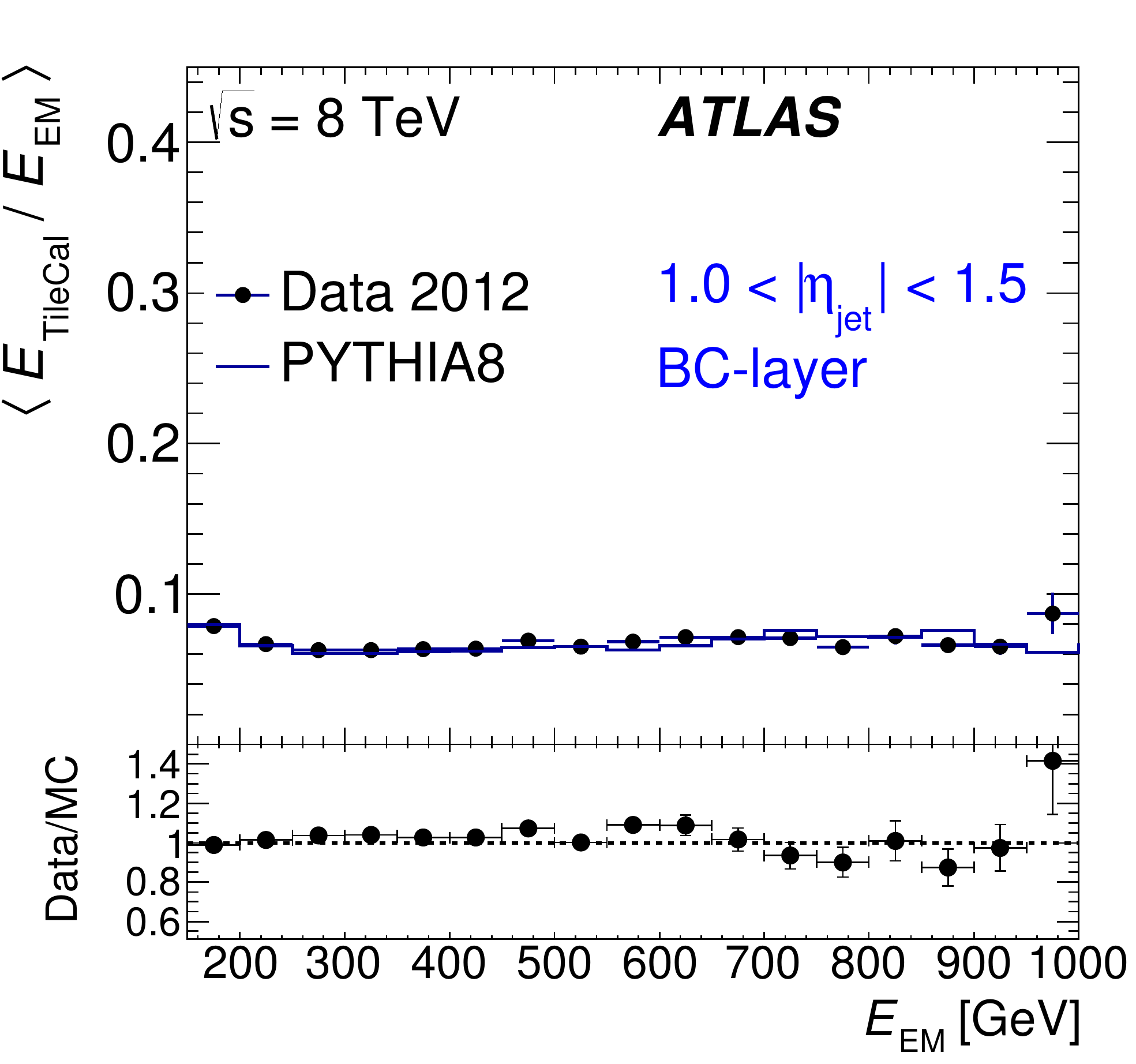}
\includegraphics[width=0.32\textwidth]{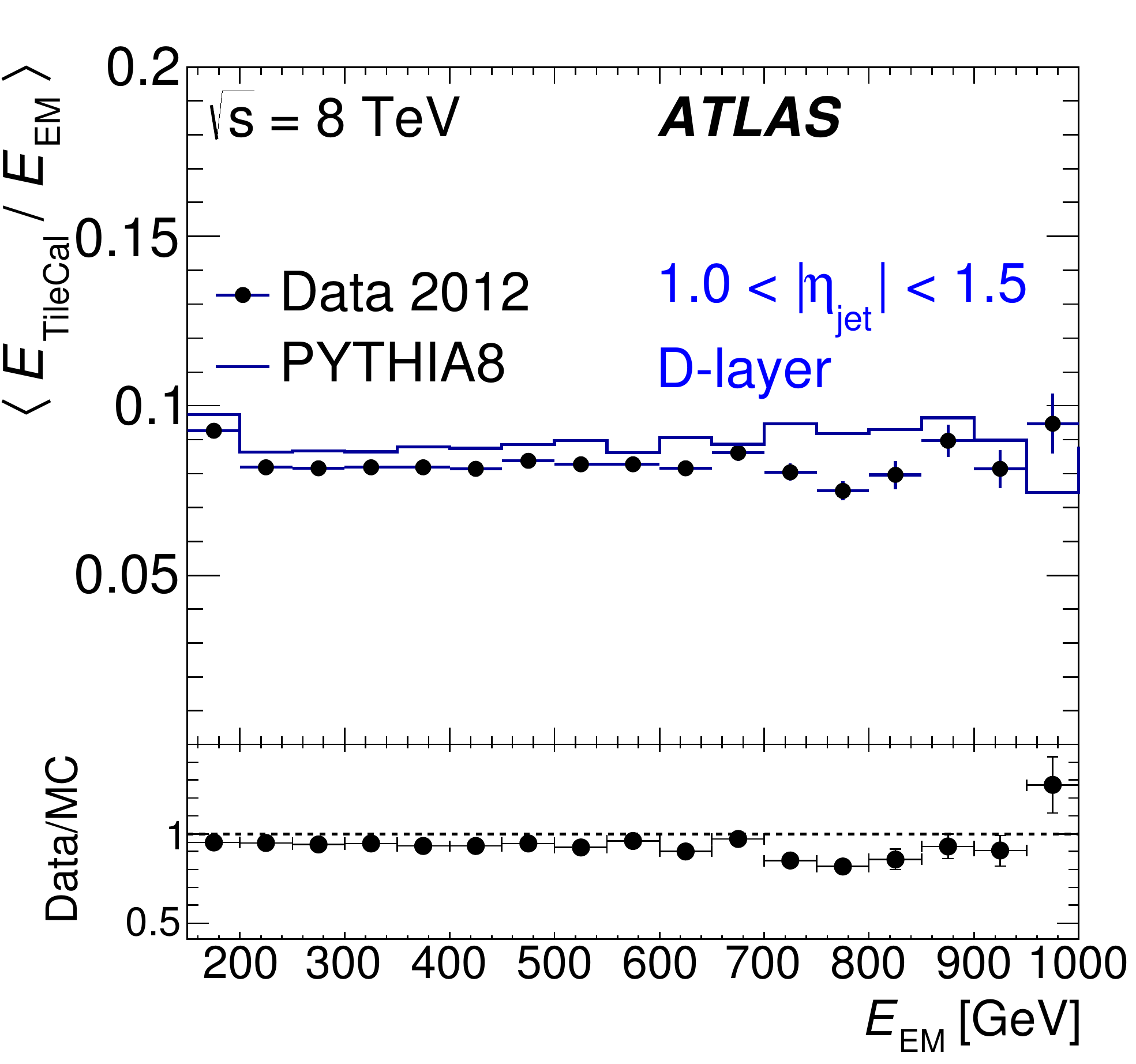}
\caption{Electromagnetic scale jet energy fraction in the TileCal for jets with $\pT > 140$\,GeV in the extended barrel ($1.0 < |\eta_\mathrm{jet}| < 1.5$) for layer A (left), layer BC (middle), and layer D (right). Error bars represent statistical uncertainties. 
\label{fig:highPtJetsEB}}
\end{figure}

Figure~\ref{fig:highPtJetsLB} shows the fraction of jet energy in each TileCal layer relative to the total energy reconstructed by the Tile and LAr calorimeters at the EM scale for both the experimental and simulated data. The energy fractions are shown in the different TileCal layers in the long barrel ($|\eta_{\mathrm{jet}}| < 0.8$). 
Similarly, Figure~\ref{fig:highPtJetsEB} shows the TileCal energy
fractions in the extended barrel region ($1.0 < |\eta_{\mathrm{jet}}| < 1.5$). Each layer has a different thickness as mentioned
  in Section~\ref{sec:intro}.

Generally the MC simulation describes the shape representing the fraction of energy deposited in each layer as function of jet energy. Good agreement is found in layer A in the long barrel and also in layer BC in the extended barrel. However, some discrepancies are observed in BC layer in the long barrel, where the MC simulation underestimates the amount of energy deposited by approximately 10\% uniformly at the EM scale. The opposite feature is observed in layer D, where the MC simulation overestimates the amount of energy deposited in this layer by 20\%. The last layer only measures approximately 1\% of the total jet energy, thus having a small impact on the total energy. Overall, better agreement is observed in the extended barrel. The energy fraction in the extended barrel is underestimated by the MC simulation in layer A. The opposite is observed in layer D.

\begin{figure}[ht]
\centering
\includegraphics[width=0.49\textwidth]{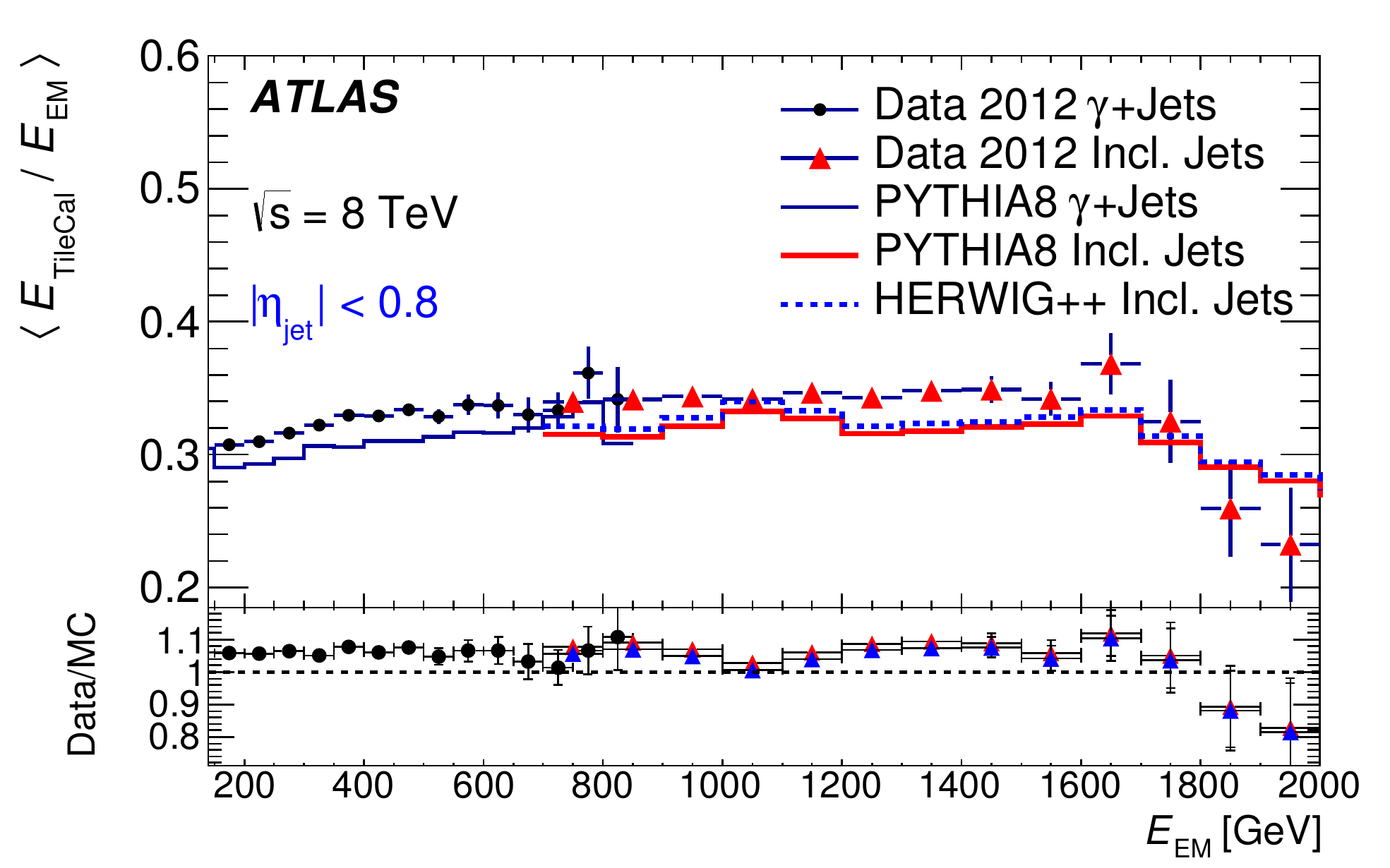}
\includegraphics[width=0.49\textwidth]{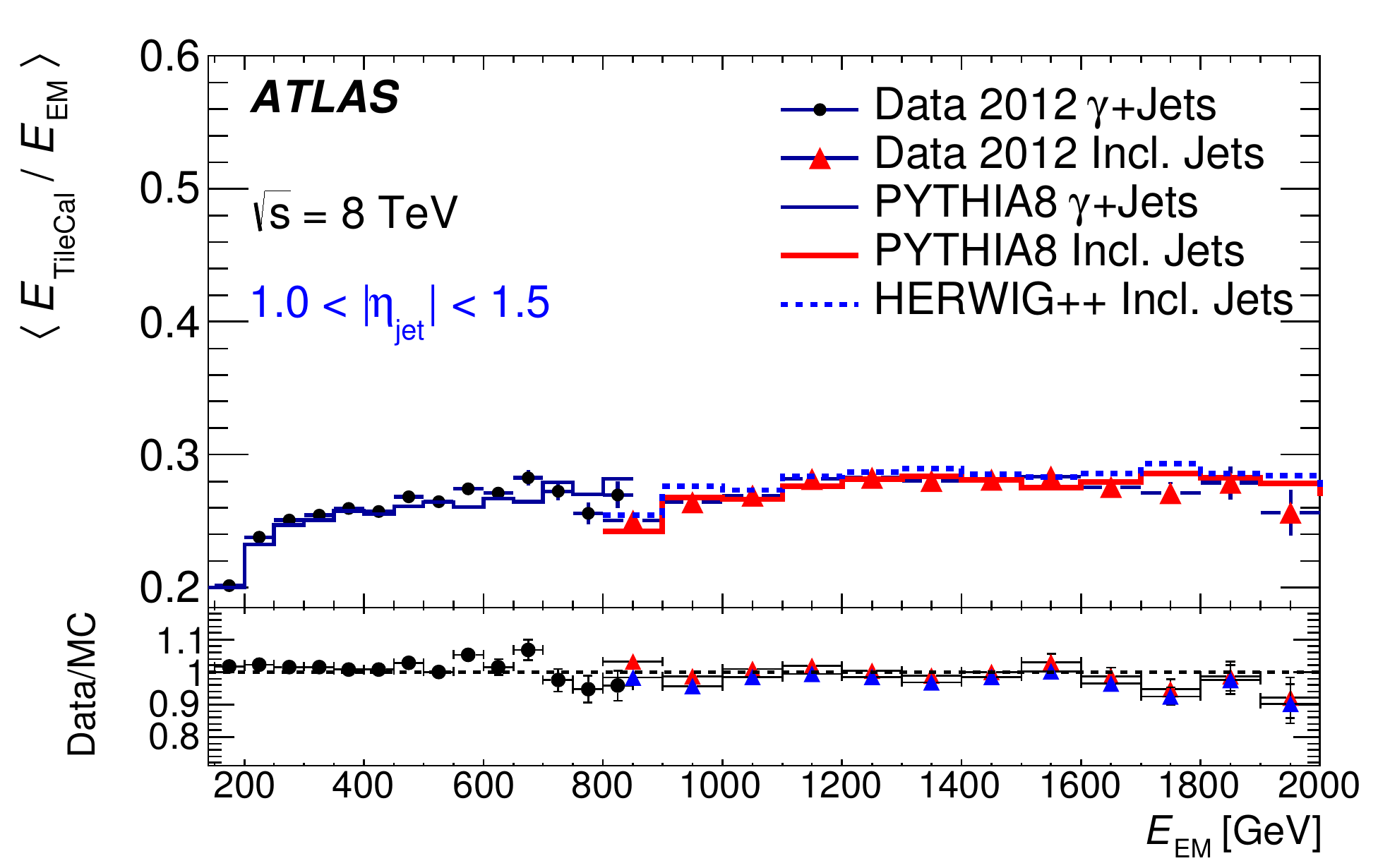}
\caption{Electromagnetic scale jet energy fraction in the TileCal for jets after combining the photon-plus-jet and inclusive jet samples for the long barrel (left) and extended barrel (right). Error bars represent statistical uncertainties. The simulation of the inclusive jets done with \PYTHIAV{8} (thick solid line) and Herwig++~\cite{bib:herwig} (dotted line) MC generators can be seen, while $\gamma$ + jets events were simulated only with \PYTHIAV{8} (thin solid line). \label{fig:highPtTotalTile}}
\end{figure}

In order to study the jet energy measured in the TileCal for large jet
$\pT$ with a larger sample, the photon-plus-jet sample is combined with a sample of fully inclusive high-$\pT$ jets without the photon requirement. The latter sample is selected using an unprescaled trigger requiring a single jet with $\pT$ above 350\,GeV. Figure~\ref{fig:highPtTotalTile} shows the jet energy fraction measured by the TileCal for jets at the electromagnetic scale in the range $\pT = 140$\,GeV to 2000\,GeV. It can be seen that the energy fraction increases from 30\% at $\pT$ = 140\,GeV to 35\% at $\pT$ = 1800\,GeV in the barrel, and from 25\% to 30\% in the extended barrel. The MC simulation describes this trend well. 
Compared with the MC simulation, the data show a larger fraction of the total jet energy in the barrel region.
A drop in the energy fraction for jets with $\pT$ > 1800\,GeV in the barrel indicates leakage of the energy behind the TileCal volume.

The total difference between the data and MC simulation is within the
expected uncertainty of {4\%}, already mentioned in
Section~\ref{sec:single_hadrons}.

\subsection{Timing performance with collision data}
\label{subsec:timePerform}

As already mentioned in Section~\ref{subsec:timingCalib}, the
time calibration is crucial for the signal reconstruction and the
ATLAS L1 and HLT trigger decisions.
Accurate time measurements of energy depositions in the TileCal are used
to distinguish non-collision background sources from hard
interactions as well as in searches for long-lived particles.

The performance of the TileCal timing is studied using
jets and muons from 2011 $pp$ collision data. The
data used in both analyses represent about 2.5\% of the full 2011
integrated luminosity, taken with 50\,ns bunch crossing spacing. 

In both analyses, the E-cells and MBTS cells are also used. Cells
with known problems related to issues such as miscalibrations or
hardware failures, or known to exhibit timing jumps
(Section~\ref{subsec:timingCalib}) are not considered. In total, approximately 2.5\% of all cells are removed. 

The time resolution of the detector is parametrised as function of
the cell energy $E$ according to:

\begin{equation}
  \label{eq:timeReso}
  \sigma = \sqrt{p_0^2 + \left(\frac{p_1}{\sqrt{E}}\right)^2 + \left(\frac{p_2}{E}\right)^2}
\end{equation}

where $p_0$ reflects the constant term accounting for miscalibrations
and detector imperfections, and $p_1$ and $p_2$ represent the
statistical and noise terms, respectively.

\subsubsection{Jet analysis}

Jets are built with the topological clustering and anti-$k_{t}$
algorithms (with radius parameter $R = 0.4$).  Only jets with $\pT > 20$\,GeV found to originate from the hard
collision's primary vertex, and satisfying basic jet cleaning criteria, are selected. One component of the recommended cleaning criteria is to include only jets with a reconstructed time\footnote{The jet time
  $|t_{\mathrm{jet}}| $ is calculated as an $E_{\mathrm{cell}}^2$-weighted average of cell times, running over all cells associated to the given jet. The cell time is measured with respect to the expected arrival time of particles coming from the interaction point.} $|t_{\mathrm{jet}}|
<10$\,ns, but to avoid any biases this cleaning cut is calculated
using only non-TileCal cells associated with the corresponding jet.

Cells selected by the topological clustering
algorithm and with energies above 500\,MeV
are used in the analysis.
The cell times are separated into several cell energy bins of
approximately 2\,GeV wide. Each distribution is fit with 
a Gaussian function and its width ($\sigma$) is considered as 
the time resolution (see also Section~\ref{subsec:timingCalib}
and Figure~\ref{fig:time_calibration}, right).
The resulting distribution of cell $\sigma$ versus energy is fit according to
Eq.~(\ref{eq:timeReso}), the results of which are discussed in
Section~\ref{subsubsec:timeCombo}.

\subsubsection{Muon analysis}

Muons are reconstructed using an algorithm that performs a global re-fitting of the muon track using the hits from both the inner detector and the muon spectrometer~\cite{bib:muonRecoRun1}. Selected muons are required to fulfil the
kinematic and detector criteria shown in Table~\ref{tab:timeMuons}.  
As all isolated muons originating from collision events are
considered in this analysis, the selection criteria
differ slightly from those presented in Section~\ref{sec:collision_muons} where
muons from $W$ boson decays were selected.

\begin{table}[htp]
\begin{center}
\begin{tabular}{| c | c | }
\hline
      Cut      &      Selection Criteria                \\
\hline\hline  
      Muon kinematics      & $p > 3$\,GeV   \\
                                                  & $\pT > 1$\,GeV   \\
\hline
      Muon track      &  Six hits in SCT, one hit in Pixel  \\         
                                & $|\eta_{\mathrm{track}}| < 2$             \\                       
\hline
      Track isolation      & $p_\mathrm{T}^{\mathrm{cone40}} < 2$\,GeV    \\
\hline
      Calorimeter isolation      &  $\left(\sum\limits_{\mathrm{cells}}^{\Delta R < 0.4} \ET \right) < 2$\,GeV (excluding cells intersected by muon tracks)    \\    
\hline
      Muon path length & $\Delta x > 0.3 r_{\mathrm{cell}}$  \\
\hline
      Time difference & $|t_{\mathrm{cell}} - \langle t_{\mathrm{cell}}\rangle| < 15$\,ns  \\
\hline
      Cell energy & $\Delta E > 540$\,MeV  \\
\hline
      Energy balance & $\alpha < 0.7$  \\
\hline
\end{tabular}
\end{center}
\caption{Selection criteria used to evaluate the TileCal time
  resolution for muons from 2011 collision data. The first four
  criteria apply to the muon track. For the track and calorimeter
  isolation criteria the sums are over the non-muon tracks and cell
  energies, respectively, within a cone of size $\Delta R = 0.4$
  centred on the passing muon. The last four criteria are used to
  select individual cells along the muon track, the variables are
  defined in the text. 
\label{tab:timeMuons}} 
\end{table}

Muon tracks are extrapolated in $\eta$ and $\phi$ through each
calorimeter layer. Muon tracks that crossed just the cell edge
are removed by requiring their path length $\Delta x$ to be at least 30\%
of the corresponding cell radial size $r_{\mathrm{cell}}$. 
Tracks with a time differing from the corresponding mean cell time\footnote{The non-zero average cell time is used since the analysis was performed on data prior to the final time calibration. Also, the reconstructed time in muon-induced signals slightly differs from that of hadrons, where it also depends on longitudinal shower development.}
 $\langle t_{\mathrm{cell}}\rangle$ 
by more than 15\,ns are also removed. These cuts appear to be sufficient
to remove muons from non-collision origins, including cosmic muons.
Moreover, only cells with energy $\Delta E$ larger than 540\,MeV are
considered in order to remove contributions from noise.\footnote{This value is
  slightly higher than in the jet analysis, as the runs 
  selected for the muon analysis had higher pile-up noise.} 
The two channels contributing to the cell reconstruction are required
to have balanced energy deposits, to ensure the time which is computed
from the average of the two channels is not biased by one purely noisy
channel. The energy balance between the two channels is defined as: 

\begin{equation*}
\alpha = \frac{|E_1 - E_2|}{E_1 + E_2}
\end{equation*} 

where $E_1, E_2$ are the energies from each channel reading the same
cell. A cut is imposed to keep cells for which $\alpha < 0.7$.

It was discovered that cells further away from the interaction
point exhibit lower values of their mean cell time. This is traced to
the residual cell time corrections which are performed using jets from collision data, as hadronic shower development is slower than passing muons. Therefore, tuning of cell times using jet data introduces a small bias towards lower cell times for more distant cells traversed by muons. 
To remove this bias from the analysis the timing of each cell is corrected by its mean time, resulting in a perfectly timed detector. 

The measured time also depends on the muon track position in the cell. In large cells, muons passing near the edge of the cell can have up to $\pm 1.5$\,ns difference relative to those passing through the cell centre. The radial track impact point in the cell also plays a role, as the light signal from muons impacting the upper half of the cell has shorter WLS fibre length to travel to the PMT\@.

Once the corrections for the mean cell times and the muon track geometry (the track position and radial impact point in the cell) are applied, the cell times are binned as function of energy. A Gaussian distribution is fit to the cell times. The standard deviation of the Gaussian distribution is taken as the time resolution for that energy bin. The time resolution as a function of cell energy is fit to Eq.~(\ref{eq:timeReso}); these results are discussed in the next subsection.

\subsubsection{Combined results}\label{subsubsec:timeCombo}
\begin{figure}[tp]
\centering
\includegraphics[width=0.5\textwidth]{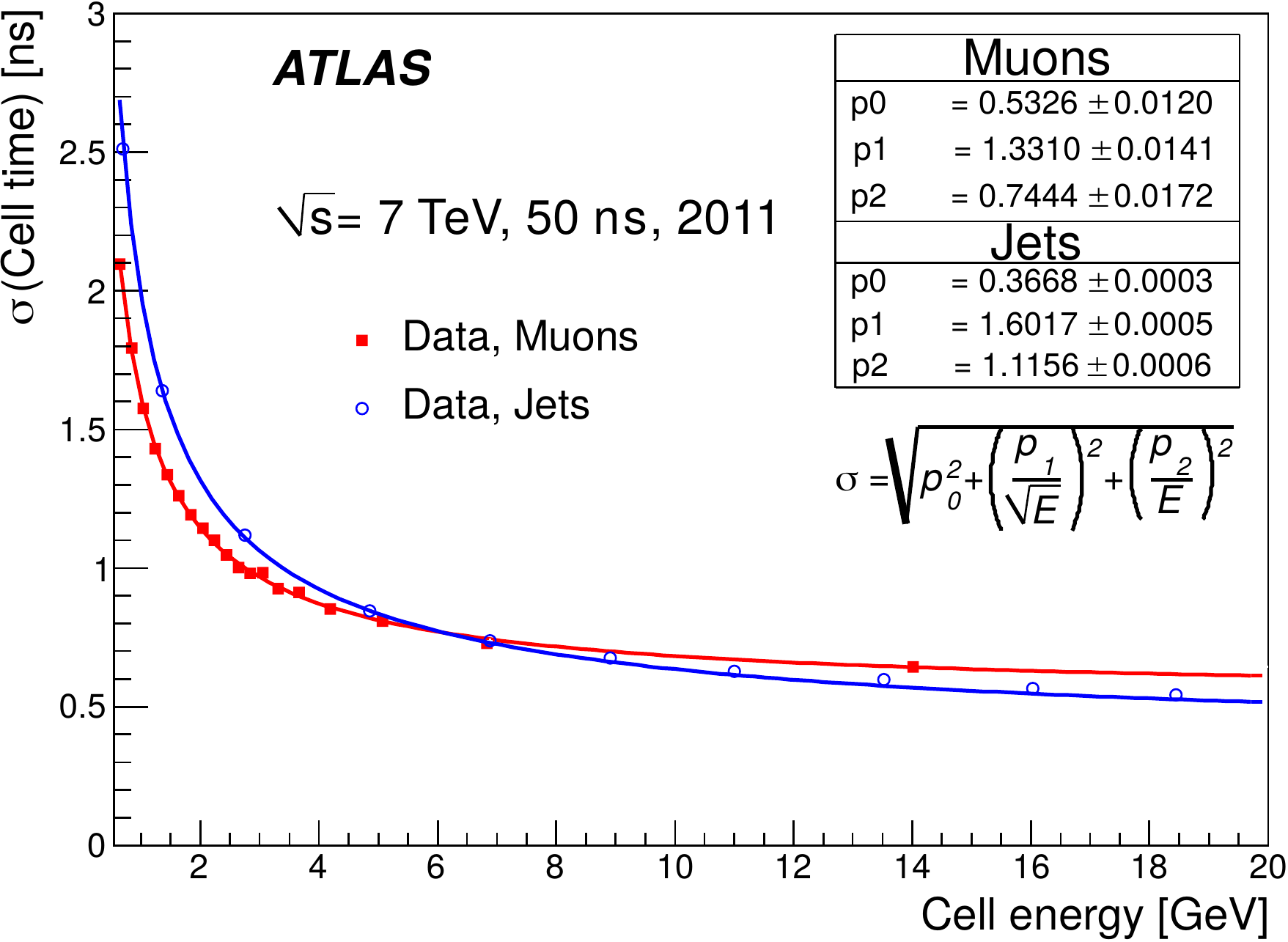}
\caption{Cell time resolution as a function of cell energy associated with jets and muons from 2011 collision data. The data are fit to Eq.~(\ref{eq:timeReso}), with the fit results for the three constants shown within the figure. The parameters $p_1$ and $p_2$ corresponds to energy expressed in GeV. Statistical errors are included on the points and on the fit parameters. The statistical uncertainties are smaller than the markers identifying the data points. \label{fig:timeReso}}
\end{figure}

The cell time resolutions as a function of cell energy associated with jets and muons are shown in Figure~\ref{fig:timeReso} along with the fit to Eq.~(\ref{eq:timeReso}). The time resolutions are similar, being slightly better for muons at lower energies, because of the slow hadronic component of low-energy jets. 
The fit for muons suffers from the small sample size at higher energies since the typical muon response per cell is of the order of 1\,GeV, depending on the cell size. The time resolution at energies above $\sim10$\,GeV is thus determined from jets and it approaches the constant term value of 0.4\,ns. Similar time resolution was obtained with single high-energy pions in beam tests~\cite{bib:emscaleTB}.

Figure~\ref{fig:timeReso} shows only cases when the cell is read out in the high--high gain mode. Using jets with cells read-out in the low--low gain, the fit result shows a similar value of the constant term $p_0$.

\subsection{Summary of performance studies}
\label{sec:performance_summary}

Muons from cosmic-ray data (2008--2010) and $W\rightarrow\mu\nu$ collision events (2011--2012), single hadrons (2010--2012) and jets (2012) were used to study the performance of the TileCal. The uniformity and time stability of the response, the level of agreement between MC simulation and experimental data, and the timing of the detector were investigated.

The cosmic muons analysis shows that the average non-uniformity of the response in each layer is approximately 2\%. The collision muon results exhibit the relative difference $\langle\Delta_{2011\rightarrow 2012} \rangle = (0.6 \pm 0.1)\%$ between the two years, indicating a good time stability of the response. The cosmic-muon results are also stable across the three years. 
Furthermore, from 2008--2010 the double ratio given in
Eq.~(\ref{eq:doubleRatioCosmics}) for the response to cosmic muons was about 0.97, indicating a systematic decrease of the EM energy scale in data by about 3\%, except for the LB-D layer. Using the muons from collision events in 2011 and 2012, this ratio was closer to 1.0 for all layers, indicating a small systematic difference between the collision and cosmic-muon results or between the two periods. Nevertheless, this difference is well within the EM scale uncertainty. Both analyses confirmed that all radial layers, except LB-D, are well intercalibrated.
The response of the LB-D layer is higher by $+4\%$ (cosmic muons) and $+3\%$ (collision muons), consistent between the two periods. Since the difference is within the expected EM scale uncertainty~\cite{bib:tileReadiness}, no correction is applied.
Nevertheless, several checks were performed to identify the origin of the difference. The response to cosmic muons was checked separately in the top and bottom parts of the calorimeter (see Section~\ref{subsec:cosmics}) and was found to be well described by MC simulation. The MC geometry and material distribution were also checked, but the detailed simulation of the optical part was not implemented. The optical non-uniformity of tiles is thus accounted for by applying additional layer-dependent weights derived in beam tests with muons passing through the calorimeter parallel to the $z$-axis~\cite{bib:emscaleTB,bib:tileReadiness}. These weights are used within Cs calibration constants (see Section~\ref{ref:subsecCs}).

The analysis of the $E / p$ of single hadrons shows good uniformity of the response across the azimuthal angle $\phi$, very good time stability, and robustness against pile-up. Good agreement between experimental data and MC simulation is observed. 
Good calibration of the cell energy to the EM scale is confirmed in this analysis. The longitudinal shower profiles are studied using high-$\pt$ jets. Compared with the MC predictions, a larger fraction of the total jet energy is deposited in the second radial layer in the barrel region. However, the total difference is within the TileCal's expected EM scale uncertainty of 4\% as determined from studies of isolated particles and in test beams. 

The time resolution of the TileCal is better than 1\,ns for energy deposits larger than a few GeV in a single cell.

\section{Conclusion}\label{sec:Conclusion}

This paper presents a description of the ATLAS Tile Calorimeter signal 
reconstruction, calibration and monitoring systems, data-quality, and 
performance during LHC Run~1.

The individual calorimeter calibration systems demonstrated their
precision to be better than 1\%. The combined calibration guarantees good
stability of the calorimeter response in time.

Robust signal reconstruction methods were developed, providing the ability to cope with varying conditions during Run~1, especially the increase in pile-up with time. 
The energy spectra for minimum-bias events with pile-up conditions in Run 1 shows good agreement between data and MC simulation for cell energies larger than a few hundreds of MeV, which is the region important for physics.

The TileCal also contributed to high-quality ATLAS data-taking with an
efficiency higher than 99\% during all three years of Run~1. 
Only 3\% of all cells were non-operational at the end of
data-taking.

The Tile Calorimeter performance was assessed with isolated
muons and hadrons as well as with jets. Cosmic-ray muons data and proton--proton collisions at the LHC at centre-of-mass energies of 7 and 8\,TeV with a total integrated luminosity of nearly 30\,fb$^{-1}$ were used in the analyses. The TileCal response was stable and uniform across the layers.
The energy scale uncertainty, which was successfully extrapolated from
the beam tests to ATLAS, is conservatively considered to be 4\%.
The MC modelling of the response to single hadrons and jets was
checked and found to be within the uncertainty.
The TileCal also demonstrated very good time resolution, below 1\,ns
for cell energy deposits above a few GeV.

Overall, the TileCal performed in accord with expectations during LHC Run~1.
Together with other ATLAS subdetectors it contributed 
to the excellent measurement of jets, $\tau$-leptons and missing transverse 
momentum, which are essential for many physics analyses including the
Higgs boson discovery and various
searches for new physics phenomena. After the successful completion of Run~1, 
extensive detector maintenance was
performed and several improvements were introduced in order to
assure TileCal's readiness for challenges imposed by Run~2 at the LHC.

\section*{Acknowledgements}

We thank CERN for the very successful operation of the LHC, as well as the
support staff from our institutions without whom ATLAS could not be
operated efficiently.

We acknowledge the support of ANPCyT, Argentina; YerPhI, Armenia; ARC, Australia; BMWFW and FWF, Austria; ANAS, Azerbaijan; SSTC, Belarus; CNPq and FAPESP, Brazil; NSERC, NRC and CFI, Canada; CERN; CONICYT, Chile; CAS, MOST and NSFC, China; COLCIENCIAS, Colombia; MSMT CR, MPO CR and VSC CR, Czech Republic; DNRF and DNSRC, Denmark; IN2P3-CNRS, CEA-DRF/IRFU, France; SRNSFG, Georgia; BMBF, HGF, and MPG, Germany; GSRT, Greece; RGC, Hong Kong SAR, China; ISF, I-CORE and Benoziyo Center, Israel; INFN, Italy; MEXT and JSPS, Japan; CNRST, Morocco; NWO, Netherlands; RCN, Norway; MNiSW and NCN, Poland; FCT, Portugal; MNE/IFA, Romania; MES of Russia and NRC KI, Russian Federation; JINR; MESTD, Serbia; MSSR, Slovakia; ARRS and MIZ\v{S}, Slovenia; DST/NRF, South Africa; MINECO, Spain; SRC and Wallenberg Foundation, Sweden; SERI, SNSF and Cantons of Bern and Geneva, Switzerland; MOST, Taiwan; TAEK, Turkey; STFC, United Kingdom; DOE and NSF, United States of America. In addition, individual groups and members have received support from BCKDF, the Canada Council, CANARIE, CRC, Compute Canada, FQRNT, and the Ontario Innovation Trust, Canada; EPLANET, ERC, ERDF, FP7, Horizon 2020 and Marie Sk{\l}odowska-Curie Actions, European Union; Investissements d'Avenir Labex and Idex, ANR, R{\'e}gion Auvergne and Fondation Partager le Savoir, France; DFG and AvH Foundation, Germany; Herakleitos, Thales and Aristeia programmes co-financed by EU-ESF and the Greek NSRF; BSF, GIF and Minerva, Israel; BRF, Norway; CERCA Programme Generalitat de Catalunya, Generalitat Valenciana, Spain; the Royal Society and Leverhulme Trust, United Kingdom.

The crucial computing support from all WLCG partners is acknowledged gratefully, in particular from CERN, the ATLAS Tier-1 facilities at TRIUMF (Canada), NDGF (Denmark, Norway, Sweden), CC-IN2P3 (France), KIT/GridKA (Germany), INFN-CNAF (Italy), NL-T1 (Netherlands), PIC (Spain), ASGC (Taiwan), RAL (UK) and BNL (USA), the Tier-2 facilities worldwide and large non-WLCG resource providers. Major contributors of computing resources are listed in Ref.~\cite{ATL-GEN-PUB-2016-002}.



\printbibliography



\clearpage
 
\begin{flushleft}
{\Large The ATLAS Collaboration}

\bigskip

M.~Aaboud$^\textrm{\scriptsize 34d}$,    
G.~Aad$^\textrm{\scriptsize 99}$,    
B.~Abbott$^\textrm{\scriptsize 125}$,    
J.~Abdallah$^\textrm{\scriptsize 8}$,    
O.~Abdinov$^\textrm{\scriptsize 13,*}$,    
B.~Abeloos$^\textrm{\scriptsize 129}$,    
D.K.~Abhayasinghe$^\textrm{\scriptsize 91}$,    
S.H.~Abidi$^\textrm{\scriptsize 164}$,    
O.S.~AbouZeid$^\textrm{\scriptsize 143}$,    
N.L.~Abraham$^\textrm{\scriptsize 153}$
% ...
\bigskip
\\

% ...
$^{8}$Department of Physics, University of Texas at Arlington, Arlington TX; United States of America.\\
% ...
$^{13}$Institute of Physics, Azerbaijan Academy of Sciences, Baku; Azerbaijan.\\
% ...
$^{34}$$^{(a)}$Facult\'e des Sciences Ain Chock, R\'eseau Universitaire de Physique des Hautes Energies - Universit\'e Hassan II, Casablanca;$^{(b)}$Centre National de l'Energie des Sciences Techniques Nucleaires (CNESTEN), Rabat;$^{(c)}$Facult\'e des Sciences Semlalia, Universit\'e Cadi Ayyad, LPHEA-Marrakech;$^{(d)}$Facult\'e des Sciences, Universit\'e Mohamed Premier and LPTPM, Oujda;$^{(e)}$Facult\'e des sciences, Universit\'e Mohammed V, Rabat; Morocco.\\
% ...
$^{91}$Department of Physics, Royal Holloway University of London, Egham; United Kingdom.\\
% ...
$^{99}$CPPM, Aix-Marseille Universit\'e, CNRS/IN2P3, Marseille; France.\\
% ...
$^{125}$Homer L. Dodge Department of Physics and Astronomy, University of Oklahoma, Norman OK; United States of America.\\
% ...
$^{129}$LAL, Universit\'e Paris-Sud, CNRS/IN2P3, Universit\'e Paris-Saclay, Orsay; France.\\
% ...
$^{143}$Santa Cruz Institute for Particle Physics, University of California Santa Cruz, Santa Cruz CA; United States of America.\\
% ...
$^{153}$Department of Physics and Astronomy, University of Sussex, Brighton; United Kingdom.\\
% ...
$^{164}$Department of Physics, University of Toronto, Toronto ON; Canada.\\
% ...
$^{a}$ Also at  Department of Physics, University of Malaya, Kuala Lumpur; Malaysia.\\
$^{b}$ Also at Borough of Manhattan Community College, City University of New York, NY; United States of America.\\
$^{c}$ Also at California State University, East Bay; United States of America.\\
$^{d}$ Also at Centre for High Performance Computing, CSIR Campus, Rosebank, Cape Town; South Africa.\\
$^{e}$ Also at CERN, Geneva; Switzerland.\\
% ...
$^{*}$ Deceased

\end{flushleft}


\end{document}